\documentclass{article}

\usepackage[english]{babel}

\usepackage[
    letterpaper,
    top=2cm,
    bottom=2cm,
    left=3cm,
    right=3cm
]{geometry}

\usepackage{amsmath}
\usepackage{amssymb}
\usepackage{amsfonts}

\usepackage{graphicx}
\usepackage{subcaption}
\usepackage{float}
\usepackage{adjustbox}
\usepackage{pdflscape}

\usepackage{booktabs}
\usepackage{tabularx}
\usepackage{xltabular}
\usepackage{longtable}
\usepackage{array}
\usepackage{makecell}
\usepackage{multirow}
\usepackage{ragged2e}
\usepackage[table]{xcolor}

\usepackage{enumitem}

\usepackage[most]{tcolorbox}

\usepackage{placeins}

\usepackage[
    colorlinks=true,
    allcolors=blue
]{hyperref}

\graphicspath{{figures/}}

\definecolor{tierbg}{HTML}{F8F3EA}
\definecolor{tierline}{HTML}{E5A85E}
\definecolor{tablehead}{HTML}{E8EEF5}
\definecolor{tableborder}{HTML}{C9D3DF}
\definecolor{textblue}{HTML}{264C73}
\definecolor{cardbg}{HTML}{FBFCFE}

\newcommand{\ratingfour}{
    \textcolor{orange}{\(\star\star\star\star\)}
}

\newcommand{\ratingthree}{
    \textcolor{orange}{\(\star\star\star\)}
}

\newtcolorbox{tierbox}[1]{
    enhanced,
    colback=white,
    colframe=tableborder,
    boxrule=0.7pt,
    arc=2mm,
    left=4pt,
    right=4pt,
    top=8pt,
    bottom=4pt,
    title=\textbf{#1},
    fonttitle=\scriptsize\bfseries,
    coltitle=black,
    attach boxed title to top left={
        xshift=4pt,
        yshift=-2mm
    },
    boxed title style={
        colback=tierbg,
        colframe=tierline,
        boxrule=0.7pt,
        arc=1.5mm,
        left=4pt,
        right=4pt,
        top=2pt,
        bottom=2pt
    }
}

\title{
Brain AI Scientist: An Evidence-Based Multi-Agent System for Translational Neuroimaging in Precision Psychiatry
}

\author{
Zeyun Zhao$^{1,2}$,
Lin Gu$^{3}$,
Ruogu Fang$^{1,2,4,*}$
}

\date{}

\begin{document}

\maketitle

\begin{center}
\small

$^{1}$ Department of Biomedical Engineering,
Vanderbilt University, Nashville, TN, USA 

\vspace{0.25em}

$^{2}$ J. Crayton Pruitt Family Department of Biomedical Engineering,
Herbert Wertheim College of Engineering,
University of Florida, Gainesville, FL, USA

\vspace{0.25em}

$^{3}$ Tohoku University,
Sendai, Japan

\vspace{0.25em}

$^{4}$ Center for Cognitive Aging and Memory,
College of Public Health and Health Professions,
University of Florida, Gainesville, FL, USA

\vspace{0.4em}

$^{*}$ Corresponding author:
Ruogu Fang, Ph.D.;
Email:
\href{mailto:ruogu.fang@vanderbilt.edu}
{ruogu.fang@vanderbilt.edu}

\end{center}


\begin{abstract}
\label{subsec:abstract}

Resting-state functional magnetic resonance imaging (rs-fMRI) has revealed widespread brain-network alterations in psychiatric disorders, yet translating these findings into individualized clinical decision support remains challenging. Here we introduce Brain AI Scientist, an autonomous multi-agent framework that integrates patient-specific fMRI, clinical information, and scientific evidence through retrieval, verification, and prioritization. Guided by Research Domain Criteria and evidence-based medicine, it builds traceable chains linking phenotypes, anatomical abnormalities, biomarkers, mechanisms, interventions, and validation strategies. In a representative autism case, Brain AI Scientist generated 37 anatomical hypotheses, retained 26 after literature verification, and derived 65 evidence-supported biomarkers from 90 unique publications to generate patient-specific treatment and validation guidance. Across 17 autism spectrum disorder cases, it outperformed GPT-3.5 baselines with and without fMRI impression in evidence grounding, personalization, clinical actionability, and biomarker–treatment reasoning. Brain AI Scientist therefore provides an auditable framework for translating patient-specific neuroimaging into individualized clinical decision support for precision psychiatry.

\end{abstract}
\section{Introduction}
\label{sec:introduction}






\begin{figure}[t]
\centering
\includegraphics[width=1.\linewidth]{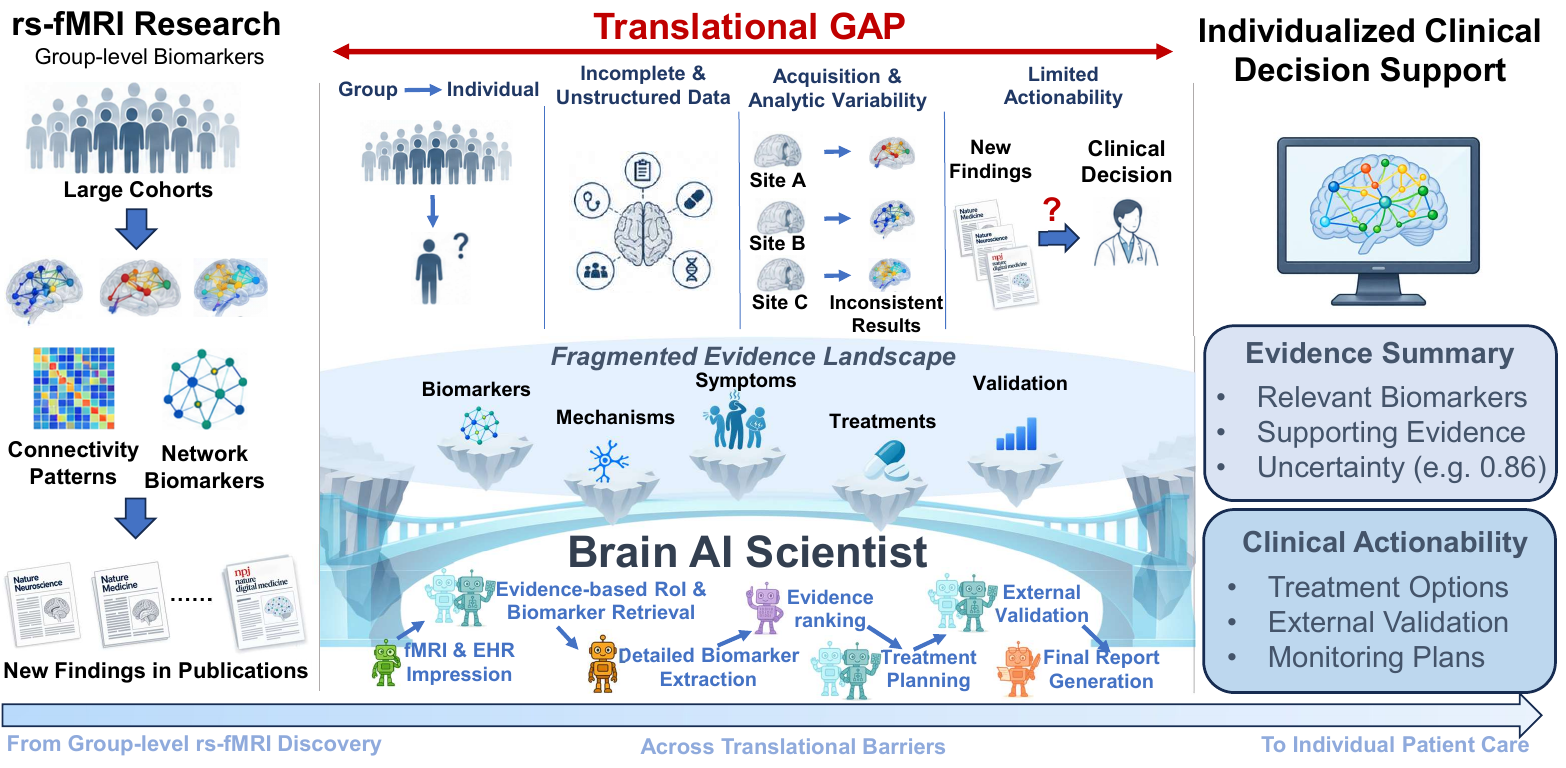}
\caption{This figure illustrates the translational challenges identified in our study, including limited individualization, missing clinical context, acquisition and analytic variability, and limited actionability, and shows how Brain AI Scientist integrates patient-specific fMRI, clinical information, evidence retrieval, treatment-relevant reasoning, and external validation to bridge these gaps.}
\label{fig:challenges}
\end{figure}

Resting-state functional magnetic resonance imaging (rs-fMRI), introduced by Biswal et al. in 1995, is a foundational tool for mapping intrinsic brain networks~\cite{biswal1995functional}. More than 25,000 studies have reported functional connectivity alterations across psychiatric and neurological disorders~\cite{biswal2025history, wei2024analyzing, beauregard1998functional, gur2002fmri, subah2021deep, agosta2012resting, pico2017emotion}, highlighting the potential of fMRI for patient stratification, prognosis, treatment selection, and longitudinal monitoring. The research domain criteria (RDoC) framework promotes a dimensional understanding of mental health and illness by relating variation in fundamental behavioral and biological systems to psychopathology~\cite{insel2010research, cuthbert2013toward}. Building on this perspective, precision psychiatry uses high-dimensional fMRI features for disease subtyping and outcome prediction, while precision neuroscience uses repeated scanning to more reliably characterize individual functional organization~\cite{biswal2025history}. Yet despite this extensive evidence base and methodological progress, fMRI biomarkers rarely inform routine clinical decisions.

To examine this translational gap, we interviewed four stakeholders across academic and clinical institutions, spanning clinical psychiatry, mental-health care delivery, translational psychiatry, and psychiatric neuroimaging, with approximately 9 to $>$30 years of professional experience. Their perspectives identified four interrelated barriers (Figure~\ref{fig:challenges}): difficulty translating group-level findings to individual patients; incomplete, fragmented, and unstructured patient data; acquisition and analytical variability that reduces confidence in imaging results; and technically complex biomarkers that are difficult for non-specialists to interpret or translate into diagnostic, treatment, or follow-up actions. These barriers define an “evidence-island” problem, in which patient-specific information, fMRI findings, mechanisms, symptoms, treatments, and validation remain fragmented across studies. Precision psychiatry therefore requires not only additional biomarkers, but also integration of these evidence sources into patient-specific interpretations that can inform clinical decision support.

Large language models (LLMs) and agentic artificial intelligence (AI) offer new opportunities for integrating complex scientific and clinical evidence. The AI Scientist paradigm shows that LLM-based agents can coordinate literature retrieval, hypothesis generation, tool use, experimentation, analysis, and reporting~\cite{lu2024ai, lu2026towards}, while persistent memory supports knowledge retention across multi-step workflows~\cite{xu2026mem, wang2024agent}. Related systems have been applied to drug discovery, mechanistic reasoning, and multi-agent scientific integration~\cite{ghareeb2026multi,gottweis2026accelerating,li2026brainpilot,aygun2026ai,ghafarollahi2025sciagents,schmidgall2025agent,ferber2026towards,zhou2026brainagent}. Such capabilities are particularly relevant as clinicians face rapidly expanding evidence and neuroscientific findings remain fragmented across domains~\cite{colmers2022keeping}. In psychiatry, LLMs have been used to derive RDoC dimensions from clinical records, link RDoC profiles to MRI biomarkers, support diagnosis and treatment planning, and translate biomarker evidence into therapeutic recommendations~\cite{mccoy2024characterizing,keri2026ai,wang2026domain,zack2026agentic}. However, these approaches generally address individual components rather than integrating patient-specific information, fMRI evidence, biological interpretation, and treatment-relevant knowledge within a unified framework for individualized and interpretable clinical guidance.

We therefore propose Brain AI Scientist, an autonomous multi-agent AI system grounded in evidence-based medicine (EBM) principles~\cite{sackett1996evidence}. The system integrates patient-specific fMRI, clinical information, and scientific evidence to support individualized decision-making in precision psychiatry. Building on the AI Scientist paradigm~\cite{lu2024ai,lu2026towards} and RDoC principles~\cite{insel2010research}, it organizes evidence along a patient profile–RoI–biomarker–mechanism–treatment–validation chain. In a representative pediatric autism case, Brain AI Scientist identified reduced dorsal attention and frontoparietal network involvement, generated 37 anatomical hypotheses, retained 26 after literature verification, and derived 65 evidence-supported biomarkers from 90 unique publications. Evidence was then ranked by methodological quality and patient relevance to generate treatment and validation hypotheses linking patient-specific fMRI abnormalities to potential interventions and measurable outcomes.

We evaluated Brain AI Scientist across 17 autism spectrum disorder (ASD) cases as a representative psychiatric neuroimaging use case, benchmarking it against GPT3.5-based baselines with and without patient-specific fMRI impressions. Across agent-, encoder-, and keyword-based Evidence-based AI Patient Guidance Score (EAPGS) evaluations, Brain AI Scientist showed the most consistent gains in overall report quality and biomarker-treatment reasoning, with additional improvements in evidence grounding, clinical actionability, and personalization. Providing fMRI information alone generally improved GPT3.5-generated guidance, but less than the full agentic workflow, indicating that imaging findings are most informative when combined with targeted evidence retrieval, biomarker interpretation, and patient-specific reasoning. Together, these results support Brain AI Scientist as a traceable and auditable framework for translating fMRI findings into individualized decision support for precision psychiatry.

\begin{figure}[!htbp]
    \centering
    \includegraphics[
        width=\linewidth,
        height=.85\textheight,
        keepaspectratio
    ]{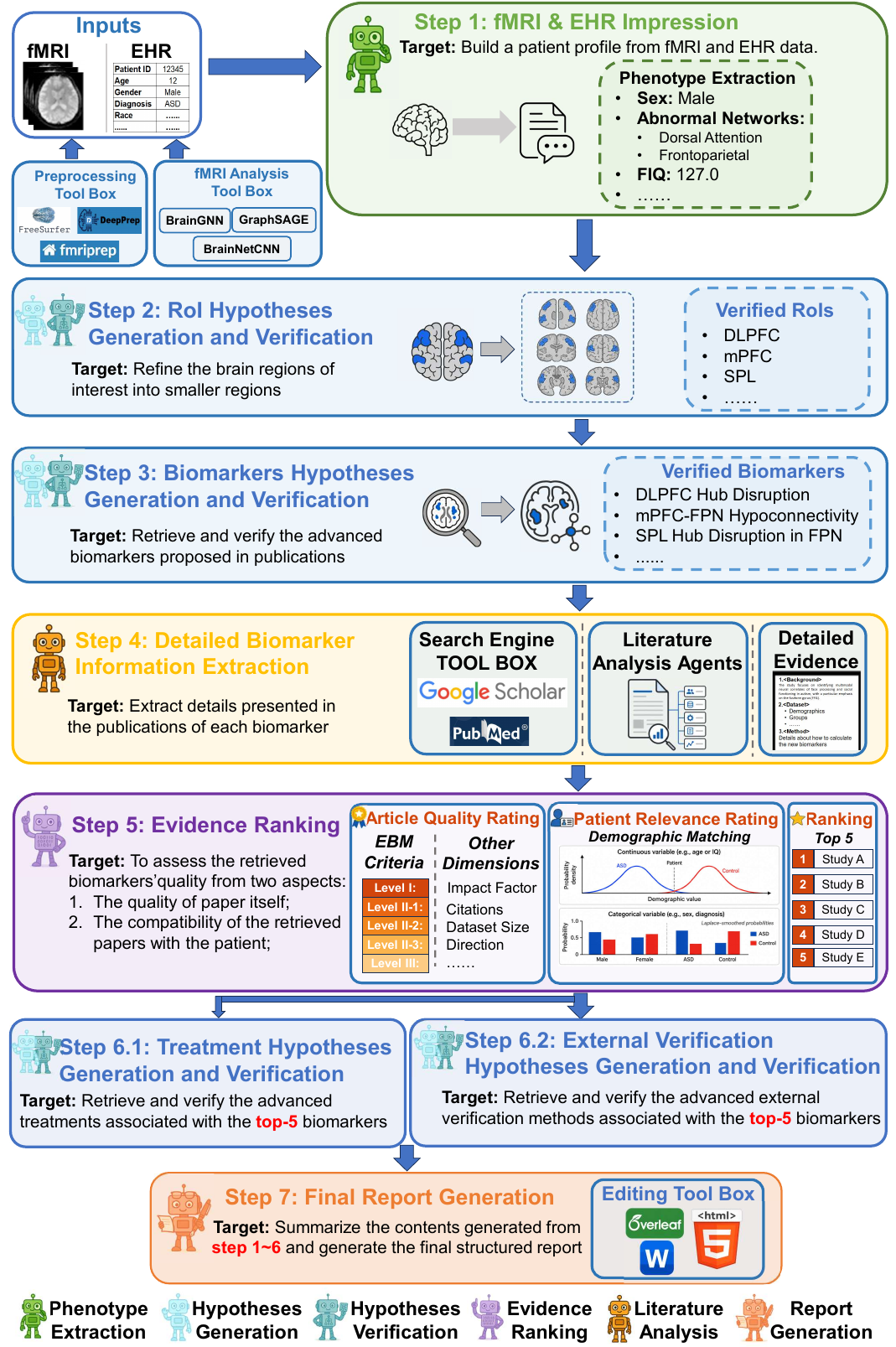}
    \caption{ Overview of the Brain AI Scientist system. The system integrates fMRI and EHR data to generate a patient-specific profile, verified ROI and biomarker hypotheses, ranked literature evidence, treatment hypotheses, and external verification strategies. These outputs are synthesized into an editable, evidence-grounded patient report.}
    \label{fig:overview}
\end{figure}

\section{Results}
\label{sec:results}

\subsection{Stakeholder interviews identified barriers to the clinical translation of fMRI}
\label{sec:stakeholder_interviews}

Although fMRI has revealed important alterations in brain organization across psychiatric and neurological disorders, these findings have rarely translated into routine clinical decision-making. To examine this translational gap, we interviewed four stakeholders across clinical psychiatry, mental-health care delivery, translational psychiatry, and psychiatric neuroimaging, with approximately 9 to $>$30 years of experience. Four interrelated barriers emerged (Figure~\ref{fig:challenges}):

\paragraph{Limited patient-level applicability.}


Group-level fMRI findings provide limited guidance for individual patients because many neuroimaging measures lack sufficient sensitivity and disorder specificity for individual-level diagnosis or treatment planning~\cite{botteron2012consensus,parkes2020towards}. Brain AI Scientist therefore constructs a demographic, clinical, cognitive, and fMRI profile, matches it to populations represented in the literature, and prioritizes evidence from more relevant cohorts rather than directly transferring published biomarkers.

\paragraph{Incomplete and unstructured clinical context.}


fMRI findings are difficult to interpret without clinical context, including symptoms, demographics, cognition, comorbidities, and medication status, particularly given mechanistic heterogeneity in precision psychiatry~\cite{feczko2019heterogeneity,van2024precision}. Because such information is often fragmented, unstructured, incomplete, or privacy-constrained~\cite{jwa2022spectrum}, Brain AI Scientist assigns heterogeneous inputs to modality-specific agents with specialized tools and integrates their outputs into a unified patient representation for subsequent evidence retrieval and reasoning.

\paragraph{Acquisition and analytic variability.}


fMRI has limited test-retest reliability and is sensitive to scan duration, participant state, motion, and analytical choices~\cite{birn2013effect,wang2017improving,luppi2024systematic}. Although site-, scanner-, and protocol-related variability cannot be removed retrospectively, Brain AI Scientist supports standardized workflows such as fMRIPrep and DeepPrep~\cite{esteban2019fmriprep,ren2025deepprep} to reduce avoidable processing variability and standardize quality-control reporting and feature extraction.

\paragraph{Limited clinical actionability.}


Stakeholders questioned the clinical value of fMRI when its cost and complexity do not yield actionable guidance~\cite{miranda2021systematic,nour2022functional}. Biomarker studies typically identify disease-associated networks~\cite{vidya2025identification,kidwell2024resting,deshpande2013identification,yahata2016small,mellema2022reproducible}, whereas intervention studies evaluate behavioral, pharmacological, or neuromodulatory treatments without directly linking effects to patient-specific fMRI abnormalities~\cite{dawson2010randomized,green2010parent,gengoux2019pivotal,liu2025transcranial,wright2025autism}. Brain AI Scientist bridges these isolated evidence streams by linking patient-specific fMRI abnormalities and candidate biomarkers to mechanistic, treatment, and validation evidence, creating a traceable biomarker-to-intervention pathway that supports treatment hypotheses, additional assessments, and external validation.

\subsection{Brain AI Scientist: A Multi-Agent System for Translational Neuroimaging}
\label{sec:brain_ai_scientist}


Brain AI Scientist operationalizes EBM as a multi-agent system transforming patient-specific fMRI and electronic health record (EHR) data into individualized, evidence-grounded biomarker, treatment, and external-validation hypotheses (Figure~\ref{fig:overview}). It integrates multimodal data into a unified clinical-imaging profile, then refines network-level abnormalities into candidate region of interest (ROI) and biomarkers through iterative hypothesis generation, literature retrieval, and verification. Supporting studies are prioritized by methodological quality and patient relevance to inform downstream treatment and validation hypotheses, while preserving links among patient findings, reasoning, and supporting evidence in a structured, traceable report.

\begin{table*}[t]
\centering
\caption{Clinical profile of a patient used in the case study.}
\label{tab:case_study_patient}
\footnotesize
\setlength{\tabcolsep}{3.5pt}
\renewcommand{\arraystretch}{1.15}

\begin{tabular*}{\textwidth}{@{\extracolsep{\fill}}llllll}
\toprule
\multicolumn{6}{@{}l}{\bfseries Demographic and diagnostic information} \\
\midrule
\textbf{Subject ID} & \textbf{Age at scan} & \textbf{Sex} & \textbf{Handedness} & \textbf{Eye status} & \textbf{Diagnosis group} \\
\textit{51169} & \textit{10 years} & \textit{Male} & \textit{Left-handed} & \textit{Eyes closed} & \textit{Autism} \\
\bottomrule
\end{tabular*}

\vspace{0.45em}

\begin{tabular*}{\textwidth}{@{\extracolsep{\fill}}lllllll}
\toprule
\multicolumn{7}{@{}l}{\bfseries Cognitive and medication profile} \\
\midrule
\textbf{FIQ} & \textbf{VIQ} & \textbf{PIQ} & \textbf{FIQ test} & \textbf{VIQ test} & \textbf{PIQ test} & \textbf{Medication status} \\
\textit{127.0} & \textit{131.0} & \textit{117.0} & \textit{WASI} & \textit{WASI} & \textit{WASI} & \textit{0} \\
\bottomrule
\end{tabular*}

\vspace{0.45em}

\begin{tabular*}{\textwidth}{@{\extracolsep{\fill}}lllllll}
\toprule
\multicolumn{7}{@{}l}{\bfseries ADOS assessment} \\
\midrule
\textbf{\makecell[l]{ADOS\\module}} &
\textbf{\makecell[l]{ADOS-G\\total}} &
\textbf{\makecell[l]{ADOS-G\\communication}} &
\textbf{\makecell[l]{ADOS-G\\social}} &
\textbf{\makecell[l]{ADOS-2\\total}} &
\textbf{\makecell[l]{ADOS-2\\severity}} &
\textbf{\makecell[l]{ADOS\\reliable}} \\
\textit{3.0} & \textit{10.0} & \textit{2.0} & \textit{8.0} & \textit{9.0} & \textit{6.0} & \textit{Yes} \\
\bottomrule
\end{tabular*}

\vspace{0.45em}

\begin{tabular*}{\textwidth}{@{\extracolsep{\fill}}llll}
\toprule
\multicolumn{4}{@{}l}{\bfseries ADI-R assessment} \\
\midrule
\textbf{\makecell[l]{ADI-R social total A}} &
\textbf{\makecell[l]{ADI-R verbal total BV}} &
\textbf{\makecell[l]{ADI-R onset total D}} &
\textbf{\makecell[l]{ADI-R reliable}} \\
\textit{23.0} & \textit{18.0} & \textit{3.0} & \textit{Yes} \\
\bottomrule
\end{tabular*}

\end{table*}


\subsubsection{Brain AI Scientist integrates multimodal data into individualized patient profiles}
\label{sec:profile_generation}

Brain AI Scientist first integrates each patient’s fMRI and clinical information into a multimodal profile capturing the clinical phenotype and principal network-level abnormalities (Figure~\ref{fig:overview}, Step~1). Standardized preprocessing, deep-learning, and graph-based analyses identify abnormal functional networks and connectivity patterns from fMRI~\cite{fischl2012freesurfer,ren2025deepprep,esteban2019fmriprep,li2021braingnn,hamilton2017inductive,kawahara2017brainnetcnn}, while demographic, diagnostic, cognitive, behavioral, comorbidity, and medication information is extracted into a structured clinical profile.

\begin{figure*}[!p]
\centering
\scriptsize

\begin{minipage}{0.98\textwidth}
\textbf{a}\par
\vspace{0.2em}

\centering
\includegraphics[width=0.78\textwidth]{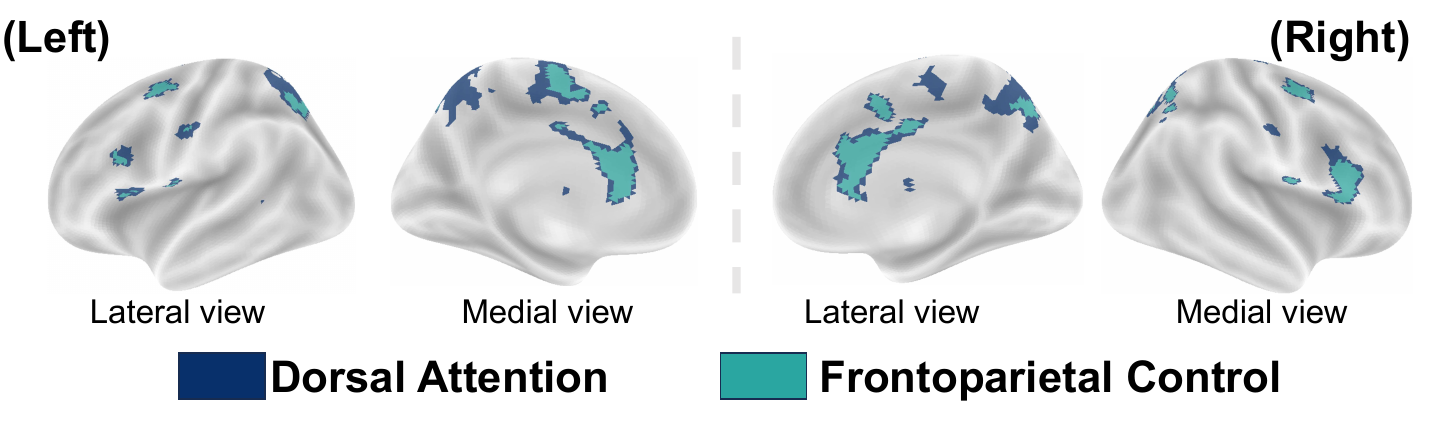}
\end{minipage}

\vspace{0.8em}

\begin{minipage}{0.98\textwidth}
\textbf{b}\par
\vspace{0.3em}

\begin{tierbox}{Tier 1: Preferred options that can be started now}
\rowcolors{2}{white}{cardbg}
\begin{tabularx}{\textwidth}{|
>{\RaggedRight\arraybackslash}p{0.16\textwidth}|
>{\RaggedRight\arraybackslash}p{0.13\textwidth}|
>{\RaggedRight\arraybackslash}X|
>{\RaggedRight\arraybackslash}X|
>{\RaggedRight\arraybackslash}X|}
\hline
\rowcolor{tablehead}
\textbf{\textcolor{textblue}{Treatment}} &
\textbf{\textcolor{textblue}{Rating / Evidence}} &
\textbf{\textcolor{textblue}{Indication Match}} &
\textbf{\textcolor{textblue}{Prescription-Level Parameters}} &
\textbf{\textcolor{textblue}{Outcome Measures}} \\
\hline

Real-time fMRI neurofeedback &
\ratingfour \newline
Moderate evidence from functional connectivity studies. &
Targets disrupted connectivity in ASD and ADHD and aligns with the patient's fMRI impression involving dorsal attention and frontoparietal systems. &
Focus on enhancing DLPFC-precuneus connectivity using clinician-supervised neurofeedback; document baseline symptoms/function, monitor adherence and tolerability, and reassess continuation based on safety and response. &
Functional connectivity changes in DLPFC and precuneus; clinician-defined baseline-to-follow-up improvement in ASD symptoms, social communication, adaptive function, or caregiver/clinician observations. \\

\hline
\end{tabularx}
\end{tierbox}

\vspace{0.6em}

\begin{tierbox}{Tier 2: Parallel or conditional options}
\rowcolors{2}{white}{cardbg}
\begin{tabularx}{\textwidth}{|
>{\RaggedRight\arraybackslash}p{0.16\textwidth}|
>{\RaggedRight\arraybackslash}p{0.13\textwidth}|
>{\RaggedRight\arraybackslash}X|
>{\RaggedRight\arraybackslash}X|
>{\RaggedRight\arraybackslash}X|}
\hline
\rowcolor{tablehead}
\textbf{\textcolor{textblue}{Treatment}} &
\textbf{\textcolor{textblue}{Rating / Evidence}} &
\textbf{\textcolor{textblue}{Indication Match}} &
\textbf{\textcolor{textblue}{Prescription-Level Parameters}} &
\textbf{\textcolor{textblue}{Outcome Measures}} \\
\hline

Video game-based neurofeedback &
\ratingthree \newline
Moderate evidence from EEG-based studies. &
Targets executive-function and attention deficits, conditionally matching the patient's attention/control-network abnormality. &
Use real-time EEG feedback during interactive tasks; confirm feasibility, document baseline symptoms/function, monitor adherence and tolerability, and reassess clinical response. &
ERP components such as P200 or LPP; EEG-derived attention metrics; clinician-defined symptom or functional improvement. \\

\hline
\end{tabularx}
\end{tierbox}

\vspace{0.8em}

\begin{tierbox}{Treatment-Biomarker Causal Chain Alignment}
\rowcolors{2}{white}{cardbg}
\begin{tabularx}{\textwidth}{|
>{\RaggedRight\arraybackslash}p{0.17\textwidth}|
>{\RaggedRight\arraybackslash}p{0.18\textwidth}|
>{\RaggedRight\arraybackslash}X|
>{\RaggedRight\arraybackslash}p{0.18\textwidth}|
>{\RaggedRight\arraybackslash}p{0.18\textwidth}|}
\hline
\rowcolor{tablehead}
\textbf{\textcolor{textblue}{Treatment}} &
\textbf{\textcolor{textblue}{Corresponding Biomarker}} &
\textbf{\textcolor{textblue}{Mechanistic Bridge}} &
\textbf{\textcolor{textblue}{Patient-Specific Evidence}} &
\textbf{\textcolor{textblue}{External Validation / Readout}} \\
\hline

Real-time fMRI neurofeedback &
Functional connectivity in DLPFC and precuneus. &
Enhances disrupted connectivity in ASD and ADHD. &
Not specified; requires clinician review. &
Functional connectivity changes; symptom improvement. \\

\hline

Video game-based neurofeedback &
ERP components and connectivity metrics. &
Improves executive function and attention. &
Not specified; requires clinician review. &
ERP changes; symptom improvement. \\

\hline
\end{tabularx}
\end{tierbox}

\end{minipage}

\caption{
Integrated visualization of patient-specific abnormalities and the corresponding ranked treatment plan.
\textbf{(a)} Reduced functional engagement/connectivity in the dorsal attention and frontoparietal networks, shown as cool-colored surface maps in each hemisphere.
\textbf{(b)} Tiered treatment options and treatment--biomarker causal chain alignment for the case-study patient.
}
\label{fig:patient_visual_treatment}

\end{figure*}

In a representative pediatric autism case (Table~\ref{tab:case_study_patient}), Brain AI Scientist identified reduced functional connectivity in the dorsal attention and frontoparietal control networks (Figure~\ref{fig:patient_visual_treatment}a). These findings were integrated with clinical, cognitive, and behavioral information to guide subsequent ROI and biomarker hypothesis generation.


\subsubsection{Brain AI Scientist generates and verifies patient-specific ROI and biomarker hypotheses}
\label{sec:roi_biomarker_generation}

Conditioned on the patient profile, a hypothesis-generation agent translated network abnormalities into finer-grained anatomical hypotheses, including candidate ROIs and laterality (Figure~\ref{fig:overview}, Step~2). Rather than directly propagating model-generated hypotheses, Brain AI Scientist applied a closed-loop evidence-verification process to reduce unsupported outputs and hallucinations. Following prior evidence-verification frameworks~\cite{lu2024ai,lu2026towards}, the system generated structured queries combining each candidate ROI, affected network, imaging phenotype, and clinical condition, then retrieved relevant publications to assess anatomical and functional consistency with the patient-specific findings. Only evidence-supported hypotheses were retained.

For the case-study patient, reduced involvement of the dorsal attention and frontoparietal networks generated 16 and 21 initial anatomical hypotheses, respectively. Evidence verification retained 10 dorsal-attention and 16 frontoparietal hypotheses. Notably, high model-assigned reliability and accuracy did not ensure retention, which required external evidence.

Verified ROIs were then used to generate more specific neuroimaging biomarker hypotheses (Figure~\ref{fig:overview}, Step~3), including alterations in regional connectivity, hub organization, and network integration. Applying the same evidence-verification procedure~\cite{freeman2025skim,liu2024retrieval} yielded 65 verified biomarkers. Literature retrieval produced 191 reference records, corresponding to 182 biomarker-paper associations after removing duplicate papers retrieved for the same biomarker and 90 unique papers overall. Representative anatomical and biomarker hypotheses, with their supporting evidence, are shown in Table~\ref{tab:hypothesis_generation}; the complete set is provided in \textit{Supplementary Table S1}.

\begin{table*}[!t]
\centering
\caption{
Generation and evidence-based verification of anatomical RoI and neuroimaging biomarker hypotheses for the case-study patient.
\textbf{Bold} text indicates hypotheses supported by retrieved literature and retained after verification.
Representative hypotheses are shown.
}
\label{tab:hypothesis_generation}

\footnotesize
\setlength{\tabcolsep}{4pt}
\renewcommand{\arraystretch}{1.2}

\begin{tabularx}{\textwidth}{
    >{\raggedright\arraybackslash}p{0.16\textwidth}
    !{\vrule width 0.5pt}
    >{\raggedright\arraybackslash}p{0.12\textwidth}
    >{\raggedright\arraybackslash}p{0.22\textwidth}
    >{\raggedright\arraybackslash}p{0.12\textwidth}
    >{\raggedright\arraybackslash}X
}
\toprule
\makecell[l]{\textbf{Patient-specific}\\ \textbf{abnormality}} &
\makecell[l]{\textbf{Candidate}\\ \textbf{network}} &
\makecell[l]{\textbf{Generated anatomical}\\ \textbf{hypotheses}} &
\makecell[l]{\textbf{Verified}\\ \textbf{RoI}} &
\makecell[l]{\textbf{Generated biomarker hypotheses}} \\
\midrule

\makecell[l]{Reduced dorsal\\attention network\\(16 initial; 10\\verified)}
&
Dorsal attention
&
IPS, FEF, SPL, \textbf{MT/V5}~\cite{herrington2007role}, SMA,
\textbf{Precuneus}~\cite{yan2018aberrant},
\textbf{IPL}~\cite{wymbs2021altered}
&
MT/V5
&
Hypoconnectivity between MT/V5 and IPS;
\textbf{Decreased functional connectivity within the dorsal attention network}~
\cite{li2025disrupted,nummenmaa2010connectivity};
\textbf{Altered directional connectivity from MT/V5 to frontal eye fields}~
\cite{zhou2024altered,bi2018abnormal};
Increased local connectivity in MT/V5;
Loss of DAN--DMN anticorrelation
\\[4pt]

&
&
&
Precuneus
&
Hypoconnectivity in DAN;
Loss of DAN--DMN anticorrelation;
Functional connectivity asymmetry in DAN;
Hyperactivity in Precuneus during task engagement;
\textbf{Reduced global efficiency in DAN}~
\cite{zhou2024altered,schmidt2017connectivity,wang2021altered}
\\[4pt]

&
&
&
IPL
&
IPL--DAN hypoconnectivity;
\textbf{IPL hub disruption}~
\cite{zhou2024altered,wang2021effect,anderkova2018theta};
Loss of IPL--DAN anticorrelation;
Reduced task-evoked IPL activation;
\textbf{IPL--DAN connectivity variability}~
\cite{su2023fnirs}
\\

\cmidrule(lr){2-5}

&
Visual
&
LGN, V1, \textbf{V2}~\cite{behrouzi2025ale}
&
V2
&
Reduced V2--V1 connectivity;
Hypoconnectivity within the visual network;
Disrupted V2--dorsal stream connectivity;
\textbf{Impaired functional integration of V2 with higher visual areas}~
\cite{ai2025mesoscale,li2025disrupted,anteraper2020functional};
\textbf{Altered V2 reactivity to visual stimuli}~
\cite{matsuoka2020increased,li2025disrupted}
\\

\midrule

Reduced frontoparietal network
(21 initial; 16 verified)
&
\multicolumn{4}{c}{\(\cdots\)}
\\

\bottomrule
\bottomrule
\end{tabularx}
\end{table*}

\subsubsection{Brain AI Scientist extracts study details and ranks biomarker evidence}
\label{sec:evidence_ranking}

For each verified biomarker, Brain AI Scientist performed full-text analysis of the retrieved studies (Figure~\ref{fig:overview}, Step~4). Instead of using title and abstract only~\cite{punuru2025queryome, hu2025enhancing}, the retrieved articles were downloaded and standardized into Markdown using MinerU~\cite{wang2024mineru}, after which nine specialized agents analyzed complementary sections to extract methodological, experimental, and clinical information. Their outputs were consolidated into eight structured report components: background, dataset, methods, results summary, results explanation, clinical inference, study limitations, and quality evaluation. Further details are provided in Section~\ref{sec:pdf_analysis}, with a complete example in \textit{Supplementary Data 1}. This multi-agent analysis captured quantitative evidence often unavailable from titles or abstracts alone.

In Step~5, extracted evidence was prioritized using a weighted framework balancing methodological credibility and patient relevance. Article quality was assessed using evidence-based medicine criteria, including study design, sample size, methodological rigor, statistical validity, reproducibility, risk of bias, and evidence directness. Patient relevance was evaluated by comparing study populations and experimental settings with the target patient across diagnosis, demographics, cognitive and behavioral characteristics, comorbidities, medication status, disease severity, and fMRI status. These assessments were combined with evidence level, recommendation strength, journal impact-factor score, proportion of significant findings, and clinical applicability into a composite score (\textit{Supplementary Table S2}), with evidence level and quality receiving the greatest weights. The top-k ranked studies were retained as the primary biomarker evidence base for downstream reasoning.

According to \textit{Supplementary Table S2}, all biomarker-discovery studies received a clinical-application score of zero, indicating that clinical translation generally fell outside their primary scope. Brain AI Scientist therefore linked this evidence with treatment-planning and external-validation studies to bridge these otherwise siloed domains and support integrated clinical guidance.

\begin{figure*}[!p]
    \centering


    \begin{subfigure}[t]{0.47\textwidth}
        \centering
        \includegraphics[width=\linewidth]
        {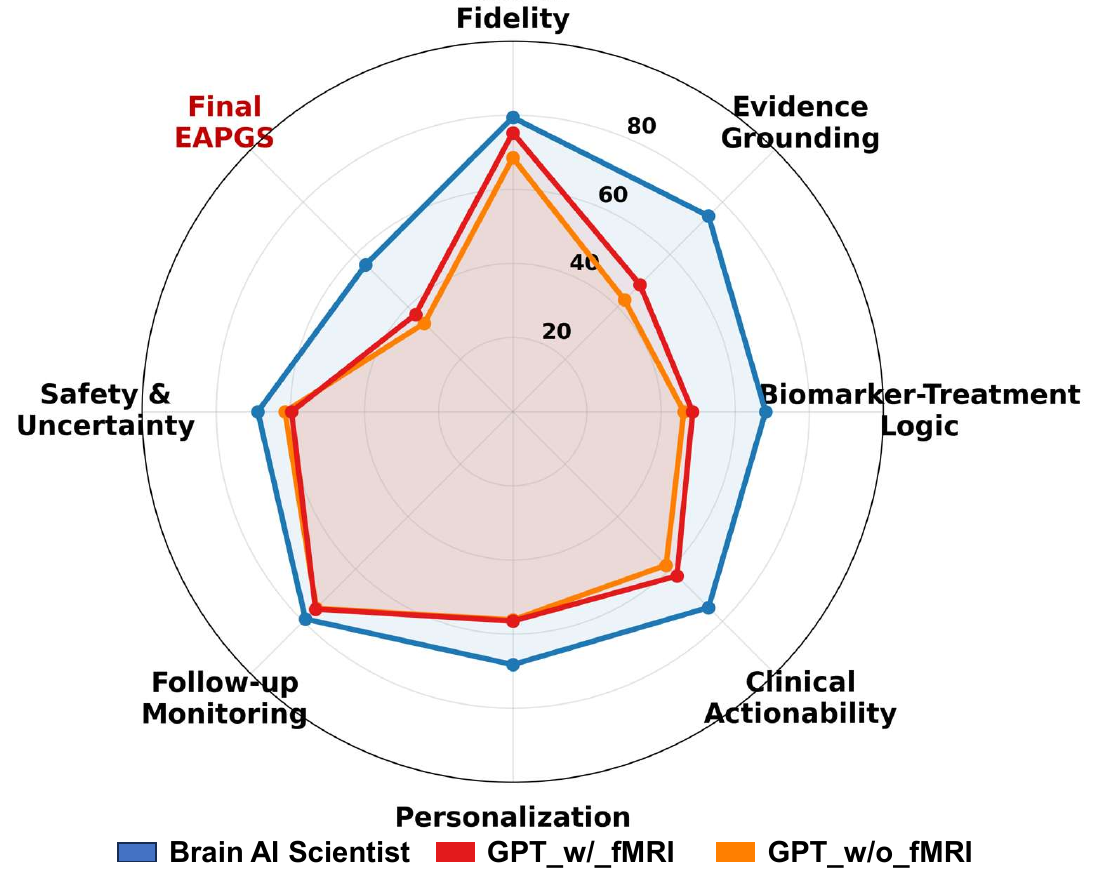}
        \caption{Agent-based evaluation}
        \label{fig:radar_agent}
    \end{subfigure}

    \vspace{0.6em}

    \begin{subfigure}[t]{0.47\textwidth}
        \centering
        \includegraphics[width=\linewidth]
        {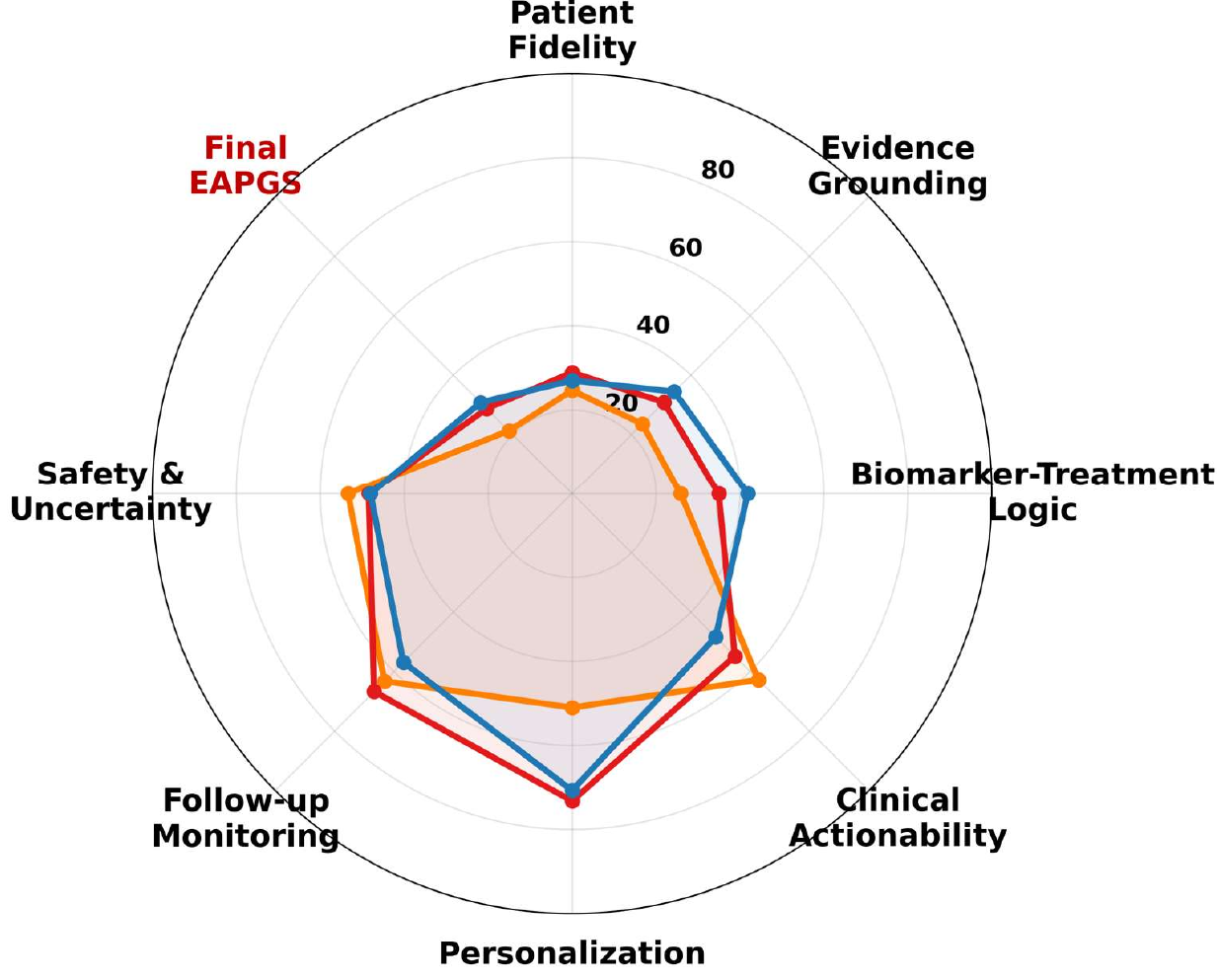}
        \caption{Encoder-based evaluation}
        \label{fig:radar_encoder}
    \end{subfigure}
    \hfill
    \begin{subfigure}[t]{0.47\textwidth}
        \centering
        \includegraphics[width=\linewidth]
        {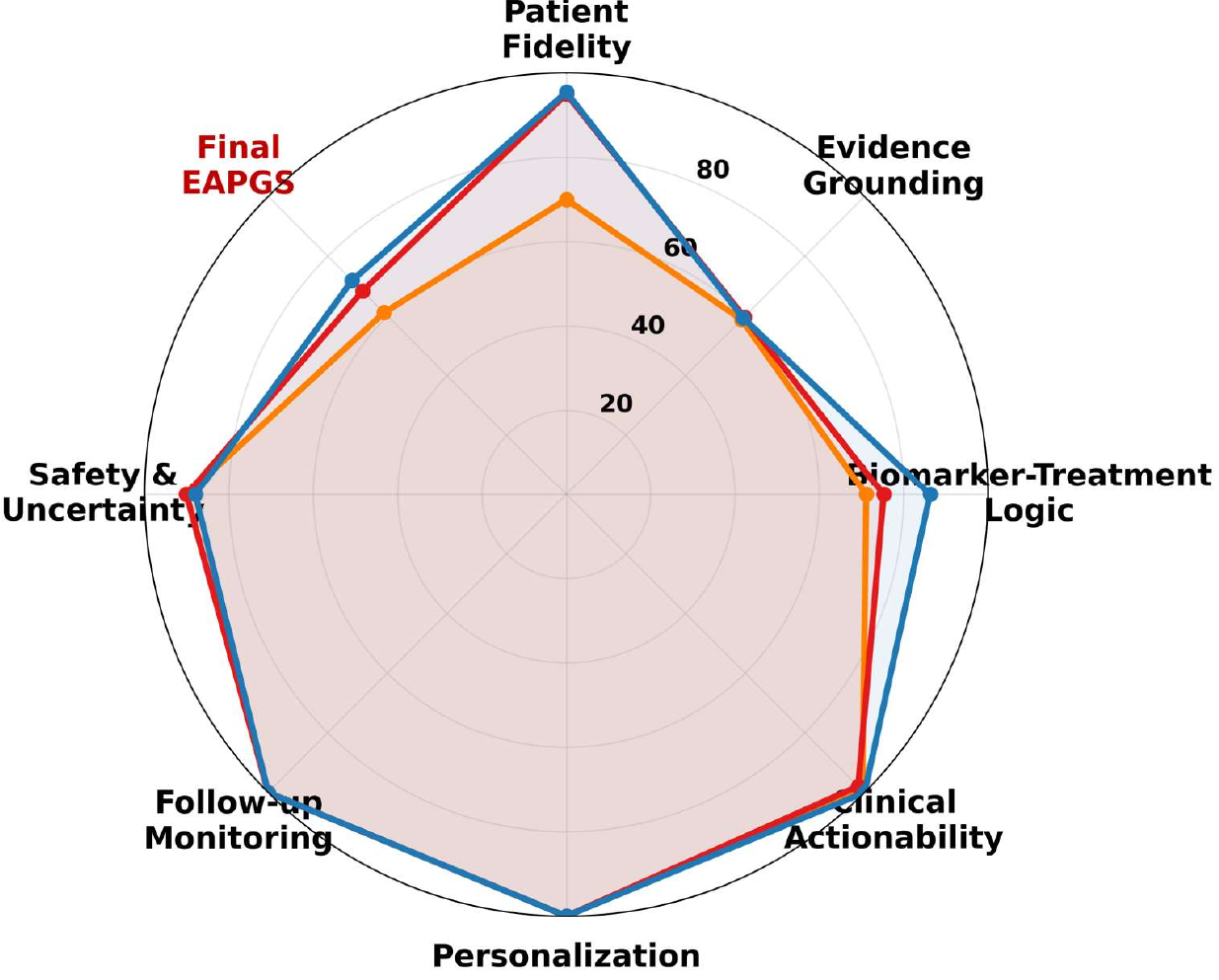}
        \caption{Keyword-based evaluation}
        \label{fig:radar_keywords}
    \end{subfigure}

    \vspace{0.8em}


    \begin{subfigure}[t]{0.95\textwidth}
        \centering
        \includegraphics[width=\linewidth]{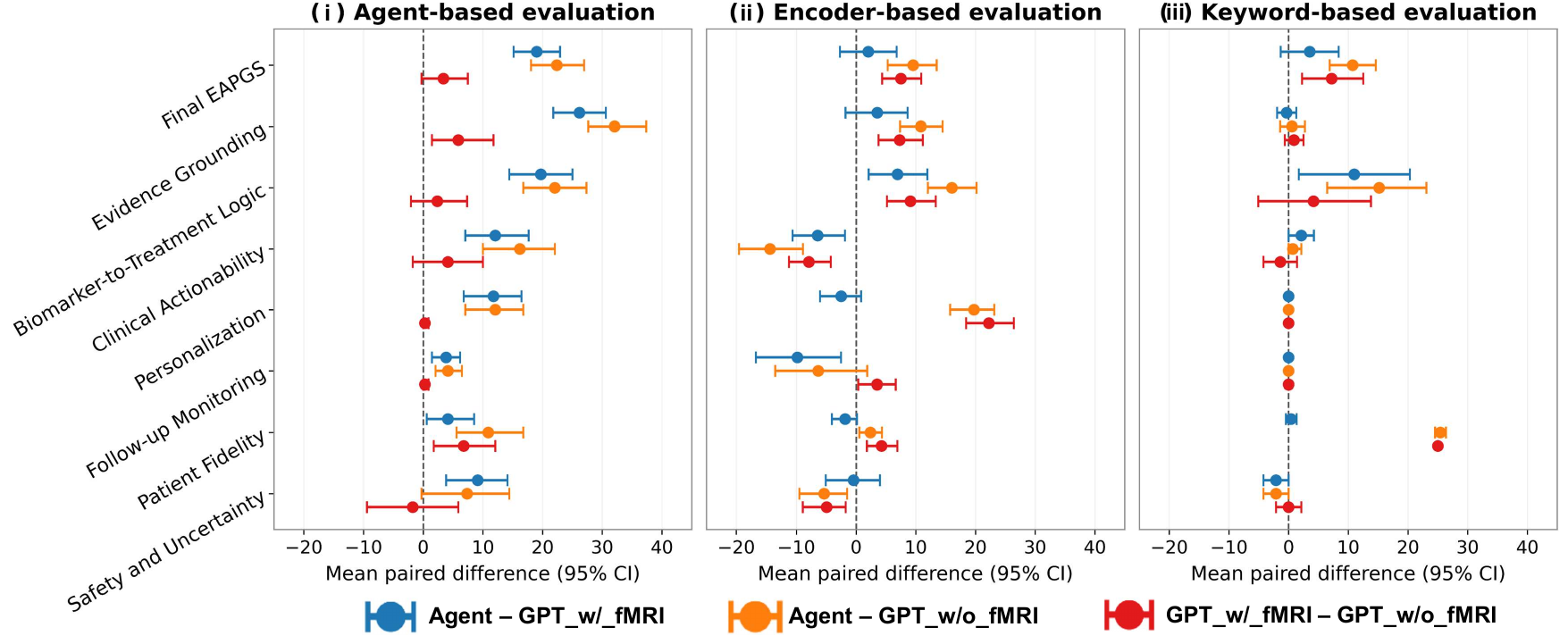}
        \caption{Patient-level paired EAPGS differences}
        \label{fig:combined_eapgs_forest}
    \end{subfigure}

    \caption{
    Comparison of report-generation methods across complementary EAPGS evaluations.
    \textbf{(a--c)} Radar plots compare GPT3.5 without image impressions, GPT3.5 with image impressions, and Brain AI Scientist using agent-, encoder-, and keyword-based evaluation, respectively.
    \textbf{(d)} Patient-level paired EAPGS differences across methods. Points show mean within-patient differences and horizontal bars show bootstrap 95\% confidence intervals. Blue compares Brain AI Scientist with GPT3.5 with image impressions, orange compares Brain AI Scientist with GPT3.5 without image impressions, and red compares GPT3.5 with versus without image impressions. Positive values favor the first method; the dashed line indicates no difference.
    }
    \label{fig:eapgs_combined}
\end{figure*}

\subsubsection{Brain AI Scientist generates biomarker-guided clinical hypotheses}
\label{sec:treatment_external_verification}

Using the verified biomarkers and top-k supporting studies, Brain AI Scientist initiated parallel treatment and external-validation pathways (Figure~\ref{fig:overview}, Steps~6.1 and~6.2). For each biomarker, both pathways integrated biomarker-specific evidence with the patient profile and applied the closed-loop verification framework described in Section~\ref{sec:roi_biomarker_generation}, retaining only literature-supported hypotheses. An example of the resulting patient-specific treatment plan and biomarker-treatment-validation alignment is shown in Figure~\ref{fig:patient_visual_treatment}b.

In the treatment pathway, a specialized agent integrated clinical characteristics, fMRI abnormalities, mechanistic findings, and supporting evidence to propose pharmacological, behavioral, cognitive, neurostimulation, or other interventions for each biomarker-patient pair. Each hypothesis linked the biomarker abnormality to a candidate treatment through an evidence-grounded neurobiological rationale and was verified through literature retrieval. Verified treatments were then organized by evidence strength, indication match, treatment parameters, and outcome measures.

In parallel, the external-validation pathway generated clinically practical strategies for assessing whether each biomarker and linked treatment were appropriate for the patient. Accessible measures included neuropsychological tests, behavioral and clinical assessments, laboratory tests, structural imaging, and repeat fMRI when feasible. Each strategy specified the expected finding, its relationship to the biomarker, and whether it would support or weaken the treatment hypothesis and was subsequently verified and linked to the corresponding biomarker-treatment pair.

\subsubsection{Brain AI Scientist produces traceable, evidence-grounded reports}
\label{sec:report_generation}


Finally, Brain AI Scientist integrated all preceding outputs into a structured patient-specific report and interactive dashboard (Figure~\ref{fig:overview}, Step~7), summarizing the multimodal profile, network abnormalities, verified ROIs, biomarker hypotheses, ranked evidence, biomarker-guided treatments, and external-validation strategies. Each conclusion was linked to relevant patient information, intermediate agent outputs, and supporting literature, creating a traceable path from the original fMRI and clinical data to final decision support. Reports could be exported as editable Word, Overleaf-compatible \LaTeX, or HTML files for evidence inspection, interpretation revision, removal of weakly supported claims, and incorporation of additional clinical information. Representative reports from the three models are provided in \textit{Supplementary Data 2}.

\subsection{Comparative Analysis of EAPGS Evaluation Strategies}

To assess whether generated reports met evidence-based AI guidance requirements, we developed the EAPGS, measuring patient fidelity, evidence grounding, biomarker-treatment logic, clinical actionability, personalization, follow-up monitoring, safety and uncertainty, and an overall final EAPGS score~\cite{bedi2025medhelm}. GPT3.5 without image impressions, GPT3.5 with image impressions, and Brain AI Scientist were evaluated using keyword-, encoder-, and agent-based scoring.

Across all three strategies, performance improved after incorporating patient-specific fMRI information and improved further with Brain AI Scientist. The largest separation occurred in the agent-based evaluation (Figure~\ref{fig:eapgs_combined}a), where Brain AI Scientist achieved the strongest overall performance. Although GPT3.5 with image impressions generally outperformed GPT3.5 without imaging, the gains were smaller than those produced by the full agentic framework, suggesting that performance benefits arose not only from fMRI information but also from specialized biomarker interpretation, targeted evidence retrieval, and patient-specific biomarker-treatment reasoning. Encoder- and keyword-based evaluations showed the same trend (Figure~\ref{fig:eapgs_combined}b,c), despite partial saturation of keyword metrics, with final EAPGS increasing stepwise across the three methods.






\subsubsection{Paired EAPGS Comparisons Across Evaluation Strategies}

To distinguish patient-specific imaging effects from those of the agentic workflow, we compared Brain AI Scientist with GPT3.5 with and without image impressions using paired patient-level EAPGS differences and bootstrap $95\%$ confidence intervals (Figure~\ref{fig:eapgs_combined}d). Under agent-based evaluation, Brain AI Scientist outperformed both baselines across most dimensions, with the largest gains in evidence grounding, biomarker-treatment logic, and clinical actionability (Figure~\ref{fig:combined_eapgs_forest} \textbf{i}). Smaller differences between GPT3.5 with and without imaging indicated that imaging alone did not explain these broader gains.

Encoder- and keyword-based evaluations were consistent, with Brain AI Scientist retaining its clearest advantages in evidence integration and biomarker reasoning (Figure~\ref{fig:combined_eapgs_forest} \textbf{ii}, \textbf{iii}). Encoder-based scoring favored Brain AI Scientist in final EAPGS, evidence grounding, and biomarker-treatment logic, while keyword-based scoring showed the same overall trend. Across both evaluations, imaging primarily improved patient fidelity, whereas the broader gains of Brain AI Scientist were concentrated in evidence grounding and biomarker-treatment reasoning.

Overall, patient-specific fMRI primarily improved patient fidelity, whereas the full agentic workflow produced broader gains in evidence grounding, biomarker-treatment reasoning, and report quality. Brain AI Scientist also showed lower cross-patient report similarity than both GPT3.5 baselines, consistent with greater individualization and stronger incorporation of patient-specific information. Detailed similarity scores and analyses are provided in \textit{Supplementary Section 3}.

\section{Discussion and Conclusion}
\label{sec:discussion}

\subsection{Discussion}
\label{sec:discussion_main}

Despite three decades of fMRI research, relatively few findings have translated into tools for individual patient decision-making~\cite{biswal1995functional,biswal2025history,nour2022functional}. This gap reflects not only limitations in biomarker reliability, but also an “evidence-island” problem in which patient information, fMRI findings, mechanisms, treatments, and validation remain fragmented. Brain AI Scientist addresses this gap by constructing a patient-specific evidence chain linking fMRI abnormalities to literature-supported biomarkers, mechanisms, treatment hypotheses, and validation strategies.

The framework integrates complementary principles from RDoC, evidence-based medicine (EBM), and precision neuroscience~\cite{insel2010research,cuthbert2013toward,biswal2025history,sackett1996evidence}. It operationalizes these principles by connecting patient-specific abnormalities to relevant evidence and clinically testable hypotheses. This design directly addresses the four barriers identified in stakeholder interviews: limited individual applicability of group-level findings, fragmented patient information, methodological variability, and limited translation of complex biomarkers into diagnosis, treatment, and follow-up guidance.

A central feature of this approach is that fMRI abnormalities are not interpreted in isolation. Instead, fMRI findings are considered alongside demographic, cognitive, behavioral, and clinical characteristics, which is particularly important given psychiatric heterogeneity and the limited individual applicability of group-derived biomarkers~\cite{feczko2019heterogeneity,parkes2020towards,van2024precision}. Patient-specific abnormalities are subsequently translated into anatomical and biomarker hypotheses that require literature support before downstream propagation. In the representative autism case, literature verification substantially filtered the initial hypotheses and generated an evidence-supported biomarker set. Although evaluation across 17 cases does not establish broad generalizability or clinical validity, it demonstrates the feasibility of constraining model-generated hypotheses through explicit evidence verification.

The treatment stage further illustrates that the translational challenge is not simply a lack of knowledge but its fragmentation across research domains. None of the biomarker-focused studies retrieved in this case explicitly proposed a clinical application, although this observation should not be generalized to the broader fMRI literature. Identifying an altered network alone does not establish treatment relevance or specify how a therapeutic hypothesis should be evaluated~\cite{miranda2021systematic,nour2022functional}. Brain AI Scientist therefore treats biomarker identification as an intermediate step, connecting verified biomarkers to separately retrieved mechanistic and intervention evidence and to measurements that can support or challenge the resulting hypotheses. The outputs are thus evidence-grounded, testable hypotheses for clinician consideration rather than autonomous treatment recommendations~\cite{sokol2025artificial}.

The comparative evaluation further suggests that patient-specific imaging information is most useful when embedded within structured evidence synthesis. Adding fMRI impressions improved GPT-3.5 patient fidelity, whereas Brain AI Scientist showed broader improvements in evidence grounding, biomarker–treatment reasoning, report quality, actionability, and personalization. 

Finally, traceability enables users to reconstruct how hypotheses were retained and to identify weak or conflicting evidence. However, traceability improves auditability rather than guaranteeing correctness: retrieved literature may be incomplete, biased, or heterogeneous, and errors may arise during retrieval or synthesis. The resulting evidence chains therefore require expert review and prospective validation across patients, diagnoses, imaging protocols, and clinical settings.

\subsection{Limitations}
\label{sec:limitations}

Several limitations warrant consideration. First, evidence appraisal and patient-to-evidence matching remain constrained by both the current ranking framework and inconsistent reporting in the literature. The USPSTF hierarchy~\cite{us1989guide} does not fully capture methodological rigor, risk of bias, reproducibility, or neuroimaging-specific clinical relevance, while many studies lack effect sizes, uncertainty estimates, or precise spatial information. Domain-specific appraisal criteria and standardized biomarker reporting could therefore improve evidence evaluation and patient-level matching.

Second, detailed evidence extraction depends on external full-text access, which may limit coverage and reproducibility. Curated domain-specific or in-house full-text knowledge bases could improve retrieval precision, evidence coverage, and extraction consistency.

Third, the system currently supports only a limited set of specialized tool calls, restricting its analytical breadth. Expanding the tool library with domain-specific capabilities developed by clinicians, neuroscientists, imaging specialists, statisticians, and other experts could improve system reliability and extend its translational capacity toward clinically actionable precision psychiatry.

\subsection{Conclusion}
\label{sec:conclusion}

Brain AI Scientist provides an operational framework for translating patient-specific fMRI abnormalities into traceable, evidence-grounded hypotheses relevant to precision psychiatry. By integrating patient context, literature verification, mechanistic reasoning, intervention evidence, and validation strategies, the framework addresses a key gap between biomarker discovery and individualized clinical interpretation. Our findings suggest that the translational value of fMRI depends not only on identifying abnormalities but also on how those abnormalities are connected to relevant evidence through structured and auditable reasoning. Prospective clinical validation will be required to determine whether this approach can ultimately support clinically useful decision-making.

\newpage
\section{Methods}
\label{sec:methods}

Brain AI Scientist comprises seven stages spanning multimodal phenotype integration, ROI and biomarker hypothesis generation and verification, full-text evidence extraction, evidence ranking, treatment and external-validation hypothesis generation, and structured report generation (Figure~\ref{fig:overview}). Overall, these stages construct a traceable patient profile–ROI–biomarker–mechanism–treatment–validation chain.

\subsection{Dataset}





Resting-state fMRI and phenotypic data were obtained from the Autism Brain Imaging Data Exchange II (ABIDE II) dataset~\cite{di2017enhancing}. ASD participants were screened for patient-specific functional abnormalities relative to site-matched healthy controls using network-level measures based on Yeo-7 atlas~\cite{yeo2011organization}. Seventeen participants with detectable rs-fMRI abnormalities were retained for downstream analysis with Brain AI Scientist. The cohort included 15 males and 2 females, with a mean age of $15.49 \pm 6.97$ years (range, 7.53-33.00 years). Mean full-scale IQ (FIQ), verbal IQ (VIQ), and performance IQ (PIQ) were $107.53 \pm 15.79$, $102.56 \pm 18.23$, and $106.75 \pm 17.69$, respectively.

\subsection{fMRI and EHR Impression}

As shown in \textit{Extended Figure 1}, the phenotype-extraction agent integrates patient-specific fMRI and EHR information into a structured multimodal profile. For fMRI processing, the agent selectively calls available neuroimaging tools, including FreeSurfer for cortical reconstruction and anatomical parcellation~\cite{fischl2012freesurfer}, and fMRIPrep and DeepPrep for standardized preprocessing and quality assessment~\cite{esteban2019fmriprep,ren2025deepprep}. Processed data are analyzed within the seven canonical Yeo networks~\cite{yeo2011organization} using statistical reference comparisons or models including BrainGNN, GraphSAGE, and BrainNetCNN~\cite{li2021braingnn,hamilton2017inductive,kawahara2017brainnetcnn}, producing a structured fMRI impression. In parallel, an EHR agent extracts relevant demographic, clinical, cognitive, behavioral, comorbidity, and medication information. These features are integrated with the fMRI impression into a unified patient profile for downstream analysis.

\subsection{Hypothesis Generation and Verification}

Brain AI Scientist decomposes patient-specific reasoning into evidence-linked hypothesis-generation and verification tasks, with biomarkers serving as an intermediate representation connecting multimodal patient features to anatomical regions, mechanisms, external validation measures, and potential interventions.

As shown in \textit{Extended Figure 2}, each generation agent receives the patient-specific fMRI impression, EHR-derived clinical information, and stage-specific inputs and iteratively generates and refines candidate hypotheses through self-reflection. Each candidate is then evaluated by a retrieval-augmented verification agent and classified as reliable, unreliable, or insufficiently supported. Reliable hypotheses are retained with supporting evidence, whereas unsupported hypotheses are discarded. When evidence is insufficient, the agent formulates a targeted literature query, retrieves relevant studies, and re-evaluates the hypothesis for up to three attempts. Only evidence-supported hypotheses are propagated to subsequent stages. This common workflow is applied to ROI identification, biomarker discovery, treatment hypothesis generation, and external verification. General prompts are provided in \textit{Supplementary Section 1.1}.

\subsubsection{Region-of-Interest Hypothesis Generation and Verification}
\label{sec:hypothesis_GV}

For each abnormal functional network identified in Step~1, the agent generates candidate anatomical regions conditioned on the patient-specific fMRI and EHR profile. Candidate ROIs are verified using evidence retrieved from PubMed, Google Scholar, Semantic Scholar, and OpenAlex. Source-specific queries are generated for each hypothesis, and retrieved studies are assessed for relevance and evidential support.

Abstract-level evidence is used for initial screening to efficiently exclude unsupported hypotheses before full-text analysis. Verified ROIs and their supporting studies are subsequently passed to biomarker generation. Representative prompts are provided in \textit{Supplementary Section~1.2}.

\subsubsection{Biomarker Hypothesis Generation and Verification}

For each verified ROI, the framework generates candidate neuroimaging biomarkers conditioned on the anatomical region, associated functional abnormality, and patient-specific fMRI and EHR features. Candidate biomarkers may include alterations in functional connectivity, regional activity, and network organization.

Relevant publications are first retrieved and screened at the abstract level according to biomarker relevance, evidence quality, and concordance with the patient phenotype. Representative prompts and examples are provided in \textit{Supplementary Section 1.3}. The top-$k$ articles then undergo full-text analysis for detailed evidence extraction and methodological appraisal, as described in Section~\ref{sec:pdf_analysis}. Verified biomarkers are subsequently prioritized according to article quality and patient relevance (Section~\ref{sec:ranking}).

\subsubsection{Treatment and External Verification Hypothesis Generation}
\label{sec:step6}

Following Steps~1-5, each verified biomarker is used to generate parallel treatment and external-validation hypotheses (Figure~\ref{fig:overview}). Both pathways integrate biomarker-specific evidence with the patient phenotype and apply the retrieval-augmented verification procedure described above.

\paragraph{Treatment hypothesis generation and verification.}

The treatment agent identifies interventions with published evidence linking the intervention, therapeutic target, or proposed mechanism to the verified biomarker or associated phenotype. For supported hypotheses, intervention studies are further analyzed to extract treatment modality, target, dose or intensity, duration, frequency, eligibility criteria, and reported outcomes. The resulting outputs are presented as patient-conditioned treatment hypotheses rather than definitive clinical recommendations. Representative prompts are provided in \textit{Supplementary Section 1.4}.

\paragraph{External verification method generation and verification.}

In parallel, the framework identifies clinically accessible measures for evaluating whether biomarker-associated clinical or mechanistic findings are concordant with the patient's phenotype. Candidate measures may include clinical scales, behavioral or cognitive assessments, laboratory tests, and electrophysiological recordings. Evidence-supported measures are retained together with the expected biomarker-phenotype relationship and supporting literature, providing clinically interpretable strategies for external validation. Representative prompts are provided in \textit{Supplementary Section 1.5}.


\subsection{Detailed Biomarker Information Extraction}
\label{sec:pdf_analysis}

The preceding stages identify candidate biomarkers, retrieve potentially relevant studies, and rank the retrieved evidence at the abstract level. Abstracts, however, provide limited information about study populations, analytical procedures, quantitative results, and potential clinical implications. We therefore perform full-text analysis of the top-$k$ articles associated with each verified biomarker to construct a structured and patient-relevant evidence representation for subsequent evidence ranking.

As illustrated in \textit{Extended Figure 3}, the workflow begins with the title of each selected article. The corresponding full text is retrieved from an available source and converted from portable document format (PDF) into machine-readable Markdown format using MinerU~\cite{wang2024mineru}. This conversion preserves the document structure required for downstream analysis while enabling the agents to process the article as structured text.

Following prior work demonstrating that divide-and-conquer multi-agent architectures can improve long-document understanding by distributing content across specialized agents and aggregating their outputs~\cite{zhao2024longagent,sun2025docagent}, we adopt a hierarchical workflow for full-text article analysis. A literature reallocator first segments each parsed manuscript into semantically coherent units and routes the relevant content to five specialized extraction agents: Background, Dataset, Method, Result Summary, and Clinical Application. Each agent operates within a predefined scope, enabling focused extraction of factual information, particularly quantitative findings and mechanistic evidence related to newly proposed biomarkers, while reducing information loss associated with long-context processing and separating evidence extraction from subsequent interpretation and synthesis.

The extracted content is then synthesized by two higher-level agents. The Result Explanation agent interprets the reported quantitative findings, including the direction and magnitude of effects, statistical associations, and their implications for the proposed biomarker. The Clinical Inference agent assesses how the reported findings may inform patient stratification, treatment hypothesis generation, external validation, or follow-up assessment. These outputs are presented as evidence-grounded interpretations rather than established causal conclusions or clinical recommendations.

Finally, a quality evaluation agent assesses the article using the outputs of the preceding agents and the criteria described in Section~\ref{sec:article_quality_rating}. The resulting structured report contains eight components: background, dataset characteristics, methods, result summary, result explanation, clinical application, clinical inference, and quality evaluation. The quality assessment further records article quality, evidence quality, and potential risk of bias. Representative prompts are provided in \textit{Supplementary Section 1.6}.

\subsection{Evidence Ranking}
\label{sec:ranking}

After verifying the proposed biomarker hypotheses, we rank the supporting evidence according to EBM principles. The framework evaluates both the methodological rigor of the underlying studies and their applicability to the patient of interest. It consists of two complementary components: (1) an article quality rating, which assesses evidence strength, reliability, and risk of bias, and (2) a patient relevance rating, which evaluates demographic, clinical, and biological applicability.

\subsubsection{Article Quality Rating}
\label{sec:article_quality_rating}

Article quality is assessed using an EBM framework that considers both study design and methodological rigor. Systematic reviews, meta-analyses, and randomized controlled trials (RCTs) are generally assigned greater weight, while observational studies, descriptive reports, and expert opinions receive lower weight because of their greater susceptibility to bias and confounding. We additionally assess randomization, blinding, sample size, completeness of follow-up, statistical validity, and risk of bias.

Evidence is classified using a hierarchy adapted from the U.S. Preventive Services Task Force (USPSTF)~\cite{us1989guide}:

\begin{itemize}
\item \textbf{Level I:} Evidence from at least one well-conducted RCT.
\item \textbf{Level II-1:} Evidence from well-designed controlled trials without randomization.
\item \textbf{Level II-2:} Evidence from well-designed cohort or case-control studies.
\item \textbf{Level II-3:} Evidence from time series or compelling uncontrolled studies.
\item \textbf{Level III:} Evidence from descriptive studies, clinical experience, expert opinion, or consensus.
\end{itemize}

Beyond study-design classification, we evaluate evidence quality, recommendation strength, imprecision, inconsistency, indirectness, publication bias, and quantitative indicators such as sample size, AUC-ROC, and effect estimates. Journal-level metrics are used only as contextual information. The corresponding prompts are provided in \textit{Supplementary Section 1.7}.

\subsubsection{Patient Relevance Rating}

After assessing overall article quality, we retained the top-k highest-ranked studies for patient-specific relevance evaluation. This evaluation was motivated by patient-similarity and evidence-generalizability principles, which link study applicability to the similarity between the study population and the target patient~\cite{stuart2015assessing}. We therefore developed a probabilistic demographic matching framework based on the conditional-independence factorization commonly used in naive Bayes models~\cite{domingos1997optimality}. The framework quantifies the correspondence between each study population and the index patient and derives a demographic matching weight for each article.

Specifically, for each retrieved article $j$, we modeled continuous demographic factors (\textit{e.g.} age, IQ) as independent normal distributions using the reported means and standard deviations in the paper. The density function can be illustrated as Equation~\ref{equ:continue}.

\begin{equation}
\label{equ:continue}
f_{v,j}(x_v^*) = \frac{1}{\sigma_{v,j}\sqrt{2\pi}}
\exp\left(
    -\frac{(x_v^* - \mu_{v,j})^2}{2\sigma_{v,j}^2}
\right).
\end{equation}
where $v$ represents a continuous variable such as age or IQ; $\mu$ and $\sigma$ represent the mean and standard deviation reported in paper $j$.

Categorical demographic variables were modeled using the category proportions reported in each article. For each article $j$ and categorical variable $b$, we defined a finite set of categories $\{1,2,\ldots,K_{b,j}\}$, with corresponding counts $\{n_{1,j}^{(b)},n_{2,j}^{(b)},\ldots,n_{K_{b,j}}^{(b)}\}$. As shown in Equation.~(\ref{equ:category_total}) and~(\ref{equ:category_probability}), we applied symmetric additive smoothing to estimate the category probabilities~\cite{jeffreys1946invariant,krichevsky1981performance}.

\begin{subequations}
\label{equ:category}

\begin{equation}
N_{j}^{(b)}
=
\sum_{k=1}^{K_{b,j}} n_{k,j}^{(b)}.
\label{equ:category_total}
\end{equation}

\begin{equation}
p_{k,j}^{(b)}
=
\frac{n_{k,j}^{(b)}+\alpha}
{N_{j}^{(b)}+\alpha K_{b,j}},
\qquad
k=1,\ldots,K_{b,j}.
\label{equ:category_probability}
\end{equation}
\end{subequations}
where $N_{j}^{(b)}$ denotes the total number of observations reported for categorical variable $b$. We set $\alpha=0.5$, corresponding to a half-count pseudocount for each category, to prevent unobserved categories from being assigned zero probability.




Finally, for a certain patient, the demographic matching weight of article j can be calculated as Equation. \ref{equ:laplace}.

\begin{equation}
\label{equ:laplace}
L_j = \prod_{v \in C} f_{v,j}(x_v^*) \times \prod_{b \in B} p_{k,j}^{(b)}
\end{equation}
where $C$ and $B$ represent the set of considered demographic variables for continuous and categorical factors, respectively. 

Based on $L_j$, the top-k most relevant articles and associated biomarkers will be used for step 6 introduced in section \ref{sec:step6} for personalized treatment planning and generating external verification methods.

\subsection{Report Generation}
From Step 1-6, Brain AI Scientist progressively transforms the patient's clinical profile and neuroimaging findings into an individual-level, evidence-based, and traceable patient profile-RoI-biomarker-mechanism-treatment-validation chain. This process includes phenotype extraction, identification and verification of clinically relevant fMRI biomarkers, appraisal of supporting evidence, biomarker-guided treatment planning, and definition of measurable follow-up outcomes. In step 7, the system synthesizes this evidence chain into a concise personalized clinical report in one of three user-selected output formats: Microsoft Word, LaTeX, or HTML. The corresponding prompts are provided in \textit{Supplementary Section 1.8}. To improve factual grounding, cross-section consistency, readability, and format validity, the report-generation workflow comprises the following stages:

\begin{enumerate}[label=\arabic*)]

    \item \textbf{Evidence-grounded section generation:}
    The verified evidence and structured notes generated in step 1 through 6 are provided to Aider together with a template corresponding to the selected output format. The report is generated section by section, including patient information, ranked treatment strategies, treatment-biomarker mechanistic alignment, treatment efficacy assessment, follow-up planning, and references. Previously generated sections remain in the model context to maintain consistency across the patient profile, biomarker evidence, and treatment recommendations. For each section, Aider receives task-specific writing guidelines and is instructed to rely exclusively on retrieved evidence and verified citations, following the general principle of retrieval-grounded generation~\cite{lewis2020retrieval,menick2022teaching}. Each section undergoes one round of self-reflection during initial drafting~\cite{shinn2023reflexion}. 

    \item \textbf{Traceable citation integration:}
    During biomarker and treatment-hypothesis verification, the system records the supporting articles, evidence snippets, and their corresponding claims. This citation metadata is passed to Aider during report generation, enabling each major clinical or scientific statement to be linked to its supporting source~\cite{menick2022teaching}. References are then formatted according to the user-selected output format, using BibTeX-based citations for LaTeX or structured reference lists for Word and HTML.

    \item \textbf{Cross-section refinement:}
    After completing the initial draft, the system performs a final section-wise self-reflection pass to remove redundancy, improve clarity, and reconcile potential inconsistencies across the patient findings, biomarker interpretations, treatment strategies, and follow-up outcomes~\cite{madaan2023self}. This stage preserves the underlying evidence chain while improving the conciseness and clinical usability of the report.

    \item \textbf{Format-specific validation and export:}
    The refined report is rendered in the user-selected format. Depending on the selected format, LaTeX outputs are compiled and inspected for syntax and citation errors, HTML outputs are checked for structural and rendering issues, and Word outputs are validated for document organization and formatting consistency. Any detected errors are returned to Aider for automatic correction through an iterative feedback-based debugging procedure~\cite{chen2024teaching}. The validated report is then exported in the requested format.

\end{enumerate}

\subsection{Evaluation Metrics}

We evaluated the clinical quality of reports generated by three methods: Brain AI Scientist, GPT3.5 with the patient-specific fMRI impression, and GPT3.5 without the fMRI impression. All methods were evaluated on the same patients using identical scoring procedures.

\subsubsection{Evidence-Based AI Patient Guidance Score}

Clinical quality was assessed using the Evidence-Based AI Patient Guidance Score (EAPGS), a multidimensional framework for evaluating AI-generated clinical guidance~\cite{bedi2025medhelm}. EAPGS comprised seven domains: Patient Fidelity (PF), Evidence Grounding (EG), Biomarker-to-Treatment Logic (BTL), Clinical Actionability (CA), Personalization (PER), Follow-up Monitoring (FM), and Safety and Uncertainty (SU). 

Each report was independently assessed using three complementary evaluation strategies: agent-based detailed article analysis, encoder-based article representation, and keyword-based evidence retrieval. The agent-based strategy performed structured patient- and article-level assessments, with the evaluation prompt provided in \textit{Supplementary Section 1.9}. The three strategies used the same domain definitions and score ranges, with detailed rubric queries, scoring criteria, and score mappings provided in \textit{Supplementary Section 2}. The encoder-based strategy measured semantic correspondence between predefined rubric queries and report or evidence text, with related keywords and semantic prototypes detailed in \textit{Supplementary Section 4}. The keyword-based strategy assessed lexical coverage and local co-occurrence of predefined clinical concepts, with rule-based keywords and sentence-matching criteria provided in \textit{Supplementary Section 5}.

For patient $i$, report-generation method $g$, evaluation strategy $m$, and domain $d$, the domain score was denoted by $S_{i,g,d}^{(m)}$ and scaled to the interval $[0,100]$. The seven domain scores were represented as

\begin{equation}
\label{equ:eapgs_domain_vector}
\mathbf{S}_{i,g}^{(m)}
=
\left(
\begin{array}{c}
S_{i,g,\mathrm{PF}}^{(m)},\;
S_{i,g,\mathrm{EG}}^{(m)},\;
S_{i,g,\mathrm{BTL}}^{(m)},\;
S_{i,g,\mathrm{CA}}^{(m)},\\
S_{i,g,\mathrm{PER}}^{(m)},\;
S_{i,g,\mathrm{FM}}^{(m)},\;
S_{i,g,\mathrm{SU}}^{(m)}
\end{array}
\right).
\end{equation}

The Final EAPGS was calculated as a weighted combination of all seven domains:

\begin{equation}
\label{equ:eapgs_final}
\begin{aligned}
\mathrm{EAPGS}_{i,g}^{(m)}
={}&\,
w_{\mathrm{PF}}S_{i,g,\mathrm{PF}}^{(m)}
+w_{\mathrm{EG}}S_{i,g,\mathrm{EG}}^{(m)}
+w_{\mathrm{BTL}}S_{i,g,\mathrm{BTL}}^{(m)} \\
&+w_{\mathrm{CA}}S_{i,g,\mathrm{CA}}^{(m)}
+w_{\mathrm{PER}}S_{i,g,\mathrm{PER}}^{(m)}
+w_{\mathrm{FM}}S_{i,g,\mathrm{FM}}^{(m)}
+w_{\mathrm{SU}}S_{i,g,\mathrm{SU}}^{(m)}.
\end{aligned}
\end{equation}

The domain weights were non-negative and summed to one:

\begin{equation}
\label{equ:eapgs_weights}
\begin{aligned}
w_{\mathrm{PF}}
+w_{\mathrm{EG}}
+w_{\mathrm{BTL}}
+w_{\mathrm{CA}}
+w_{\mathrm{PER}}
+w_{\mathrm{FM}}
+w_{\mathrm{SU}}
&=1, \\
w_d &\geq 0.
\end{aligned}
\end{equation}

Because all domain scores were bounded between 0 and 100 and the weights summed to one, the Final EAPGS also remained within $[0,100]$. Scores obtained from the three evaluation strategies were analyzed separately and were not averaged across strategies.

For each report-generation method and evaluation strategy, the mean score for domain $d$ and the mean Final EAPGS across $N$ patients were calculated as

\begin{equation}
\label{equ:eapgs_group_mean}
\overline{S}_{g,d}^{(m)}
=
\frac{1}{N}
\sum_{i=1}^{N}
S_{i,g,d}^{(m)},
\qquad
\overline{\mathrm{EAPGS}}_{g}^{(m)}
=
\frac{1}{N}
\sum_{i=1}^{N}
\mathrm{EAPGS}_{i,g}^{(m)}.
\end{equation}

\subsubsection{Patient-Level Paired Comparisons}

Because all report-generation methods were evaluated on the same patients, method comparisons were performed using within-patient paired differences. For report-generation methods $g_1$ and $g_2$, the paired difference for domain $d$ under evaluation strategy $m$ was defined as

\begin{equation}
\label{equ:eapgs_paired_difference}
\Delta_{i,d}^{(m)}(g_1,g_2)
=
S_{i,g_1,d}^{(m)}
-
S_{i,g_2,d}^{(m)}.
\end{equation}

The mean paired difference was calculated as

\begin{equation}
\label{equ:eapgs_mean_paired_difference}
\overline{\Delta}_{d}^{(m)}(g_1,g_2)
=
\frac{1}{N}
\sum_{i=1}^{N}
\Delta_{i,d}^{(m)}(g_1,g_2).
\end{equation}

The same procedure was applied to the final EAPGS. Positive values indicated that the first method in the comparison achieved a higher score than the second method. Two-sided bootstrap 95\% confidence intervals were estimated by resampling patients with replacement, thereby preserving the paired study design.

\section{Data Availability}

Resting-state fMRI and phenotypic data used in this study were obtained from the publicly available Autism Brain Imaging Data Exchange II (ABIDE II) dataset. Access to the original data is subject to the data-use terms and governance requirements established by the ABIDE data providers. In this study, ASD participants were screened for patient-specific functional abnormalities relative to site-matched healthy controls using network-level measures based on the Yeo-7 atlas, and 17 participants with detectable rs-fMRI abnormalities were retained for downstream analysis with Brain AI Scientist. No new clinical or neuroimaging data were collected specifically for this study.

\section{Code Availability}

The code implementing the Brain AI Scientist framework, including the core workflows for patient-specific neuroimaging interpretation, hypothesis generation, literature retrieval, evidence verification, and downstream analysis, will be made publicly available on GitHub upon publication. The repository will include documentation and instructions necessary to reproduce the principal analyses reported in this study.

\newpage
\section{Acknowledgements}

We thank Dr. Randy Hirschtick, Dr. Tshekedi Dennis, Dr. Dana Wang, and Dr. Carol A Mathews for helpful discussions and clinical insights related to this work.

\section{Funding statement}

No funding.

\section*{Author contributions}

Z.Z., L.G. and R.F. conceived and designed the study. Z.Z. developed the Brain AI Scientist framework, implemented the code, conducted the experiments and analyses, and drafted the manuscript. L.G. and R.F. contributed to study conceptualization, supervision, scientific guidance, resource provision and interpretation of the results. R.F. served as the principal investigator and supervised the overall project. All authors reviewed, revised and approved the final manuscript.

\section{Competing interests}

No competing interests.

\section{Extended data}
\setcounter{figure}{0}

\begin{figure}[htbp]
\centering
\includegraphics[width=.7\linewidth]{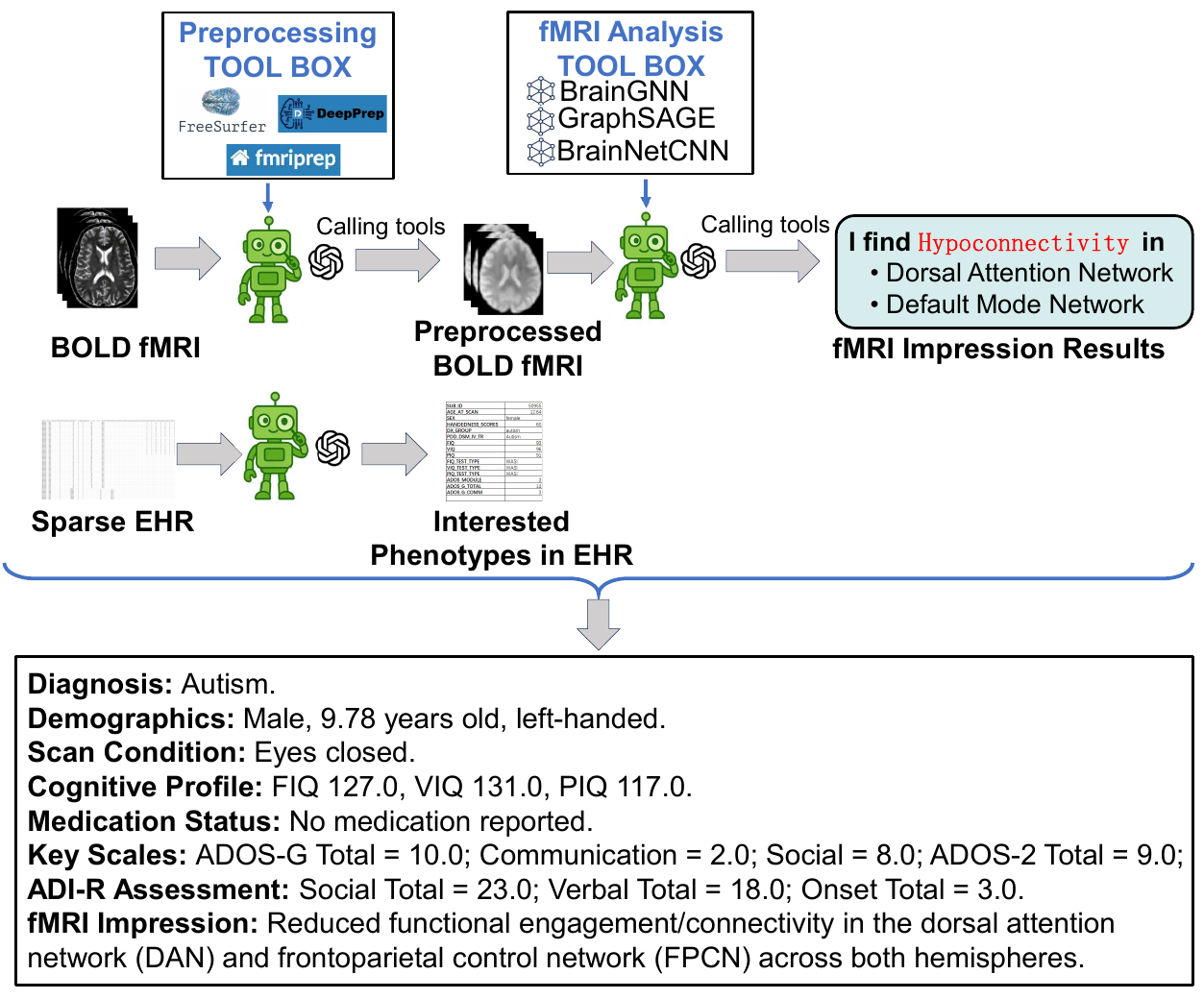}
\caption{Extraction and integration of patient-specific phenotypes from fMRI and EHR data. BOLD fMRI is processed and analyzed using dedicated neuroimaging tools to generate an fMRI impression, while sparse EHR data are parsed to extract clinically relevant phenotypes. These outputs are integrated into a structured patient profile comprising diagnosis, demographics, scan conditions, cognitive measures, medication status, clinical scales, and fMRI abnormalities.}
\label{fig:Impression}
\end{figure}

\begin{figure}[htbp]
\centering
\includegraphics[width=.9\linewidth]{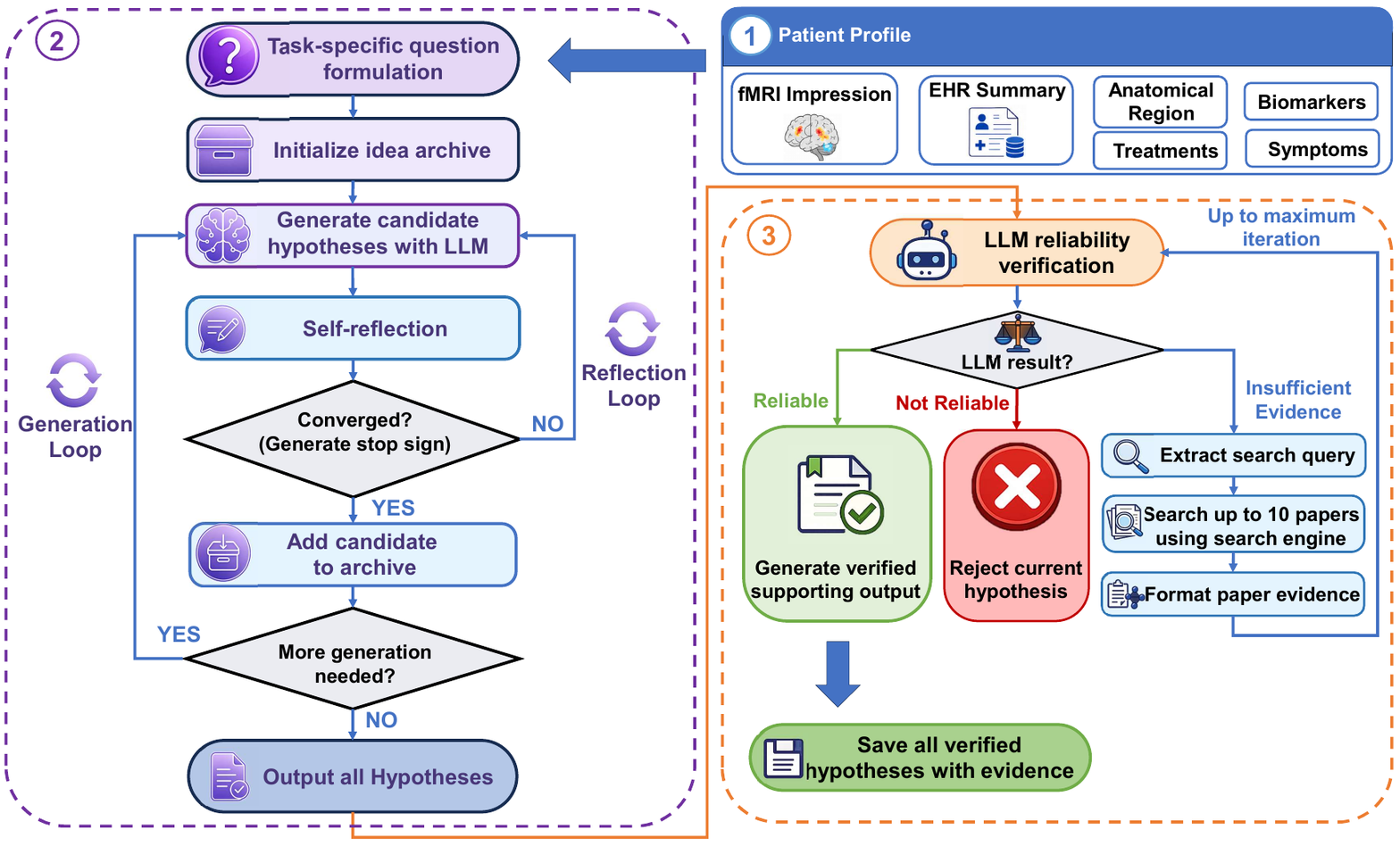}
\caption{Generation and evidence-based verification of anatomical RoI and neuroimaging biomarker hypotheses. Patient-specific inputs guide iterative LLM-based hypothesis generation and self-reflection. Candidate hypotheses are then verified against retrieved literature; supported hypotheses are retained with evidence, unsupported hypotheses are rejected, and insufficient evidence triggers targeted literature retrieval and re-evaluation.}
\label{fig:hypothesis}
\end{figure}

\begin{figure}[htbp]
\centering
\includegraphics[width=.9\linewidth]{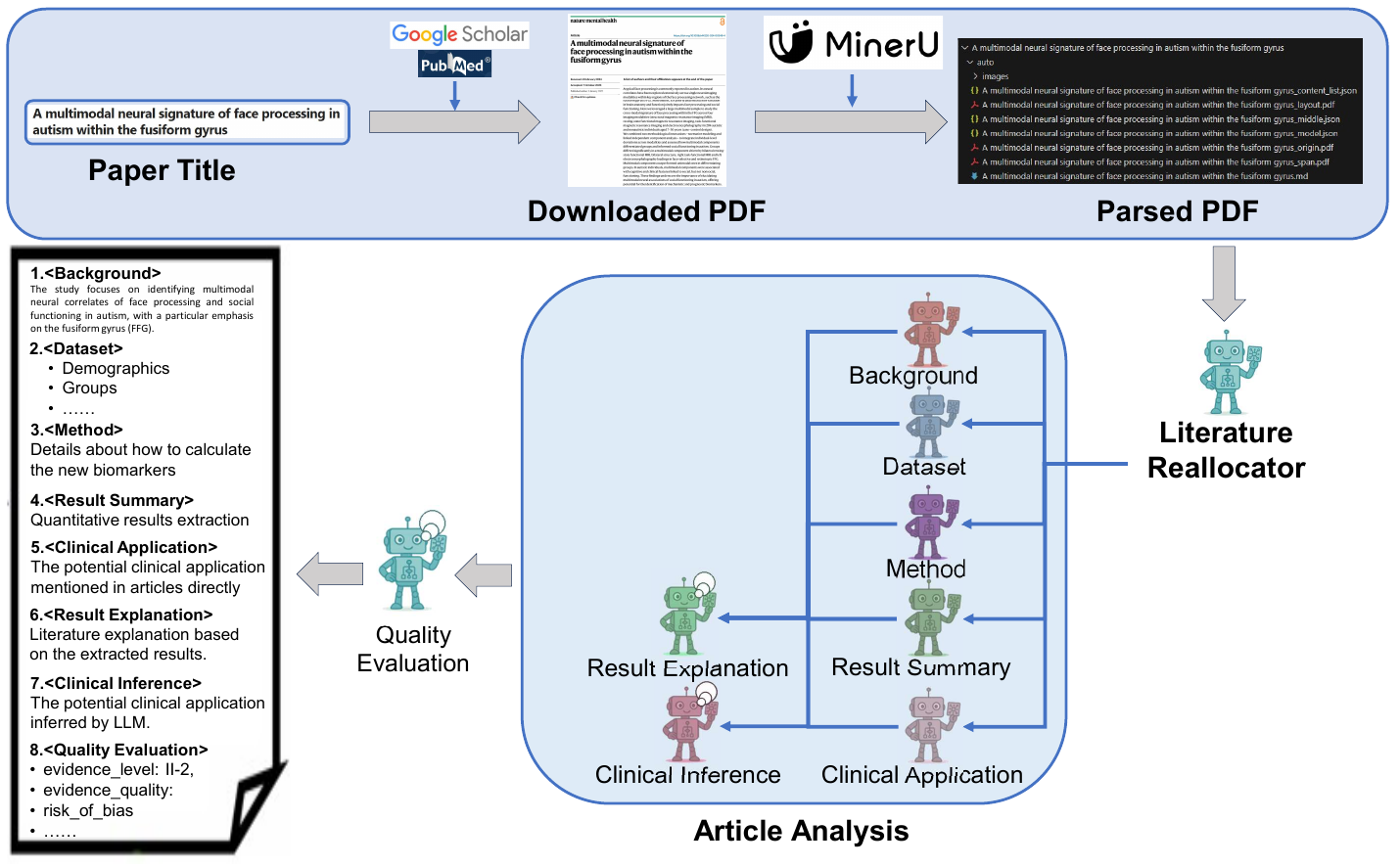}
\caption{Multi-agent workflow for full-text biomarker evidence extraction. Selected articles are retrieved and parsed into structured text using MinerU. A literature-reallocation agent routes article content to specialized extraction agents, whose outputs are synthesized for result interpretation, clinical inference, and study-quality evaluation.}
\label{fig:pdf}
\end{figure}

\section*{Supplementary Tables}


\begin{xltabular}{\textwidth}{
    >{\RaggedRight\arraybackslash}p{0.15\textwidth}
    !{\vrule width 0.4pt}
    >{\RaggedRight\arraybackslash}p{0.11\textwidth}
    >{\RaggedRight\arraybackslash}p{0.20\textwidth}
    >{\RaggedRight\arraybackslash}p{0.15\textwidth}
    >{\RaggedRight\arraybackslash}X
}

\caption{
Generation and evidence-based verification of anatomical RoI and
neuroimaging biomarker hypotheses for the case-study patient.
\textbf{Bold} text indicates hypotheses supported by retrieved literature
and retained after verification.
An em dash indicates that no corresponding biomarker hypotheses were
reported in the biomarker-generation table.
}
\label{tab:hypothesis_generation_complete} \\

\toprule
\textbf{Patient-specific abnormality} &
\textbf{Candidate network} &
\textbf{Generated anatomical hypotheses} &
\textbf{Verified RoI} &
\textbf{Generated biomarker hypotheses} \\
\midrule
\endfirsthead

\multicolumn{5}{l}{\textit{Table \thetable{} continued}}\\

\toprule
\textbf{Patient-specific abnormality} &
\textbf{Candidate network} &
\textbf{Generated anatomical hypotheses} &
\textbf{Verified RoI} &
\textbf{Generated biomarker hypotheses} \\
\midrule
\endhead

\midrule
\multicolumn{5}{r}{\textit{Continued on next page}}\\
\endfoot

\bottomrule
\endlastfoot


Reduced dorsal attention network
(16 initial; 10 verified)
&
Dorsal attention
&
IPS, FEF, SPL,
\textbf{MT/V5}~\cite{herrington2007role},
SMA,
\textbf{Precuneus}~\cite{yan2018aberrant},
\textbf{IPL}~\cite{wymbs2021altered}
&
Middle Temporal Area (MT/V5)
&
Hypoconnectivity between MT/V5 and Intraparietal Sulcus;
\textbf{Decreased functional connectivity within the Dorsal Attention Network}~
\cite{li2025disrupted,nummenmaa2010connectivity};
\textbf{Altered directional connectivity from MT/V5 to Frontal Eye Fields}~
\cite{zhou2024altered,bi2018abnormal,nummenmaa2010connectivity};
Increased local connectivity (hyperconnectivity) in MT/V5;
Loss of anticorrelation between DAN and Default Mode Network (DMN)
\\[4pt]

&
&
&
Precuneus
&
Hypoconnectivity in DAN;
Loss of anticorrelation between DAN and Default Mode Network (DMN);
Functional connectivity asymmetry in DAN;
Hyperactivity in Precuneus within DAN during task engagement;
\textbf{Reduced global efficiency in DAN network}~
\cite{zhou2024altered,schmidt2017connectivity,wang2021altered}
\\[4pt]

&
&
&
Inferior Parietal Lobule (IPL)
&
IPL-DAN Hypoconnectivity;
\textbf{IPL Hub Disruption}~
\cite{zhou2024altered,wang2021effect,anderkova2018theta};
Loss of IPL-DAN Anti-correlation;
Reduced Task-Evoked IPL Activation;
\textbf{IPL-DAN Connectivity Variability}~
\cite{su2023fnirs,su2020differences,wang2021altered}
\\[4pt]


\cmidrule(lr){2-5}

&
Visual
&
LGN, V1,
\textbf{V2}~\cite{behrouzi2025ale}
&
Secondary Visual Cortex (V2)
&
Reduced V2 Connectivity with V1;
Hypoconnectivity within the Visual Network;
Disrupted V2-Dorsal Stream Connectivity;
\textbf{Impaired Functional Integration of V2 with Higher Visual Areas}~
\cite{ai2025mesoscale,li2025disrupted,anteraper2020functional};
\textbf{Altered V2 Reactivity to Visual Stimuli}~
\cite{matsuoka2020increased,li2025disrupted,anteraper2020functional}
\\[4pt]


\cmidrule(lr){2-5}

&
Limbic
&
\textbf{ACC}~\cite{zhou2016functional},
\textbf{Hippocampus}~\cite{li2025hippocampal},
\textbf{Amygdala}~\cite{rausch2016altered}
&
ACC
&
---
\\[4pt]

&
&
&
Hippocampus
&
---
\\[4pt]

&
&
&
Amygdala
&
---
\\[4pt]


\cmidrule(lr){2-5}

&
Frontoparietal
&
\textbf{DLPFC}~\cite{christakou2013disorder},
\textbf{IFG}~\cite{kim2014abnormal},
\textbf{PPC}~\cite{shafritz2008neural}
&
Dorsolateral Prefrontal Cortex (DLPFC)
&
\textbf{DLPFC-Frontoparietal Connectivity}~
\cite{Vasantha2022DiscoveryON};
\textbf{DLPFC Activation during Task Switching}~
\cite{Haweel2021ACD,Haweel2021AND,Jiang2024TaskswitchingMU};
\textbf{DLPFC Hub Disruption}~
\cite{Vasantha2022DiscoveryON,Parmigiani2025FromDP};
Reduced DLPFC-PPC Anticorrelation;
Dynamic Functional Connectivity Variability
\\[4pt]

&
&
&
Inferior Frontal Gyrus (IFG)
&
Reduced IFG--DLPFC Connectivity;
IFG Activation During Inhibitory Control Tasks;
\textbf{Disrupted IFG--PPC Functional Connectivity}~
\cite{May2020FrontoparietalNI,Liu2025MappingTF,Navaei2025ComparingFA};
IFG Hub Disruption;
\textbf{Reduced IFG--ACC Connectivity}~
\cite{Wang2025FunctionalCA,Kato2024ReducedRF}
\\[4pt]

&
&
&
Posterior Parietal Cortex (PPC)
&
\textbf{PPC--DLPFC Hypoconnectivity}~
\cite{Guan2024DualViewCA,Zhu2022AlteredIB};
\textbf{PPC Hub Disruption}~
\cite{Libovner2020RepeatedET,Shafieizadegan2024InvestigatingEC};
\textbf{PPC--MPFC Functional Connectivity}~
\cite{Navaei2025ComparingFA,Shafieizadegan2024InvestigatingEC,Alves2024MulticlassCO};
\textbf{PPC Activation during Cognitive Tasks}~
\cite{May2020FrontoparietalNI,Jigo2018NeuralDO,Libovner2020RepeatedET};
\textbf{PPC--Precuneus Connectivity}~
\cite{Xiao2023AtypicalFC,Ye2025ReducedLF,Polk2021DisruptedFC}
\\[4pt]


\midrule

Reduced frontoparietal network
(21 initial; 16 verified)
&
Frontoparietal
&
\textbf{DLPFC}~\cite{chantiluke2015disorder},
\textbf{IPL}~\cite{wymbs2021altered},
\textbf{ACC}~\cite{zhou2016functional},
\textbf{SPL}~\cite{travers2015motor},
\textbf{mPFC}~\cite{chantiluke2015inverse},
\textbf{PPC}~\cite{shafritz2008neural},
\textbf{VLPFC}~\cite{pitskel2014neural},
\textbf{Precuneus}~\cite{yan2018aberrant},
\textbf{Precentral Gyrus}~\cite{nebel2014precentral},
\textbf{Supramarginal Gyrus}~\cite{salmi2013brains},
\textbf{MFG}~\cite{lepping2022visuomotor},
\textbf{IPS}~\cite{massaccesi2021neural},
OFC
&
Dorsolateral Prefrontal Cortex (DLPFC)
&
\textbf{DLPFC-Frontoparietal Hypoconnectivity}~
\cite{May2020FrontoparietalNI,Fatemi2024QuantitativePO,Weerasekera2022DorsolateralPC};
\textbf{DLPFC Hub Disruption}~
\cite{Vasantha2022DiscoveryON,Aydin2025AssessingTI,Fatemi2024QuantitativePO,
Moxon-Emre2021ModulationOD,Khadem-Reza2023AlteredRF};
\textbf{Frontoparietal Network Anticorrelation Loss}~
\cite{Vasantha2022DiscoveryON,Fatemi2024QuantitativePO,
Polk2021DisruptedFC,Aydin2025AssessingTI};
\textbf{DLPFC-Frontoparietal Network Synchronization Deficit}~
\cite{May2020FrontoparietalNI,Vasantha2022DiscoveryON};
\textbf{DLPFC-Frontoparietal Connectivity Variability}~
\cite{Vasantha2022DiscoveryON,May2020FrontoparietalNI,
Saha2023EigenvectorCC,Polk2021DisruptedFC}
\\[4pt]

&
&
&
Inferior Parietal Lobule (IPL)
&
\textbf{Frontoparietal Hypoconnectivity}~
\cite{May2020FrontoparietalNI,Saha2023EigenvectorCC};
IPL-Default Mode Network Hyperconnectivity;
Task-Induced IPL Deactivation Deficits;
Graph-theory Based Hub Disruption;
\textbf{Dynamic Functional Connectivity Variability}~
\cite{May2020FrontoparietalNI,Lee2017AutismSD}
\\[4pt]

&
&
&
Anterior Cingulate Cortex (ACC)
&
Frontoparietal Hypoconnectivity;
ACC Hyperactivation;
\textbf{Disrupted ACC-DLPFC Coupling}~
\cite{Huang2025TheFC,Zilverstand2017fMRINT,Liu2016BetelQD};
\textbf{ACC Hub Disruption}~
\cite{Dawson2023SexualDI,Liu2016BetelQD,Guo2019AnteriorCC,Miguel2025DistinctiveOO};
Loss of ACC-FPN Anticorrelation
\\[4pt]

&
&
&
Superior Parietal Lobule (SPL)
&
Reduced SPL-DLPFC Connectivity;
Aberrant SPL Activation During Cognitive Tasks;
Decreased SPL Functional Connectivity Variability;
\textbf{Disrupted SPL Hub Function}~
\cite{Lepping2021VisuomotorBN,May2020FrontoparietalNI};
Altered SPL-DMN Anticorrelation
\\[4pt]

&
&
&
Medial Prefrontal Cortex (mPFC)
&
Frontoparietal Hypoconnectivity;
\textbf{mPFC Hub Disruption}~
\cite{May2020FrontoparietalNI,Lepping2021VisuomotorBN,
cong2023altered,shafieizadegan2025age};
\textbf{Frontoparietal Anti-correlation Loss}~
\cite{May2020FrontoparietalNI,Vasantha2022DiscoveryON,Yoon2022AlterationsIS};
\textbf{Increased mPFC-Insula Connectivity}~
\cite{May2020FrontoparietalNI,Margolis2019SalienceNC,cong2023altered};
\textbf{Reduced mPFC-DLPFC Connectivity}~
\cite{Vasantha2022DiscoveryON,May2020FrontoparietalNI}
\\[4pt]

&
&
&
Posterior Parietal Cortex (PPC)
&
\textbf{PPC-DLPFC Hypoconnectivity}~
\cite{Vasantha2022DiscoveryON,May2020FrontoparietalNI,cong2023altered};
PPC-SMA Functional Connectivity;
\textbf{PPC Hub Disruption}~
\cite{King2022ContrastingFN,Huang2026BrainComputerIT,
Yoon2022AlterationsIS,cong2023altered};
PPC-MTL Dysconnectivity;
PPC-ACC Functional Integration
\\[4pt]

&
&
&
Ventrolateral Prefrontal Cortex (VLPFC)
&
\textbf{VLPFC-Frontoparietal Hypoconnectivity}~
\cite{May2020FrontoparietalNI};
\textbf{VLPFC Hub Disruption}~
\cite{May2020FrontoparietalNI,Cai2019InhibitionrelatedMO};
VLPFC-Frontoparietal Dynamic Connectivity;
VLPFC-Frontoparietal Task-Evoked Activation;
\textbf{VLPFC Emotional Regulation Network}~
\cite{Guazzo2025ExecutiveFA,Liu2022TheCA,Morawetz2024RestingstateEC}
\\[4pt]

&
&
&
Precuneus
&
\textbf{Executive Function Deficit Marker}~
\cite{Minnigulova2025AtypicalSO,May2020FrontoparietalNI};
\textbf{Self-Referential Processing Disruption}~
\cite{Minnigulova2025AtypicalSO,May2020FrontoparietalNI};
\textbf{Attention Network Dysfunction}~
\cite{wang2021altered,Wan2019EffectsOT,Harikumar202344FC};
\textbf{Working Memory Capacity Indicator}~
\cite{May2020FrontoparietalNI,Yuk2020FrontoparietalNC};
\textbf{Cognitive Control Network Instability}~
\cite{Yuk2020FrontoparietalNC,May2020FrontoparietalNI,
wang2021altered,Wan2019EffectsOT}
\\[4pt]

&
&
&
Precentral Gyrus
&
\textbf{Frontoparietal Connectivity Variability}~
\cite{Nebel2014PrecentralGF,Li2020DynamicFC,May2020FrontoparietalNI};
\textbf{Hypoconnectivity between Precentral Gyrus and DLPFC}~
\cite{May2020FrontoparietalNI,Nebel2014PrecentralGF,Li2020DynamicFC};
\textbf{Altered Precentral Gyrus Activation during Motor Tasks}~
\cite{May2020FrontoparietalNI,Nebel2014PrecentralGF,Bhoyroo2021AssociationBM};
\textbf{Disrupted Frontoparietal Network Hub Integrity}~
\cite{May2020FrontoparietalNI,Nebel2014PrecentralGF};
Loss of Anticorrelation with Default Mode Network
\\[4pt]

&
&
&
Supramarginal Gyrus
&
\textbf{Frontoparietal Hypoconnectivity}~
\cite{Kim2019616DL,Zhao2021ChangesOE};
\textbf{Supramarginal Gyrus Activation Deficit}~
\cite{May2020FrontoparietalNI,Kim2019616DL};
\textbf{Disrupted Frontoparietal Hub Function}~
\cite{May2020FrontoparietalNI,Kim2019616DL,Zhao2021ChangesOE};
Loss of Anticorrelation in Frontoparietal Network;
\textbf{Altered Frontoparietal Network Synchronization}~
\cite{Hoffmann2016PreservedSD,May2020FrontoparietalNI}
\\[4pt]

&
&
&
Middle Frontal Gyrus (MFG)
&
Reduced MFG-Posterior Parietal Cortex Connectivity;
\textbf{Hyperconnectivity between MFG and Default Mode Network}~
\cite{May2020FrontoparietalNI,Shafieizadegan2024InvestigatingEC,
Kidwell2024ARF,Zhao2021ChangesOE};
\textbf{Altered MFG-Dorsolateral Prefrontal Cortex Connectivity}~
\cite{May2020FrontoparietalNI,Zheng2025TemporalIS,Kidwell2024ARF};
\textbf{Decreased MFG Activation During Cognitive Control Tasks}~
\cite{Xu2023TheRO,Cai2022TheNM,Dong2024AssociationBE};
Impaired MFG-Medial Prefrontal Cortex Connectivity
\\[4pt]

&
&
&
Intraparietal Sulcus (IPS)
&
Reduced IPS-DLPFC connectivity;
\textbf{Altered IPS activation during working memory tasks}~
\cite{Ma2022DynamicFC,You2013AtypicalMO,Li2012AnFS};
IPS hub disruption in network efficiency;
Increased IPS-precuneus connectivity;
Loss of IPS-IFG anticorrelation
\\[4pt]


\cmidrule(lr){2-5}

&
Visual
&
V1, LGN, V2--V5
&
---
&
---
\\[4pt]


\cmidrule(lr){2-5}

&
Limbic
&
Amygdala,
\textbf{Hippocampus}~\cite{li2025hippocampal},
\textbf{Parahippocampal Gyrus}~\cite{monk2009abnormalities},
\textbf{Anterior Insula}~\cite{guo2019partially},
\textbf{PCC}~\cite{polk2022disrupted}
&
Hippocampus
&
\textbf{Hippocampal Volume Reduction}~
\cite{dufour2022prenatal,cooper2017reduced};
\textbf{Hippocampal-Prefrontal Dysconnectivity}~
\cite{kember2024specialization,zheng2017aldh2};
\textbf{Altered Hippocampal-Default Mode Network (DMN) Connectivity}~
\cite{shafieizadegan2025age,cong2023altered,kember2024specialization};
Hippocampal Hyperactivation during Memory Tasks;
Decreased Hippocampal Functional Connectivity with Amygdala
\\[4pt]

&
&
&
Parahippocampal Gyrus
&
\textbf{Parahippocampal Gyrus Connectivity}~
\cite{lee2016abnormalities,makhdoum2025disturbances,mciver2025impact};
Amygdala-Parahippocampal Connectivity;
\textbf{Hub Disruption in Limbic Network}~
\cite{lee2016abnormalities,deng2024impaired,Qian2021SharedAD};
\textbf{Altered Functional Connectivity Dynamics}~
\cite{lee2016abnormalities,nagai2024functional,cooper2017reduced};
\textbf{Parahippocampal Gyrus Activity during Memory Tasks}~
\cite{lee2016abnormalities,cooper2017reduced,gaigg2015episodic}
\\[4pt]

&
&
&
Anterior Insula
&
Anterior Insula -- Limbic Network Hypoconnectivity;
\textbf{Anterior Insula Hub Disruption}~
\cite{caria2015anterior,deng2024impaired,Xie2022DisruptedCC,Li2025DynamicAS};
Anterior Insula to Prefrontal Cortex Hypoconnectivity;
\textbf{Anterior Insula -- Default Mode Network Dysconnectivity}~
\cite{caria2015anterior,Li2025DynamicAS,deng2024impaired};
Dynamic Connectivity Variability in Anterior Insula
\\[4pt]

&
&
&
Posterior Cingulate Cortex (PCC)
&
PCC-Limbic Network Hypoconnectivity;
\textbf{PCC-Medial Prefrontal Cortex (mPFC) Connectivity}~
\cite{shafieizadegan2025age,Vasantha2022DiscoveryON};
PCC-Dorsal Attention Network (DAN) Anticorrelation;
\textbf{PCC Connectivity Variability}~
\cite{Polk2021DisruptedFC,shafieizadegan2025age,
Gao2019TheLN,Jung2014DefaultMN};
\textbf{PCC-Insula Connectivity}~
\cite{Polk2021DisruptedFC,shafieizadegan2025age,
Vasantha2022DiscoveryON,Gao2019TheLN}
\\[4pt]

\end{xltabular}

\begin{table*}[t]
\centering

\caption{Weighted component scores and overall ranking of studies in the evidence-ranking framework.}
\label{tab:component_scores}

\scriptsize
\setlength{\tabcolsep}{3pt}
\renewcommand{\arraystretch}{1.10}

\begin{adjustbox}{max width=\textwidth}

\begin{tabular}{
>{\raggedright\arraybackslash}l
*{8}{>{\centering\arraybackslash}c}
}

\toprule

\textbf{Article} &
\makecell{\textbf{Evidence}\\\textbf{level}} &
\makecell{\textbf{Evidence}\\\textbf{quality}} &
\makecell{\textbf{Recommendation}\\\textbf{strength}} &
\makecell{\textbf{Sample}\\\textbf{size}} &
\makecell{\textbf{Journal IF}\\\textbf{score}} &
\makecell{\textbf{Significant result}\\\textbf{proportion}} &
\makecell{\textbf{Clinical}\\\textbf{application}} &
\makecell{\textbf{Total}\\\textbf{score}} \\

\midrule

Margolis et al.~\cite{margolis2019salience}
& 1.600 & 0.400 & 0.100 & 0.202 & 0.300 & 0.100 & 0.000 & 2.702 \\

Lee et al.~\cite{lee2017autism}
& 1.600 & 0.400 & 0.100 & 0.300 & 0.200 & 0.100 & 0.000 & 2.700 \\

Ma et al.~\cite{ma2022dynamic}
& 1.600 & 0.400 & 0.100 & 0.300 & 0.200 & 0.100 & 0.000 & 2.700 \\

Wang et al.~\cite{wang2021altered}
& 1.600 & 0.400 & 0.100 & 0.300 & 0.200 & 0.100 & 0.000 & 2.700 \\

Lee et al.~\cite{lee2016abnormalities}
& 1.600 & 0.400 & 0.100 & 0.300 & 0.200 & 0.100 & 0.000 & 2.700 \\

Nebel et al.~\cite{nebel2014precentral}
& 1.600 & 0.400 & 0.100 & 0.300 & 0.200 & 0.100 & 0.000 & 2.700 \\

Zhou et al.~\cite{zhou2024altered}
& 1.600 & 0.400 & 0.100 & 0.276 & 0.300 & 0.000 & 0.000 & 2.676 \\

Bi et al.~\cite{bi2018abnormal}
& 1.600 & 0.400 & 0.100 & 0.276 & 0.200 & 0.100 & 0.000 & 2.676 \\

Cai et al.~\cite{cai2022neural}
& 1.800 & 0.400 & 0.100 & 0.040 & 0.200 & 0.100 & 0.000 & 2.640 \\

Yuk et al.~\cite{yuk2020frontoparietal}
& 1.600 & 0.400 & 0.100 & 0.276 & 0.200 & 0.050 & 0.000 & 2.626 \\

Xiao et al.~\cite{xiao2023atypical}
& 1.600 & 0.400 & 0.100 & 0.172 & 0.300 & 0.050 & 0.000 & 2.622 \\

Moxon-Emre et al.~\cite{moxon2021modulation}
& 1.800 & 0.400 & 0.100 & 0.060 & 0.200 & 0.050 & 0.000 & 2.610 \\

Gaigg et al.~\cite{gaigg2015episodic}
& 1.600 & 0.400 & 0.100 & 0.075 & 0.300 & 0.100 & 0.000 & 2.575 \\

Liu et al.~\cite{liu2016betel}
& 1.600 & 0.400 & 0.100 & 0.074 & 0.300 & 0.100 & 0.000 & 2.574 \\

Zhu et al.~\cite{zhu2022altered}
& 1.600 & 0.400 & 0.100 & 0.162 & 0.200 & 0.100 & 0.000 & 2.562 \\

Gao et al.~\cite{gao2019language}
& 1.600 & 0.400 & 0.100 & 0.102 & 0.300 & 0.050 & 0.000 & 2.552 \\

Kato et al.~\cite{kato2024reduced}
& 1.600 & 0.400 & 0.100 & 0.165 & 0.200 & 0.050 & 0.000 & 2.515 \\

Su et al.~\cite{su2020differences}
& 1.600 & 0.400 & 0.100 & 0.093 & 0.200 & 0.100 & 0.000 & 2.493 \\

Yoon et al.~\cite{yoon2022alterations}
& 1.600 & 0.400 & 0.100 & 0.186 & 0.100 & 0.100 & 0.000 & 2.486 \\

You et al.~\cite{you2013atypical}
& 1.600 & 0.400 & 0.100 & 0.093 & 0.200 & 0.050 & 0.000 & 2.443 \\

Li et al.~\cite{li2012fmri}
& 1.600 & 0.400 & 0.100 & 0.016 & 0.200 & 0.100 & 0.000 & 2.416 \\

Zhao et al.~\cite{Zhao2021ChangesOE}
& 1.600 & 0.400 & 0.100 & 0.300 & 0.000 & 0.000 & 0.000 & 2.400 \\

Zilverstand et al.~\cite{zilverstand2017fmri}
& 1.800 & 0.200 & 0.000 & 0.032 & 0.200 & 0.100 & 0.000 & 2.332 \\

Alves et al.~\cite{alves2024multiclass}
& 1.600 & 0.400 & 0.100 & 0.080 & 0.000 & 0.000 & 0.000 & 2.180 \\

Liu et al.~\cite{liu2022common}
& 1.200 & 0.200 & 0.000 & 0.220 & 0.200 & 0.050 & 0.000 & 1.870 \\

Deng et al.~\cite{deng2024impaired}
& 1.200 & 0.200 & 0.000 & 0.000 & 0.200 & 0.100 & 0.000 & 1.700 \\

Vasantha et al.~\cite{vasantha2022discovery}
& 1.200 & 0.200 & 0.000 & 0.000 & 0.000 & 0.000 & 0.000 & 1.400 \\

\bottomrule

\end{tabular}

\end{adjustbox}

\end{table*}

\newpage

\section{Prompts}

\subsection{General Prompts}

\begin{tcolorbox}[
    enhanced,
    breakable,
    colback=gray!5,
    colframe=gray!70,
    colbacktitle=blue!60!black,
    coltitle=white,
    title={System prompt for literature-based reliability assessment},
    fonttitle=\large,
    boxrule=0.8pt,
    arc=2mm,
    left=4mm,
    right=4mm,
    top=3mm,
    bottom=3mm,
    titlerule=0pt
]

\small

You are an experienced neuroscience researcher and clinical psychological doctor who is looking to provide reliable and rich messages based on the fMRI findings and public research papers to the field.

\medskip

You have a tentative diagnosis and you want to check if it is reliable or not, i.e., whether there is significant support for the diagnosis in the existing literature or in the well-researched literature.

\medskip

Be a harsh critic for reliability. Ensure there is sufficient supporting evidence in public papers for this diagnosis.

\medskip

You will be given access to the Semantic Scholar API, which you may use to survey the literature and find relevant papers to help you make your decision.

\medskip

The top 10 results for any search query will be presented to you with the abstracts.

\medskip

You will be given \{\texttt{num\_rounds}\} rounds to decide on the paper, but you do not need to use them all.

\medskip

At any round, you may exit early and decide on the reliability of the diagnosis.

\medskip

Decide a paper idea is \textbf{not reliable} if, after sufficient searching, you have not found a paper that significantly overlaps with your diagnosis.

\medskip

Decide a paper idea is \textbf{reliable} if you have found a paper that significantly overlaps with your diagnosis.

\medskip

The more papers that significantly overlap with your diagnosis, the more reliable this diagnosis will be.

\medskip

\{\texttt{task\_description}\}

\end{tcolorbox}


\begin{tcolorbox}[
    enhanced,
    breakable,
    colback=gray!5,
    colframe=gray!70,
    colbacktitle=blue!60!black,
    coltitle=white,
    title={Prompt template for literature-based reliability verification},
    fonttitle=\large,
    boxrule=0.8pt,
    arc=2mm,
    left=4mm,
    right=4mm,
    top=3mm,
    bottom=3mm,
    titlerule=0pt
]

\small

Round \{\texttt{current\_round}\}/\{\texttt{num\_rounds}\}.

\medskip

You have this diagnosis:

\medskip

\texttt{"""}

\{\texttt{diagnosis}\}

\texttt{"""}

\medskip

The results of the last query are (empty on first round):

\medskip

\texttt{"""}

\{\texttt{last\_query\_results}\}

\texttt{"""}

\medskip

Respond in the following format:

\medskip

\texttt{THOUGHT:}

\texttt{<THOUGHT>}

\medskip

\texttt{RESPONSE:}

\texttt{<JSON>}

\medskip

In \texttt{<THOUGHT>}:

\begin{enumerate}
    \item Briefly reason over the idea and identify any query that could help you make your decision.

    \item When you mention papers, you \textbf{MUST} follow the exact format:

    \medskip

    \texttt{**Paper <id>** (<full paper title>) <your description...>}

    \begin{itemize}
        \item \texttt{<id>} must be a pure number (integer) with no extra text.
        \item The title must be inside parentheses right after the id.
        \item After the title, continue your sentence with your explanation of its relevance.
    \end{itemize}

    \item Example of \textbf{CORRECT} format:

    \medskip

    \texttt{**Paper 103** (Aberrant dynamic functional connectivity in retinal detachment) shows increased VN-DMN connectivity originating from V1.}

    \item Example of \textbf{INCORRECT} formats (do \textbf{NOT} use):

    \medskip

    \texttt{**Paper 103 from Round 2** (Title...)}

    \texttt{**Paper \#103** (Title...)}

    \texttt{Paper 103 (Title...)}

    \item If you have made your decision, add
    \texttt{"Decision made: reliable."} or
    \texttt{"Decision made: not reliable."}
    at the end of \texttt{THOUGHT}.
\end{enumerate}

\medskip

In \texttt{<JSON>}, respond in JSON format with \textbf{ONLY} the following field:

\begin{itemize}
    \item \texttt{"Query":} An optional search query to search the literature
    (e.g., \texttt{attention is all you need}). You must make a query if you
    have not decided this round.
\end{itemize}

\medskip

This JSON will be automatically parsed, so ensure the format is precise.

\medskip

\textbf{Remember:} If you mention papers in the \texttt{THOUGHT}, you must use the correct
\texttt{**Paper <id>** (<title>)} format, otherwise the output will be considered invalid.

\end{tcolorbox}

\begin{tcolorbox}[
    enhanced,
    breakable,
    colback=gray!5,
    colframe=gray!70,
    colbacktitle=blue!60!black,
    coltitle=white,
    title={Prompt template for the first hypothesis generation},
    fonttitle=\large,
    boxrule=0.8pt,
    arc=2mm,
    left=4mm,
    right=4mm,
    top=3mm,
    bottom=3mm,
    titlerule=0pt
]

\small

\{\texttt{task\_description}\}

\medskip

\texttt{<experiment.py>}

\{\texttt{code}\}

\texttt{</experiment.py>}

\medskip

Here are the ideas that you have already generated:

\medskip

\texttt{'''}

\{\texttt{prev\_ideas\_string}\}

\texttt{'''}

\medskip

Come up with the next impactful and creative idea for research experiments and directions you can feasibly investigate with the code provided.

\medskip

Note that you will not have access to any additional resources or datasets.

\medskip

Make sure any idea is not overfit to the specific training dataset or model, and has wider significance.

\medskip

Respond in the following format:

\medskip

\texttt{THOUGHT:}

\texttt{<THOUGHT>}

\medskip

\texttt{NEW IDEA JSON:}

\texttt{<JSON>}

\medskip

In \texttt{<THOUGHT>}, first briefly discuss your intuitions and motivations for the idea. Detail your high-level plan, necessary design choices, and ideal outcomes of the experiments. Justify how the idea is different from the existing ones.

\medskip

In \texttt{<JSON>}, provide the new idea in JSON format with the following fields:

\begin{itemize}
    \item \texttt{"Name":} A shortened descriptor of the idea. Lowercase, no spaces, underscores allowed.
    
    \item \texttt{"Title":} A title for the idea, which will be used for the report writing.
    
    \item \texttt{"Experiment":} An outline of the implementation, e.g., which functions need to be added or modified and how results will be obtained.
    
    \item \texttt{"Interestingness":} A rating from 1 to 10 (lowest to highest).
    
    \item \texttt{"Feasibility":} A rating from 1 to 10 (lowest to highest).
    
    \item \texttt{"Novelty":} A rating from 1 to 10 (lowest to highest).
\end{itemize}

\medskip

Be cautious and realistic on your ratings.

\medskip

This JSON will be automatically parsed, so ensure the format is precise.

\medskip

You will have \{\texttt{num\_reflections}\} rounds to iterate on the idea, but do not need to use them all.

\end{tcolorbox}

\begin{tcolorbox}[
    enhanced,
    breakable,
    colback=gray!5,
    colframe=gray!70,
    colbacktitle=blue!60!black,
    coltitle=white,
    title={Prompt template for idea reflection and refinement},
    fonttitle=\large,
    boxrule=0.8pt,
    arc=2mm,
    left=4mm,
    right=4mm,
    top=3mm,
    bottom=3mm,
    titlerule=0pt
]

\small

Round \{\texttt{current\_round}\}/\{\texttt{num\_reflections}\}.

\medskip

In your thoughts, first carefully consider the quality, novelty, and feasibility of the idea you just created.

\medskip

Include any other factors that you think are important in evaluating the idea.

\medskip

Ensure the idea is clear and concise, and the JSON is the correct format.

\medskip

Do not make things overly complicated.

\medskip

In the next attempt, try and refine and improve your idea.

\medskip

Stick to the spirit of the original idea unless there are glaring issues.

\medskip

Respond in the same format as before:

\medskip

\texttt{THOUGHT:}

\texttt{<THOUGHT>}

\medskip

\texttt{NEW IDEA JSON:}

\texttt{<JSON>}

\medskip

If there is nothing to improve, simply repeat the previous JSON \textbf{EXACTLY} after the thought and include \texttt{"I am done"} at the end of the thoughts but before the JSON.

\medskip

\textbf{ONLY INCLUDE} \texttt{"I am done"} \textbf{IF YOU ARE MAKING NO MORE CHANGES.}

\end{tcolorbox}

\begin{tcolorbox}[
    enhanced,
    breakable,
    colback=gray!5,
    colframe=gray!70,
    colbacktitle=blue!60!black,
    coltitle=white,
    title={System prompt for novelty assessment},
    fonttitle=\large,
    boxrule=0.8pt,
    arc=2mm,
    left=4mm,
    right=4mm,
    top=3mm,
    bottom=3mm,
    titlerule=0pt
]

\small

You are an ambitious AI PhD student who is looking to publish a paper that will contribute significantly to the field.

\medskip

You have an idea and you want to check if it is novel or not, i.e., whether it does not overlap significantly with existing literature or has not already been well explored.

\medskip

Be a harsh critic for novelty. Ensure there is a sufficient contribution in the idea for a new conference or workshop paper.

\medskip

You will be given access to the Semantic Scholar API, which you may use to survey the literature and find relevant papers to help you make your decision.

\medskip

The top 10 results for any search query will be presented to you with the abstracts.

\medskip

You will be given \{\texttt{num\_rounds}\} rounds to decide on the paper, but you do not need to use them all.

\medskip

At any round, you may exit early and decide on the novelty of the idea.

\medskip

Decide a paper idea is \textbf{novel} if, after sufficient searching, you have not found a paper that significantly overlaps with your idea.

\medskip

Decide a paper idea is \textbf{not novel} if you have found a paper that significantly overlaps with your idea.

\medskip

\{\texttt{task\_description}\}

\medskip

\texttt{<experiment.py>}

\{\texttt{code}\}

\texttt{</experiment.py>}

\end{tcolorbox}

\begin{tcolorbox}[
    enhanced,
    breakable,
    colback=gray!5,
    colframe=gray!70,
    colbacktitle=blue!60!black,
    coltitle=white,
    title={Prompt template for novelty verification},
    fonttitle=\large,
    boxrule=0.8pt,
    arc=2mm,
    left=4mm,
    right=4mm,
    top=3mm,
    bottom=3mm,
    titlerule=0pt
]

\small

Round \{\texttt{current\_round}\}/\{\texttt{num\_rounds}\}.

\medskip

You have this idea:

\medskip

\texttt{"""}

\{\texttt{idea}\}

\texttt{"""}

\medskip

The results of the last query are (empty on first round):

\medskip

\texttt{"""}

\{\texttt{last\_query\_results}\}

\texttt{"""}

\medskip

Respond in the following format:

\medskip

\texttt{THOUGHT:}

\texttt{<THOUGHT>}

\medskip

\texttt{RESPONSE:}

\texttt{<JSON>}

\medskip

In \texttt{<THOUGHT>}, first briefly reason over the idea and identify any query that could help you make your decision.

\medskip

If you have made your decision, add \texttt{"Decision made: novel."} or \texttt{"Decision made: not novel."} to your thoughts.

\medskip

In \texttt{<JSON>}, respond in JSON format with \textbf{ONLY} the following field:

\begin{itemize}
    \item \texttt{"Query":} An optional search query to search the literature
    (e.g., \texttt{attention is all you need}). You must make a query if you have not decided this round.
\end{itemize}

\medskip

A query will work best if you are able to recall the exact name of the paper you are looking for, or the authors.

\medskip

This JSON will be automatically parsed, so ensure the format is precise.

\end{tcolorbox}

\begin{tcolorbox}[
    enhanced,
    breakable,
    colback=gray!5,
    colframe=gray!70,
    colbacktitle=blue!60!black,
    coltitle=white,
    title={System prompt for biomedical literature search strategy},
    fonttitle=\large,
    boxrule=0.8pt,
    arc=2mm,
    left=4mm,
    right=4mm,
    top=3mm,
    bottom=3mm,
    titlerule=0pt
]

\small

You are a senior biomedical literature search strategist.

\medskip

Before writing queries, you \textbf{MUST} first interpret the raw EHR into meaningful, search-ready labels.

\medskip

\textbf{INTERPRETATION RUBRIC (DO NOT HALLUCINATE)}

\begin{itemize}

    \item \textbf{Age:} \(\geq 18\) $\Rightarrow$ \texttt{"adult"}.

    \item \textbf{ADOS Module:} 4 $\Rightarrow$ typically adult, verbally fluent.

    \item \textbf{FIQ buckets} (descriptive; \textbf{NOT} diagnostic):
    \begin{itemize}
        \item \(\geq 130\): very high
        \item 120--129: high
        \item 110--119: above average
        \item 90--109: average
        \item 80--89: below average
        \item \(<80\): low
    \end{itemize}

    \item \textbf{SRS-2 T-score buckets:}
    \begin{itemize}
        \item \(\geq 76\): severe
        \item 66--75: moderate-to-severe
        \item 60--65: mild-to-moderate
        \item \(<60\): within-typical/at-risk
    \end{itemize}

    \item \textbf{BDI} (assume BDI-II unless specified):
    \begin{itemize}
        \item 29+: severe
        \item 20--28: moderate
        \item 14--19: mild
        \item 0--13: minimal
    \end{itemize}

    \item \textbf{Eye status at scan \& medication status:} usually \textbf{NOT} directly searchable; keep for manual screening.

\end{itemize}

\medskip

\textbf{TIERING GUIDELINES}

\begin{itemize}

    \item \texttt{tier1}: ASD/``autism spectrum disorder'', adult/high-functioning/high IQ (if applicable), limbic system/limbic network, resting-state fMRI/functional connectivity.

    \item \texttt{tier2}: ADOS (mention ``module 4'' if applicable), SRS, depression/BDI, sex.

    \item \texttt{tier3}: exact numeric scores (FIQ/SRS/ADOS/BDI), handedness, medication, eye status.

\end{itemize}

\medskip

\textbf{STEPWISE QUERY STRATEGY}

\begin{itemize}

    \item \textbf{Stage 1:} broad/high recall with \texttt{tier1} cores.

    \item \textbf{Stage 2+:} add refiners gradually
    (ADOS $\rightarrow$ ``module 4'' $\rightarrow$ SRS $\rightarrow$ depression $\rightarrow$ sex $\rightarrow$ \ldots).

    \item \textbf{Relaxation stage:} if too few results, remove the tightest refiners first.

\end{itemize}

\medskip

\textbf{GOOGLE SCHOLAR HYGIENE RULES (REQUIRED)}

\begin{itemize}

    \item Quote multi-word phrases:
    \texttt{"functional connectivity"},
    \texttt{"resting-state fMRI"},
    \texttt{"limbic network"},
    \texttt{"limbic system"},
    \texttt{"module 4"}.

    \item Use \texttt{OR} in \textbf{UPPERCASE}; group synonyms in parentheses.

    \item \textbf{NEVER} include numeric equality like \texttt{"SRS=79"} or \texttt{"BDI=16"} (keep numbers only for manual screening).

    \item At each stage, add at most 1--2 new terms.

\end{itemize}

\medskip

\textbf{PUBMED BOOLEAN GROUPING (REQUIRED)}

\begin{itemize}

    \item Parenthesize every \texttt{OR} group at the top level, e.g.:

    \medskip

    \texttt{("autism spectrum disorder" OR ASD)}

    \texttt{("limbic system" OR "limbic network")}

    \texttt{("resting-state fMRI" OR rs-fMRI)}

\end{itemize}

\medskip

\textbf{OUTPUT RULES}

\begin{itemize}

    \item Use \textbf{ONLY} terms from \texttt{TERM\_WHITELIST} supplied by the user message (do not invent).

    \item PubMed \texttt{title\_abstract\_variant} \textbf{MUST} apply \texttt{[Title/Abstract]} field tags to key terms/phrases.

    \item Return \textbf{VALID JSON only} (no markdown).

    \item English only.

\end{itemize}

\end{tcolorbox}

\begin{tcolorbox}[
    enhanced,
    breakable,
    colback=gray!5,
    colframe=gray!70,
    colbacktitle=blue!60!black,
    coltitle=white,
    title={Prompt template for EHR-guided literature query generation},
    fonttitle=\large,
    boxrule=0.8pt,
    arc=2mm,
    left=4mm,
    right=4mm,
    top=3mm,
    bottom=3mm,
    titlerule=0pt
]

\small

\textbf{EHR PROFILE (key=value lines):}

\texttt{\$ehr\_text}

\medskip

\textbf{Optional imaging finding:}

\texttt{\$finding\_text}

\medskip

\textbf{TERM\_WHITELIST (you MUST restrict all query terms to this set):}

\texttt{\$term\_whitelist}

\medskip

\textbf{Tasks:}

\begin{enumerate}

    \item Build \texttt{"derived\_profile"} (adult? high IQ? scale severities?).

    \item Build \texttt{"tiers"}: tier1/2/3 with \texttt{"term"} + \texttt{"rationale"}.

    \item Build \texttt{"queries"}: PubMed \& Scholar, step1 \(\rightarrow\) step2 + title/abstract variant for PubMed.

    \item Build \texttt{"query\_plan"}: 4--10 stages.

    \textbf{HARD RULE:} at each stage, you \textbf{MUST} add only 1--2 new terms
    (\(\mathrm{length(add\_terms)} \in \{1,2\}\)).

    \begin{itemize}

        \item \textbf{Stage1:} minimal core
        (e.g., ASD + adult + ``functional connectivity'') if \texttt{'adult'} is present in whitelist; otherwise omit \texttt{'adult'}.

        \item \textbf{Stage2:} add
        \texttt{("limbic system" OR "limbic network")}
        if present.

        \item \textbf{Stage3:} add
        \texttt{"resting-state fMRI"} (or \texttt{rs-fMRI})
        if present.

        \item \textbf{Stage4+:} add one clinical/demographic cue at a time
        (ADOS \(\rightarrow\) ``module 4'' \(\rightarrow\) SRS \(\rightarrow\) depression \(\rightarrow\) male/female \(\rightarrow\) ``high-functioning''/``high IQ''), but only if present in whitelist.

        \item Each stage must include:
        \texttt{stage}, \texttt{goal}, \texttt{strategy}, \texttt{add\_terms},
        \texttt{relax\_terms}, \texttt{pubmed\_query}, \texttt{scholar\_query},
        \texttt{rationale}.

        \item Queries at a stage must equal the previous stage's logic \textbf{PLUS} only the terms in \texttt{add\_terms}
        (unless \texttt{relax\_terms} explicitly remove prior terms).

        \item PubMed queries \textbf{MUST} parenthesize every OR group as specified.

        \item Scholar queries must quote multi-word phrases and \textbf{MUST NOT}
        contain numeric equality such as \texttt{"SRS=79"} or \texttt{"BDI=16"}.

    \end{itemize}

    \item \texttt{"manual\_screening\_checklist"}: list tier3 items as human-match criteria
    (do not put numbers into queries).

\end{enumerate}

\medskip

Return JSON with this schema:

\medskip

{\ttfamily
\{
\par
\quad "derived\_profile": \{
\par
\qquad "population": str,
\par
\qquad "cognition": str,
\par
\qquad "autism\_scales": str,
\par
\qquad "ados": str,
\par
\qquad "depression": str,
\par
\qquad "imaging\_finding": str
\par
\quad \},
\par
\quad "tiers": \{
\par
\qquad "tier1": [\{"term": str, "rationale": str\}],
\par
\qquad "tier2": [\{"term": str, "rationale": str\}],
\par
\qquad "tier3": [\{"term": str, "rationale": str\}]
\par
\quad \},
\par
\quad "queries": \{
\par
\qquad "pubmed": \{
\par
\qquad\quad "step1\_core": str,
\par
\qquad\quad "step2\_add\_scales": str,
\par
\qquad\quad "title\_abstract\_variant": \{
\par
\qquad\qquad "step1\_core\_ta": str,
\par
\qquad\qquad "step2\_add\_scales\_ta": str
\par
\qquad\quad \}
\par
\qquad \},
\par
\qquad "scholar": \{
\par
\qquad\quad "step1\_core": str,
\par
\qquad\quad "step2\_add\_scales": str
\par
\qquad \}
\par
\quad \},
\par
\quad "query\_plan": [
\par
\qquad \{
\par
\qquad\quad "stage": str,
\par
\qquad\quad "goal": str,
\par
\qquad\quad "strategy": str,
\par
\qquad\quad "add\_terms": [str],
\par
\qquad\quad "relax\_terms": [str],
\par
\qquad\quad "pubmed\_query": str,
\par
\qquad\quad "scholar\_query": str,
\par
\qquad\quad "rationale": str
\par
\qquad \}
\par
\quad ],
\par
\quad "manual\_screening\_checklist": [str]
\par
\}
}

\end{tcolorbox}

\begin{tcolorbox}[
    enhanced,
    breakable,
    colback=gray!5,
    colframe=gray!70,
    colbacktitle=blue!60!black,
    coltitle=white,
    title={Revision prompt for query-plan stage constraints},
    fonttitle=\large,
    boxrule=0.8pt,
    arc=2mm,
    left=4mm,
    right=4mm,
    top=3mm,
    bottom=3mm,
    titlerule=0pt
]

\small

Your previous output violated the \textbf{HARD RULE}:

\begin{itemize}
    \item At each stage, \texttt{add\_terms} \textbf{MUST} contain only 1 or 2 items (never \(>2\)).
\end{itemize}

\medskip

Given the JSON below, \textbf{REVISE ONLY} the \texttt{"query\_plan"} so that each stage adds 1--2 new terms, keeping other fields unchanged.

\medskip

Ensure each stage's queries equal the previous stage plus \textbf{ONLY} those \texttt{add\_terms}
(unless \texttt{relax\_terms} are used).

\medskip

Return \textbf{VALID JSON} with the same top-level keys, but with a corrected \texttt{"query\_plan"}.

\end{tcolorbox}

\begin{tcolorbox}[
    enhanced,
    breakable,
    colback=gray!5,
    colframe=gray!70,
    colbacktitle=blue!60!black,
    coltitle=white,
    title={Revision prompt for query hygiene},
    fonttitle=\large,
    boxrule=0.8pt,
    arc=2mm,
    left=4mm,
    right=4mm,
    top=3mm,
    bottom=3mm,
    titlerule=0pt
]

\small

Your queries violate Google Scholar hygiene:

\begin{itemize}

    \item Quote multi-word phrases
    (e.g., \texttt{"functional connectivity"},
    \texttt{"resting-state fMRI"},
    \texttt{"limbic network"},
    \texttt{"limbic system"},
    \texttt{"module 4"}).

    \item Do \textbf{NOT} include numeric equality filters
    (e.g., \texttt{"SRS=79"}, \texttt{"BDI=16"}). Remove such patterns.

    \item Keep \texttt{OR} in \textbf{UPPERCASE} and group synonyms in parentheses.

\end{itemize}

\medskip

\textbf{REVISE ONLY} the query strings under \texttt{"queries"} and all
\texttt{"scholar\_query"}/\texttt{"pubmed\_query"} fields in \texttt{"query\_plan"}.

\medskip

Return \textbf{VALID JSON} with the same top-level structure.

\end{tcolorbox}

\begin{tcolorbox}[
    enhanced,
    breakable,
    colback=gray!5,
    colframe=gray!70,
    colbacktitle=blue!60!black,
    coltitle=white,
    title={Revision prompt for PubMed Boolean grouping},
    fonttitle=\large,
    boxrule=0.8pt,
    arc=2mm,
    left=4mm,
    right=4mm,
    top=3mm,
    bottom=3mm,
    titlerule=0pt
]

\small

Your PubMed queries must parenthesize OR groups.

\medskip

For example:

\medskip

\texttt{("autism spectrum disorder" OR ASD)}

\texttt{("limbic system" OR "limbic network")}

\texttt{("resting-state fMRI" OR rs-fMRI)}

\medskip

\textbf{REVISE ONLY} the PubMed queries in \texttt{"queries"} and every
\texttt{"pubmed\_query"} in \texttt{"query\_plan"} to ensure all OR groups are parenthesized correctly.

\medskip

Do not change other keys.

\medskip

Return \textbf{VALID JSON} with the same top-level structure.

\end{tcolorbox}

\begin{tcolorbox}[
    enhanced,
    breakable,
    colback=gray!5,
    colframe=gray!70,
    colbacktitle=blue!60!black,
    coltitle=white,
    title={Revision prompt for term-whitelist compliance},
    fonttitle=\large,
    boxrule=0.8pt,
    arc=2mm,
    left=4mm,
    right=4mm,
    top=3mm,
    bottom=3mm,
    titlerule=0pt
]

\small

Some \texttt{add\_terms} are not in the \texttt{TERM\_WHITELIST}.

\medskip

\textbf{REVISE ONLY} the \texttt{"query\_plan"} to use \textbf{ONLY} terms from the whitelist.

\medskip

Do not change other keys.

\medskip

Keep the per-stage \texttt{add\_terms} size in \(\{1,2\}\).

\medskip

Return \textbf{VALID JSON} with the same top-level structure.

\end{tcolorbox}

\subsection{Anatomical Region Geration Prompts}
\label{sec:anatomical_prompts}
\begin{tcolorbox}[
    enhanced,
    colback=gray!5,
    colframe=gray!70,
    colbacktitle=blue!60!black,
    coltitle=white,
    title={System prompt},
    fonttitle=\bfseries,
    boxrule=0.8pt,
    arc=2mm,
    left=5mm,
    right=5mm,
    top=3mm,
    bottom=3mm
]
\texttt{`system':} You are an experienced fMRI analyst and neuroscience researcher who is looking to provide reliable and rich messages based on the fMRI findings for \{Condition\} patient.
\end{tcolorbox}

\begin{tcolorbox}[
    enhanced,
    breakable,
    colback=gray!5,
    colframe=gray!70,
    colbacktitle=blue!60!black,
    coltitle=white,
    title={Prompt template for anatomical region generation agent},
    fonttitle=\large,
    boxrule=0.8pt,
    arc=2mm,
    left=4mm,
    right=4mm,
    top=3mm,
    bottom=3mm,
    titlerule=0pt
]

\small

\texttt{'fMRI findings':} A patient with \{Condition\} shows fMRI abnormalities in:
\{\texttt{abnormal\_networks}\}.

\medskip

\textbf{FOCUS:} Identify the brain structures (anatomical regions) within each abnormal network appearing in \texttt{'fMRI findings'} for patient with \{Condition\}.

\medskip

\textbf{Requirements:}

\begin{itemize}
    \item Use peer-reviewed literature to map each network to its known anatomical components.
    \item Maintain accuracy and specificity (e.g., ``dorsolateral prefrontal cortex'' instead of just ``frontal lobe'').
    \item Use DSM-5/ICD-11 terminology where relevant for structural-symptom associations.
\end{itemize}

Here are the ideas that you have already generated:

\medskip

\texttt{'''}

\texttt{\{prev\_ideas\_string\}}

\texttt{'''}

\medskip

Come up with multiple impactful and reliable ideas about the anatomical brain regions related to the functional networks mentioned in \texttt{'fMRI findings'}.

\medskip

\textbf{Respond in the following format:}

\medskip

\texttt{THOUGHT:}

\texttt{<THOUGHT>}

\medskip

\texttt{NEW IDEA JSON:}

\texttt{<JSON>}

\medskip

In \texttt{<THOUGHT>}, first briefly discuss your intuitions and motivations for the idea. Detail your evidence and necessary description. Justify how the idea is different from the existing ones.

\medskip

In \texttt{<JSON>}, provide the new idea in JSON format with the following fields:

\medskip

\texttt{- "Functional Network1": abnormal functional networks mentioned in 'fMRI findings'}

\quad \texttt{+ "RegionA": anatomical regions related to Functional Network1}

\quad \texttt{+ "Reliability": A rating from 1 to 10 (lowest to highest)}

\quad \texttt{+ "Accuracy": A rating from 1 to 10 (lowest to highest)}

\medskip

\quad \texttt{+ "RegionB": anatomical regions related to Functional Network1}

\quad \texttt{+ "Reliability": A rating from 1 to 10 (lowest to highest)}

\quad \texttt{+ "Accuracy": A rating from 1 to 10 (lowest to highest)}

\medskip

\quad \texttt{+ "RegionC": anatomical regions related to Functional Network1}

\quad \texttt{+ "Reliability": A rating from 1 to 10 (lowest to highest)}

\quad \texttt{+ "Accuracy": A rating from 1 to 10 (lowest to highest)}

\medskip

\quad \texttt{+ "Region...": anatomical regions related to Functional Network1}

\quad \texttt{+ "Reliability": A rating from 1 to 10 (lowest to highest)}

\quad \texttt{+ "Accuracy": A rating from 1 to 10 (lowest to highest)}

\bigskip

\texttt{- "Functional Network2": abnormal functional networks mentioned in 'fMRI findings'}

\quad \texttt{+ "RegionA": anatomical regions related to Functional Network2}

\quad \texttt{+ "Reliability": A rating from 1 to 10 (lowest to highest)}

\quad \texttt{+ "Accuracy": A rating from 1 to 10 (lowest to highest)}

\medskip

\quad \texttt{+ "RegionB": anatomical regions related to Functional Network2}

\quad \texttt{+ "Reliability": A rating from 1 to 10 (lowest to highest)}

\quad \texttt{+ "Accuracy": A rating from 1 to 10 (lowest to highest)}

\medskip

\quad \texttt{+ "RegionC": anatomical regions related to Functional Network2}

\quad \texttt{+ "Reliability": A rating from 1 to 10 (lowest to highest)}

\quad \texttt{+ "Accuracy": A rating from 1 to 10 (lowest to highest)}

\medskip

\quad \texttt{+ "Region...": anatomical regions related to Functional Network2}

\quad \texttt{+ "Reliability": A rating from 1 to 10 (lowest to highest)}

\quad \texttt{+ "Accuracy": A rating from 1 to 10 (lowest to highest)}

\bigskip

\texttt{- "Functional Network...": abnormal functional networks mentioned in 'fMRI findings'}

\quad \texttt{+ "RegionA": anatomical regions related to Functional Network...}

\quad \texttt{+ "Reliability": A rating from 1 to 10 (lowest to highest)}

\quad \texttt{+ "Accuracy": A rating from 1 to 10 (lowest to highest)}

\medskip

\quad \texttt{+ "RegionB": anatomical regions related to Functional Network...}

\quad \texttt{+ "Reliability": A rating from 1 to 10 (lowest to highest)}

\quad \texttt{+ "Accuracy": A rating from 1 to 10 (lowest to highest)}

\medskip

\quad \texttt{+ "RegionC": anatomical regions related to Functional Network...}

\quad \texttt{+ "Reliability": A rating from 1 to 10 (lowest to highest)}

\quad \texttt{+ "Accuracy": A rating from 1 to 10 (lowest to highest)}

\medskip

\quad \texttt{+ "Region...": anatomical regions related to Functional Network...}

\quad \texttt{+ "Reliability": A rating from 1 to 10 (lowest to highest)}

\quad \texttt{+ "Accuracy": A rating from 1 to 10 (lowest to highest)}

\bigskip

Be cautious and realistic on your ratings.

This JSON will be automatically parsed, so ensure the format is precise.

You will have \{\texttt{num\_reflections}\} rounds to iterate on the idea, but do not need to use them all.

\end{tcolorbox}

\begin{tcolorbox}[
    enhanced,
    breakable,
    colback=gray!5,
    colframe=gray!70,
    colbacktitle=blue!60!black,
    coltitle=white,
    title={Prompt template for literature evidence extraction},
    fonttitle=\large,
    boxrule=0.8pt,
    arc=2mm,
    left=4mm,
    right=4mm,
    top=3mm,
    bottom=3mm,
    titlerule=0pt
]

\small

The results of the last query are as follows (the first round is empty):

\medskip

\texttt{"""}

\texttt{\{last\_query\_results\}}

\texttt{"""}

\medskip

Please reply in the following format:

\medskip

\texttt{Thinking:}

\texttt{<Thinking>}

\medskip

\texttt{Reply:}

\texttt{<JSON>}

\medskip

In \texttt{<Thinking>}, extract the papers mentioned in
\{\texttt{last\_query\_results}\} that support the final decision, and extract
the relevant content of the papers from \{\texttt{last\_query\_results}\}
intact, and record them in \texttt{<JSON>} according to the original structure.

\medskip

In \texttt{<JSON>}, please reply in JSON format:

Store according to the structure in
\{\texttt{last\_query\_results}\}.

\medskip

This JSON will be parsed automatically, so please make sure the format is accurate.

\end{tcolorbox}

\begin{tcolorbox}[
    enhanced,
    breakable,
    colback=gray!5,
    colframe=gray!70,
    colbacktitle=blue!60!black,
    coltitle=white,
    title={Prompt template for anatomical region aggregation},
    fonttitle=\large,
    boxrule=0.8pt,
    arc=2mm,
    left=4mm,
    right=4mm,
    top=3mm,
    bottom=3mm,
    titlerule=0pt
]

\small

You are given multiple JSONs representing anatomical brain regions involved in functional networks (Visual Network, Limbic Network, and Frontoparietal Network) for ASD patients.

\medskip

Each JSON lists anatomical regions under each network, along with a ``Reliability'' and ``Accuracy'' score (range 1--10).

\medskip

Your task is to:

\begin{enumerate}
    \item \textbf{Merge} all JSONs into a \textbf{single unified JSON}.
    
    \item \textbf{Remove duplicated anatomical regions} (based on name), retaining the \textbf{highest Reliability and Accuracy} score if duplicate entries exist.
    
    \item \textbf{Preserve all distinct anatomical regions} across all networks.
    
    \item Ensure \textbf{correct JSON format}, identical to the structure shown below.
    
    \item Do not invent new regions. Only use entries from the existing JSONs.
\end{enumerate}

\medskip

Return your output in the format:

\medskip

\texttt{NEW IDEA JSON:}

\texttt{<JSON>}

\medskip

In \texttt{<JSON>}, provide the new idea in JSON format with the following fields:

\medskip

\texttt{- "Functional Network...": abnormal functional networks mentioned in 'fMRI findings'}

\quad \texttt{+ "RegionA": anatomical regions related to Functional Network...}

\quad \texttt{+ "Reliability": A rating from 1 to 10 (lowest to highest)}

\quad \texttt{+ "Accuracy": A rating from 1 to 10 (lowest to highest)}

\medskip

\quad \texttt{+ "RegionB": anatomical regions related to Functional Network...}

\quad \texttt{+ "Reliability": A rating from 1 to 10 (lowest to highest)}

\quad \texttt{+ "Accuracy": A rating from 1 to 10 (lowest to highest)}

\medskip

\quad \texttt{+ "RegionC": anatomical regions related to Functional Network...}

\quad \texttt{+ "Reliability": A rating from 1 to 10 (lowest to highest)}

\quad \texttt{+ "Accuracy": A rating from 1 to 10 (lowest to highest)}

\medskip

\quad \texttt{+ "Region...": anatomical regions related to Functional Network...}

\quad \texttt{+ "Reliability": A rating from 1 to 10 (lowest to highest)}

\quad \texttt{+ "Accuracy": A rating from 1 to 10 (lowest to highest)}

\bigskip

The following part is a sample of results. Please make sure the generated results have a similar structure:

\medskip

{\ttfamily
\{
\par
\quad "dorsal attention": \{
\par
\qquad "Intraparietal Sulcus": \{
\par
\qquad\quad "Reliability": 9,
\par
\qquad\quad "Accuracy": 9
\par
\qquad \},
\par
\qquad "Frontal Eye Fields": \{
\par
\qquad\quad "Reliability": 9,
\par
\qquad\quad "Accuracy": 9
\par
\qquad \},
\par
\qquad "Superior Parietal Lobule": \{
\par
\qquad\quad "Reliability": 8,
\par
\qquad\quad "Accuracy": 8
\par
\qquad \}
\par
\quad \}
\par
\}
}

\end{tcolorbox}

\subsection{Biomarker Generation Prompts}

\begin{tcolorbox}[
    enhanced,
    breakable,
    colback=gray!5,
    colframe=gray!70,
    colbacktitle=blue!60!black,
    coltitle=white,
    title={System prompt for biomarker generation agent},
    fonttitle=\large,
    boxrule=0.8pt,
    arc=2mm,
    left=4mm,
    right=4mm,
    top=3mm,
    bottom=3mm,
    titlerule=0pt
]

\small

\texttt{'system':} You are an experienced neuroscience researcher and clinical psychological doctor who is looking to provide reliable and rich messages based on the fMRI findings and public research papers for \{Condition\} patient.

\end{tcolorbox}

\begin{tcolorbox}[
    enhanced,
    breakable,
    colback=gray!5,
    colframe=gray!70,
    colbacktitle=blue!60!black,
    coltitle=white,
    title={Prompt template for functional biomarker generation},
    fonttitle=\large,
    boxrule=0.8pt,
    arc=2mm,
    left=4mm,
    right=4mm,
    top=3mm,
    bottom=3mm,
    titlerule=0pt
]

\small

\texttt{'Clinical Scenario':} A patient with \{Condition\} shows fMRI abnormalities in the \{\texttt{abnormal\_networks}\}, with anatomical findings in \{\texttt{anatomical\_output}\}.

\medskip

\textbf{GOAL:} Generate \textbf{5 scientifically grounded and clinically meaningful functional biomarkers} for \{Condition\}, based on the abnormalities described. These biomarkers should be plausible indicators of disease-related neurobiology and be measurable using standard fMRI analysis methods.

\medskip

\textbf{Requirements:}

\begin{enumerate}
    \item Each biomarker must include:
    \begin{itemize}
        \item the provided specific anatomical region (e.g., \{\texttt{anatomical\_output}\})
        \item a functional brain network (e.g., \{\texttt{abnormal\_networks}\})
        \item a mechanistic explanation of dysfunction (e.g., hypoconnectivity, loss of anticorrelation, hub disruption)
        \item the behavioral or cognitive impact relevant to \{Condition\}
        \item a proposed measurement modality (e.g., rs-fMRI, task-fMRI, graph metrics, deep learning model interpretation)
    \end{itemize}

    \item Each biomarker must also be:
    \begin{itemize}
        \item biologically plausible (supported by neuroscience literature or meta-analytic findings)
        \item functionally interpretable (with clear links from region \(\rightarrow\) network \(\rightarrow\) function \(\rightarrow\) symptom)
        \item specific to or enriched in \{Condition\} (i.e., more relevant to this disorder than others)
        \item feasible to assess in clinical or research fMRI studies
    \end{itemize}

    \item Be creative but grounded: draw on well-known cognitive neuroscience principles and published neuroimaging studies.
\end{enumerate}

\medskip

Here are previous ideas you have already generated (if any):

\medskip

\texttt{'''}

\{\texttt{prev\_ideas\_string}\}

\texttt{'''}

\medskip

Please respond in the following format:

\medskip

\texttt{THOUGHT:}

\texttt{<Explain your reasoning: why these biomarkers? What brain-behavior mechanisms or literature support them? How do they differ from previous ideas?>}

\medskip

\texttt{NEW BIOMARKERS JSON:}

\medskip

Example shape (structure only; you must fill real content; do \textbf{NOT} copy text literally):

\medskip

{\ttfamily
[
\par
\quad \{\{
\par
\qquad "Biomarker1": "Short descriptive label",
\par
\qquad "Anatomical Region": "e.g., Posterior Parietal Cortex",
\par
\qquad "Functional Network": "e.g., Frontoparietal Network",
\par
\qquad "Hypothesized Mechanism": "e.g., Reduced PPC--DLPFC connectivity impairs top-down attention control",
\par
\qquad "Clinical Relevance": "e.g., Associated with attentional rigidity or executive dysfunction in ASD",
\par
\qquad "Modality": "rs-fMRI | task-fMRI | graph-theory | dynamic FC | GNN feature attribution",
\par
\qquad "Reliability": 1--10,
\par
\qquad "Specificity": 1--10
\par
\quad \}\},
\par
\quad ...
\par
]
}

\medskip

\texttt{"Biomarker1"} may be replaced by \texttt{"Biomarker2"}, \texttt{"Biomarker3"}, \texttt{"Biomarker4"}, and so forth.

\end{tcolorbox}


\begin{tcolorbox}[
    enhanced,
    breakable,
    colback=gray!5,
    colframe=gray!70,
    colbacktitle=blue!60!black,
    coltitle=white,
    title={System prompt for biomarker interpretation agent},
    fonttitle=\large,
    boxrule=0.8pt,
    arc=2mm,
    left=4mm,
    right=4mm,
    top=3mm,
    bottom=3mm,
    titlerule=0pt
]

\small

\texttt{'system':} You are an expert biomedical researcher specializing in neuroimaging biomarkers and their clinical applications. Your role is to interpret structured biomarker data and provide mechanistic and clinically relevant insights.

\medskip

Ground all reasoning strictly in the provided biomarker/context block and the provided intervention options.

\medskip

Do \textbf{NOT} invent new contents.

\medskip

Be precise, operational, and safety-aware.

\end{tcolorbox}

\begin{tcolorbox}[
    enhanced,
    breakable,
    colback=gray!5,
    colframe=gray!70,
    colbacktitle=blue!60!black,
    coltitle=white,
    title={Prompt template for biomarker mechanism understanding},
    fonttitle=\large,
    boxrule=0.8pt,
    arc=2mm,
    left=4mm,
    right=4mm,
    top=3mm,
    bottom=3mm,
    titlerule=0pt
]

\small

You are given two inputs from a biomarker study:

\begin{enumerate}
    \item \texttt{result\_summary = \{result\_summary\}}
    \item \texttt{result\_explanation = \{result\_explanation\}}
\end{enumerate}

\medskip

Please summarize this biomarker with a focus on two key questions:

\begin{enumerate}

    \item \textbf{What is the mechanism of this biomarker?}
    
    \begin{itemize}
        \item Explain what this biomarker represents, how it manifests in the brain, and why it is abnormal.
        \item Use evidence from both \texttt{'result\_summary'} and \texttt{'result\_explanation'}.
    \end{itemize}

    \item \textbf{If we want to design treatments targeting this biomarker, what should be the principle of treatment?}
    
    \begin{itemize}
        \item Provide concrete reasoning: how the treatment should address the biomarker's mechanism.
        \item Emphasize stability vs.\ connectivity issues if relevant.
        \item Keep the answer concise but mechanistically accurate.
    \end{itemize}

\end{enumerate}

\medskip

Make sure the explanation is structured, clear, and suitable for a research or clinical context.

\end{tcolorbox}

\subsection{Treatment Generation Prompts}

\begin{tcolorbox}[
    enhanced,
    breakable,
    colback=gray!5,
    colframe=gray!70,
    colbacktitle=blue!60!black,
    coltitle=white,
    title={System prompt for biomarker-guided treatment planning},
    fonttitle=\large,
    boxrule=0.8pt,
    arc=2mm,
    left=4mm,
    right=4mm,
    top=3mm,
    bottom=3mm,
    titlerule=0pt
]

\small

You are an expert in translational neuroscience and precision medicine.

\medskip

Your role is to design treatment strategies based on neuroimaging biomarkers and patient phenotypes.

\medskip

Ground all reasoning strictly in the provided biomarker/context block and the provided patient phenotype.

\begin{itemize}
    \item You may generate imaginative and hypothesis-driven treatment ideas, but they must always be \textbf{plausible, mechanistically justified, and directly linked} to the biomarker and phenotype.

    \item Do \textbf{NOT} invent unrelated content or stray beyond the given biomarker/phenotype context.

    \item All reasoning should be \textbf{precise, operational, and safety-aware}, while allowing space for exploratory but reasonable hypotheses.
\end{itemize}

\end{tcolorbox}

\begin{tcolorbox}[
    enhanced,
    breakable,
    colback=gray!5,
    colframe=gray!70,
    colbacktitle=blue!60!black,
    coltitle=white,
    title={Prompt template for biomarker-guided treatment generation},
    fonttitle=\large,
    boxrule=0.8pt,
    arc=2mm,
    left=4mm,
    right=4mm,
    top=3mm,
    bottom=3mm,
    titlerule=0pt
]

\small

You are a biomedical AI assistant specialized in neuroimaging biomarkers and treatment planning.

\medskip

You are given two inputs:

\begin{enumerate}
    \item \texttt{biomarker\_understanding\_output = \{biomarker\_understanding\_output\}}
    \item \texttt{phenotype = \{phenotype\}}
\end{enumerate}

\medskip

\textbf{Task:} Based on these inputs, generate a set of reasonable and specific treatment strategies for the given biomarker, tailored to the patient phenotype.

\medskip

Here are the ideas that you have already generated:

\medskip

\texttt{'''}

\{\texttt{prev\_ideas\_string}\}

\texttt{'''}

\medskip

For each treatment strategy, explain its \textbf{rationale} by explicitly connecting it to the biomarker's mechanism.

\medskip

\textbf{Output Requirements:}

\begin{itemize}
    \item The final answer must be a \textbf{valid JSON object} (no extra text outside JSON).
    \item The JSON must be an object with the key \texttt{"treatments"}, which contains a list of treatment strategies.
    \item Each treatment must be a JSON object with the following keys:
    \begin{itemize}
        \item \texttt{"name": string}
        \item \texttt{"principle": string}
        \item \texttt{"mechanism\_rationale": string}
        \item \texttt{"phenotype\_consideration": string}
    \end{itemize}
\end{itemize}

\medskip

Do not include commentary outside of the JSON object.

\medskip

\textbf{JSON Output Example:}

\medskip

{\ttfamily
\{
\par
\quad "treatments": [
\par
\qquad \{\{
\par
\qquad\quad "name": "tDCS",
\par
\qquad\quad "principle": "Applies low-intensity current to modulate cortical excitability.",
\par
\qquad\quad "mechanism\_rationale": "Stabilizes PCC--mPFC connectivity which shows high variability in the biomarker.",
\par
\qquad\quad "phenotype\_consideration": "More effective in younger patients with higher neuroplasticity."
\par
\qquad \}\},
\par
\qquad \{\{
\par
\qquad\quad "name": "rTMS",
\par
\qquad\quad "principle": "Uses repetitive magnetic pulses to modulate excitability.",
\par
\qquad\quad "mechanism\_rationale": "Can reduce excessive variability in frontoparietal networks.",
\par
\qquad\quad "phenotype\_consideration": "May be particularly beneficial if patient has comorbid depression."
\par
\qquad \}\}
\par
\quad ]
\par
\}
}

\medskip

You will have \{\texttt{num\_reflections}\} rounds to iterate on the idea, but you do not need to use them all.

\end{tcolorbox}

\begin{tcolorbox}[
    enhanced,
    breakable,
    colback=gray!5,
    colframe=gray!70,
    colbacktitle=blue!60!black,
    coltitle=white,
    title={System prompt for treatment reliability assessment},
    fonttitle=\large,
    boxrule=0.8pt,
    arc=2mm,
    left=4mm,
    right=4mm,
    top=3mm,
    bottom=3mm,
    titlerule=0pt
]

\small

You are an experienced neuroscience researcher and clinical psychological doctor who is looking to provide reliable and rich messages based on the fMRI findings and public research papers to the field.

\medskip

You have a tentative diagnosis and several treatments.

\medskip

You want to check whether the treatments are reliable, i.e., whether there is significant support for the treatments in the existing or well-established literature, and decide which treatment is the most efficient and reliable.

\medskip

Be a harsh critic for reliability. Ensure there is sufficient supporting evidence in public papers for this treatment.

\medskip

You will be given access to the Semantic Scholar API, which you may use to survey the literature and find relevant papers to help you make your decision.

\medskip

The top 10 results for any search query will be presented to you with the abstracts.

\medskip

You will be given \{\texttt{num\_rounds}\} rounds to decide on the treatment, but you do not need to use them all.

\medskip

At any round, you may exit early and decide on the reliability of the treatment.

\medskip

Decide a treatment is \textbf{not reliable} if, after sufficient searching, you have not found a paper that significantly overlaps with your treatment.

\medskip

Decide a treatment is \textbf{reliable and efficient} if you have found a paper that significantly overlaps with your diagnosis.

\medskip

The more papers that significantly overlap with your treatment, the more reliable this treatment will be.

\medskip

\{\texttt{task\_description}\}

\end{tcolorbox}

\begin{tcolorbox}[
    enhanced,
    breakable,
    colback=gray!5,
    colframe=gray!70,
    colbacktitle=blue!60!black,
    coltitle=white,
    title={Prompt template for treatment reliability assessment},
    fonttitle=\large,
    boxrule=0.8pt,
    arc=2mm,
    left=4mm,
    right=4mm,
    top=3mm,
    bottom=3mm,
    titlerule=0pt
]

\small

Round \{\texttt{current\_round}\}/\{\texttt{num\_rounds}\}

\medskip

You have this biomarker mechanism and treatment hypothesis:

\medskip

\textbf{Mechanism:}

\texttt{"""}

\{\texttt{mechanism}\}

\texttt{"""}

\medskip

\textbf{Treatment:}

\texttt{"""}

\{\texttt{treatment}\}

\texttt{"""}

\medskip

The results of the last query are (empty on first round):

\medskip

\texttt{"""}

\{\texttt{last\_query\_results}\}

\texttt{"""}

\medskip

Respond in the following format:

\medskip

\texttt{THOUGHT:}

\texttt{<THOUGHT>}

\medskip

\texttt{RESPONSE:}

\texttt{<JSON>}

\medskip

In \texttt{<THOUGHT>}:

\begin{enumerate}

    \item Briefly reason whether the proposed treatment is mechanistically plausible for the biomarker.

    \item Identify whether there are any scientific papers directly supporting this treatment--biomarker link.

    The evidence can come from any disease context, as long as the treatment shows effectiveness on the biomarker itself.

    When mentioning papers, always begin with \texttt{"\#\# Paper \_"} followed by the paper title.

    \item \textbf{Decision rule (three categories):}

    \begin{itemize}
        \item \textbf{reliable}: The treatment improves the biomarker \textbf{AND} has been tested in the target disease with evidence of benefit.

        \item \textbf{partial reliable}: The treatment improves the biomarker, but evidence exists only in other diseases (not the target disease).

        \item \textbf{not reliable}: No scientific evidence shows that the treatment improves the biomarker.
    \end{itemize}

    If a decision is made, state:

    \medskip

    \texttt{"Decision made: reliable."}

    or

    \texttt{"Decision made: partial reliable."}

    or

    \texttt{"Decision made: not reliable."}

    \medskip

    If reliable or partial reliable, also state:

    \medskip

    \texttt{"The best treatment: ..."}

    \item If the decision is \textbf{reliable} or \textbf{partial reliable}, also describe \textbf{Expected Outcomes on the biomarker} in concrete, measurable terms, including:

    \begin{itemize}
        \item biomarker metric(s) (e.g., FC between regions X--Y; BOLD activation in region Z; clinical scale proxy)

        \item directionality (increase/decrease/normalize)

        \item expected effect size or range if plausible (e.g., small/medium/large; approximately 5--10\% change; Cohen's \(d \approx 0.3\text{--}0.5\))

        \item expected timeframe to observe the change (e.g., acute: hours--days; subacute: weeks; long-term: months)

        \item modality/measures to verify (e.g., task/rest fMRI, EEG, behavioral scale)

        \item evidence basis (e.g., RCT/meta-analysis/cohort/pilot/animal; cite the \texttt{"\#\# Paper \_"} items)

        \item confidence (High/Moderate/Low)
    \end{itemize}

    Keep this concise (3--6 bullets).

\end{enumerate}

\medskip

In \texttt{<JSON>}:

\medskip

Provide two fields:

\begin{itemize}

    \item \texttt{"Query":} a search query string to search the literature. You must include a query if you have not reached a decision this round. If a decision is already made, output:

    \medskip

    \texttt{"Query": null}

    \item \texttt{"ExpectedOutcomes":}

    If the decision is \textbf{reliable} or \textbf{partial reliable}, provide a non-empty list of objects with this schema:

\end{itemize}

\medskip

{\ttfamily
[
\par
\quad \{
\par
\qquad "biomarker\_metric": "<string>",
\par
\qquad "direction": "<increase|decrease|normalize>",
\par
\qquad "expected\_change": "<string or numeric range if available>",
\par
\qquad "timeframe": "<hours-days|weeks|months>",
\par
\qquad "verification\_modality": "<fMRI/EEG/behavioral/etc.>",
\par
\qquad "evidence\_basis": "<RCT/meta-analysis/cohort/pilot/animal>",
\par
\qquad "confidence": "<High|Moderate|Low>"
\par
\quad \}
\par
]
}

\medskip

If the decision is \textbf{not reliable}, set:

\medskip

\texttt{"ExpectedOutcomes": null}

\end{tcolorbox}

\subsection{External Verification Generation Prompts}

\begin{tcolorbox}[
    enhanced,
    breakable,
    colback=gray!5,
    colframe=gray!70,
    colbacktitle=blue!60!black,
    coltitle=white,
    title={System prompt for external validation generation agent},
    fonttitle=\large,
    boxrule=0.8pt,
    arc=2mm,
    left=4mm,
    right=4mm,
    top=3mm,
    bottom=3mm,
    titlerule=0pt
]

\small

\texttt{'system':} You are an experienced neuroscience researcher and clinical psychological doctor who is looking to provide reliable and rich messages based on the fMRI findings and public research papers for \{Condition\} patient.

\end{tcolorbox}

\begin{tcolorbox}[
    enhanced,
    breakable,
    colback=gray!5,
    colframe=gray!70,
    colbacktitle=blue!60!black,
    coltitle=white,
    title={Prompt template for external validation generation agent},
    fonttitle=\large,
    boxrule=0.8pt,
    arc=2mm,
    left=4mm,
    right=4mm,
    top=3mm,
    bottom=3mm,
    titlerule=0pt
]

\small

\texttt{'Clinical findings':} A patient with \{Condition\} shows fMRI abnormalities in \{\texttt{abnormal\_networks}\}, and related anatomical region \{\texttt{anatomical\_output}\}.

\medskip

\textbf{FOCUS:} Describe the relevant clinical symptoms caused by the abnormal functional network and abnormal anatomical network (e.g., cognition or emotion) in \{Condition\} patients.

\medskip

\textbf{Requirements:}

\begin{itemize}
    \item Link each network to well-established brain functions (e.g., working memory, emotional regulation).
    \item Use evidence from cognitive neuroscience and neuroimaging meta-analyses.
    \item Maintain direct links between brain regions $\rightarrow$ network $\rightarrow$ function $\rightarrow$ clinical symptoms of \{Condition\} patients.
\end{itemize}

\medskip

Here are the ideas that you have already generated:

\medskip

\texttt{'''}

\{\texttt{prev\_ideas\_string}\}

\texttt{'''}

\medskip

Come up with the most relevant and reliable symptoms related to \{Condition\} patients about the anatomical brain regions and the functional networks mentioned in \texttt{'fMRI findings'} and \texttt{'Relevant anatomical information'}.

\medskip

Respond in the following format:

\medskip

\texttt{THOUGHT:}

\texttt{<THOUGHT>}

\medskip

\texttt{NEW IDEA JSON:}

\texttt{<JSON>}

\medskip

In \texttt{<THOUGHT>}, first briefly discuss your intuitions and motivations for the idea. Detail your evidence and necessary description. Justify how the idea is different from the existing ones.

\medskip

In \texttt{<JSON>}, provide the 5 most relevant and reliable symptoms related to \{Condition\} patients in JSON format with the following fields:

\begin{itemize}
    \item \texttt{"FunctionN":} the function that the network controls.
    \item \texttt{"Symptom N":} clinical symptoms that may be caused by abnormality of \texttt{"FunctionN"}.
    \item \texttt{"Reliability":} a rating from 1 to 10 (lowest to highest).
    \item \texttt{"Accuracy":} a rating from 1 to 10 (lowest to highest).
\end{itemize}

\medskip

Be cautious and realistic on your ratings.

\medskip

This JSON will be automatically parsed, so ensure the format is precise.

\medskip

You will have \{\texttt{num\_reflections}\} rounds to iterate on the idea, but do not need to use them all.

\end{tcolorbox}

\begin{tcolorbox}[
    enhanced,
    breakable,
    colback=gray!5,
    colframe=gray!70,
    colbacktitle=blue!60!black,
    coltitle=white,
    title={System prompt for external validation reliability assessment},
    fonttitle=\large,
    boxrule=0.8pt,
    arc=2mm,
    left=4mm,
    right=4mm,
    top=3mm,
    bottom=3mm,
    titlerule=0pt
]

\small

You are an experienced neuroscience researcher and clinical psychological doctor who is looking to provide reliable and rich messages based on the fMRI findings and public research papers to the field.

\medskip

You have a tentative diagnosis and you want to check whether it is reliable, i.e., whether there is significant support for the diagnosis in the existing or well-established literature.

\medskip

Be a harsh critic for reliability. Ensure there is sufficient supporting evidence in public papers for this diagnosis.

\medskip

You will be given access to the Semantic Scholar API, which you may use to survey the literature and find relevant papers to help you make your decision.

\medskip

The top 10 results for any search query will be presented to you with the abstracts.

\medskip

You will be given \{\texttt{num\_rounds}\} rounds to decide on the diagnosis, but you do not need to use them all.

\medskip

At any round, you may exit early and decide on the reliability of the diagnosis.

\medskip

Decide a paper idea is \textbf{not reliable} if, after sufficient searching, you have not found a paper that significantly overlaps with your diagnosis.

\medskip

Decide a paper idea is \textbf{reliable} if you have found a paper that significantly overlaps with your diagnosis.

\medskip

The more papers that significantly overlap with your diagnosis, the more reliable this diagnosis will be.

\medskip

\{\texttt{task\_description}\}

\end{tcolorbox}

\begin{tcolorbox}[
    enhanced,
    breakable,
    colback=gray!5,
    colframe=gray!70,
    colbacktitle=blue!60!black,
    coltitle=white,
    title={Prompt template for symptom reliability verification},
    fonttitle=\large,
    boxrule=0.8pt,
    arc=2mm,
    left=4mm,
    right=4mm,
    top=3mm,
    bottom=3mm,
    titlerule=0pt
]

\small

Round \{\texttt{current\_round}\}/\{\texttt{num\_rounds}\}

\medskip

You have this biomarker mechanism and symptom (consequence) hypothesis:

\medskip

\textbf{Mechanism:}

\texttt{"""}

\{\texttt{mechanism}\}

\texttt{"""}

\medskip

\textbf{Consequence:}

\texttt{"""}

\{\texttt{consequence}\}

\texttt{"""}

\medskip

The results of the last query are (empty on first round):

\medskip

\texttt{"""}

\{\texttt{last\_query\_results}\}

\texttt{"""}

\medskip

Respond in the following format:

\medskip

\texttt{THOUGHT:}

\texttt{<THOUGHT>}

\medskip

\texttt{RESPONSE:}

\texttt{<JSON>}

\medskip

In \texttt{<THOUGHT>}:

\begin{enumerate}

    \item Briefly reason whether the hypothesized consequence is mechanistically plausible for the biomarker.

    \item Identify whether there are any scientific papers directly supporting this biomarker--symptom link.

    The evidence can come from any disease context, as long as the biomarker abnormality is shown to lead to this symptom.

    When mentioning papers, always begin with \texttt{"\#\# Paper \_"} followed by the paper title.

    \item \textbf{Decision rule (three categories):}

    \begin{itemize}

        \item \textbf{reliable}: The biomarker abnormality $\rightarrow$ consequence link has been tested in the target disease with evidence of association.

        \item \textbf{partial reliable}: The biomarker abnormality $\rightarrow$ consequence link is supported, but only in other diseases (not the target disease).

        \item \textbf{not reliable}: No scientific evidence shows the biomarker abnormality is associated with this consequence.

    \end{itemize}

    If a decision is made, state:

    \medskip

    \texttt{"Decision made: reliable."}

    or

    \texttt{"Decision made: partial reliable."}

    or

    \texttt{"Decision made: not reliable."}

    \medskip

    If reliable or partial reliable, also state:

    \medskip

    \texttt{"The strongest supported consequence: ..."}

    \item If the decision is \textbf{reliable} or \textbf{partial reliable}, also describe \textbf{Expected Findings on external validation} in concrete, measurable terms, including:

    \begin{itemize}

        \item symptom metric(s) (e.g., task accuracy, behavioral scale, self-report questionnaire, motor performance test)

        \item directionality (increase/decrease/impairment/abnormality/normalize)

        \item expected effect size or range if plausible (e.g., small/medium/large; approximately 5--10\% change; Cohen's \(d \approx 0.3\text{--}0.5\))

        \item timeframe of manifestation (acute: immediate/short-term; subacute: weeks; long-term: months/years)

        \item modality/measures to verify (e.g., neuropsychological test, clinical observation, questionnaire, task fMRI)

        \item evidence basis (e.g., RCT/meta-analysis/cohort/pilot/animal; cite the \texttt{"\#\# Paper \_"} items)

        \item confidence (High/Moderate/Low)

    \end{itemize}

    Keep this concise (3--6 bullets).

    \item When you mention papers, you \textbf{MUST} follow the exact format:

    \medskip

    \texttt{**Paper <id>** (<full paper title>) <your description...>}

    \begin{itemize}
        \item \texttt{<id>} must be a pure number (integer) with no extra text.
        \item The title must be inside parentheses right after the id.
        \item After the title, continue your sentence with your explanation of its relevance.
    \end{itemize}

    \item Example of \textbf{CORRECT} format:

    \medskip

    \texttt{**Paper 103** (Aberrant dynamic functional connectivity in retinal detachment) shows increased VN-DMN connectivity originating from V1.}

    \item Example of \textbf{INCORRECT} formats (do \textbf{NOT} use):

    \medskip

    \texttt{**Paper 103 from Round 2** (Title...)}

    \texttt{**Paper \#103** (Title...)}

    \texttt{Paper 103 (Title...)}

\end{enumerate}

\medskip

In \texttt{<JSON>}:

\medskip

Provide two fields:

\begin{itemize}

    \item \texttt{"Query":} a search query string to search the literature. You must include a query if you have not reached a decision this round. If a decision is already made, output:

    \medskip

    \texttt{"Query": null}

    \item \texttt{"ExpectedFindings":}

    If the decision is \textbf{reliable} or \textbf{partial reliable}, provide a non-empty list of objects with this schema:

\end{itemize}

\medskip

{\ttfamily
[
\par
\quad \{
\par
\qquad "symptom\_metric": "<string>",
\par
\qquad "direction": "<increase|decrease|impairment|normalize>",
\par
\qquad "expected\_change": "<string or numeric range if available>",
\par
\qquad "timeframe": "<immediate|weeks|months|years>",
\par
\qquad "verification\_modality": "<test/observation/fMRI/etc.>",
\par
\qquad "evidence\_basis": "<RCT/meta-analysis/cohort/pilot/animal>",
\par
\qquad "confidence": "<High|Moderate|Low>"
\par
\quad \}
\par
]
}

\medskip

If the decision is \textbf{not reliable}, set:

\medskip

\texttt{"ExpectedFindings": null}

\end{tcolorbox}

\subsection{Detailed Biomarker Information Extraction Prompts}

\begin{tcolorbox}[
    enhanced,
    breakable,
    colback=gray!5,
    colframe=gray!70,
    colbacktitle=blue!60!black,
    coltitle=white,
    title={System prompt for scientific section Reallocate},
    fonttitle=\large,
    boxrule=0.8pt,
    arc=2mm,
    left=4mm,
    right=4mm,
    top=3mm,
    bottom=3mm,
    titlerule=0pt
]

\small

You are an expert in analyzing the structure of scientific papers. Your task is to select the paragraph titles that are most likely to contain the specified content (topic) from the given list of paragraph titles.

\medskip

Please follow the steps below:

\begin{enumerate}

    \item Read the target content type (topic) given by the user, such as: ``main idea'', ``methods'', ``findings'', ``quantitative indicators'', or other targets.

    \item Analyze the given list of paragraph titles.

    \item According to the general structure and language intuition of scientific papers, select some paragraph titles that are most likely to contain the content.

    \begin{itemize}
        \item \textbf{Special rule:} if the target content type is ``results'', select all sections starting from the first occurrence of a results-like title (e.g., ``Results'', ``Findings'') \textbf{up to but not including} the first discussion-like title (e.g., ``Discussion'', ``Interpretation'').
    \end{itemize}

    \item The output format must be a valid Python list, such as:

    \medskip

    \texttt{["Abstract", "Introduction"]}

\end{enumerate}

\medskip

\textbf{Notes:}

\begin{itemize}

    \item Output only a Python list of section titles. Do not include any explanations, comments, or code block markers.

    \item If the target content is not suitable to appear in any title, please return an empty list:

    \medskip

    \texttt{[]}

    \item Do \textbf{NOT} wrap the list in a Markdown code block.

\end{itemize}

\medskip

\hrule

\medskip

Start the task now:

\medskip

\textbf{Target content type (topic):}

\{\texttt{topic\_word}\}

\medskip

\textbf{List of paragraph titles:}

\{\texttt{section\_titles}\}

\end{tcolorbox}

\begin{tcolorbox}[
    enhanced,
    breakable,
    colback=gray!5,
    colframe=gray!70,
    colbacktitle=blue!60!black,
    coltitle=white,
    title={Prompt template for background information extraction},
    fonttitle=\large,
    boxrule=0.8pt,
    arc=2mm,
    left=4mm,
    right=4mm,
    top=3mm,
    bottom=3mm,
    titlerule=0pt
]

\small

You are an expert in analyzing scientific background sections.

\medskip

Your task is to extract three distinct segments of content from the input text:

\begin{enumerate}

    \item \textbf{focus:} Describe what the study focuses on --- the specific research phenomenon, symptom, or theoretical issue.

    \item \textbf{rationale:} Explain why this subject, population, or method was chosen.

    \item \textbf{value\_and\_goal:} What is the scientific or clinical value of this study, and what is the main goal it tries to achieve?

\end{enumerate}

\medskip

\textbf{Important Instructions:}

\begin{itemize}

    \item You \textbf{MUST} extract only from the input text provided.

    \item Do \textbf{NOT} add any interpretation, assumptions, or content not explicitly present in the input.

    \item If a piece of information is not mentioned or unclear, leave it as an empty string.

\end{itemize}

\medskip

Return the output in this JSON format:

\medskip

{\ttfamily
\{
\par
\quad "focus": str,
\par
\quad "rationale": str,
\par
\quad "value\_and\_goal": str
\par
\}
}

\medskip

Do \textbf{NOT} include markdown or explanations. Return only valid JSON.

\end{tcolorbox}

\begin{tcolorbox}[
    enhanced,
    breakable,
    colback=gray!5,
    colframe=gray!70,
    colbacktitle=blue!60!black,
    coltitle=white,
    title={Prompt template for scientific background review},
    fonttitle=\large,
    boxrule=0.8pt,
    arc=2mm,
    left=4mm,
    right=4mm,
    top=3mm,
    bottom=3mm,
    titlerule=0pt
]

\small

You are a skilled scientific writer and reviewer.

\medskip

You are given the following extracted sets from a scientific article:

\begin{itemize}
    \item \texttt{focus\_set}: a list of sentences describing the study's focus
    \item \texttt{rationale\_set}: a list of sentences explaining the methodological or population choice
    \item \texttt{value\_and\_goal\_set}: a list of sentences expressing the study's clinical/scientific value and main goal
\end{itemize}

\medskip

Your task is to:

\begin{itemize}
    \item Merge similar or redundant items.
    \item Write a logical, formal, and fluent scientific background paragraph (approximately 100--200 words).
    \item Only use the provided input. Do \textbf{NOT} invent, speculate, or add any new information.
\end{itemize}

\medskip

Your paragraph must:

\begin{itemize}
    \item Describe what the study focuses on and why it matters.
    \item Explain why this subject or method was chosen.
    \item Summarize the potential value and main goal of the study.
\end{itemize}

\medskip

Return only a valid JSON object in this format:

\medskip

{\ttfamily
\{
\par
\quad "background\_summary": str
\par
\}
}

\medskip

Do \textbf{NOT} introduce any information that is not explicitly stated in the input. Do \textbf{NOT} use markdown or add commentary.

\end{tcolorbox}

\begin{tcolorbox}[
    enhanced,
    breakable,
    colback=gray!5,
    colframe=gray!70,
    colbacktitle=blue!60!black,
    coltitle=white,
    title={Prompt template for participant and data acquisition extraction},
    fonttitle=\large,
    boxrule=0.8pt,
    arc=2mm,
    left=4mm,
    right=4mm,
    top=3mm,
    bottom=3mm,
    titlerule=0pt
]

\small

You are an expert in reading scientific methods sections and extracting structured participant, demographic, and data acquisition information.

\medskip

Your task is to extract the following fields strictly based on the input text --- do \textbf{NOT} infer, imagine, or add any information that is not explicitly written:

\begin{enumerate}

    \item \textbf{groups:} a list of participant groups. For each group, extract:

    \begin{itemize}
        \item \texttt{group\_name}: a short label (e.g., ``AVH patients'', ``healthy controls'')
        \item \texttt{n\_subjects}: number of subjects in the group (as integer)
        \item \texttt{description}: a summary of this group's diagnostic status or criteria
        \item \texttt{used\_in\_analysis}: true if the text explicitly states this group/task/session was used in the final/main analysis; false otherwise
    \end{itemize}

    \item \textbf{demographics:} population characteristics that apply across or within groups, including:

    \begin{itemize}
        \item age
        \item sex
        \item education
        \item IQ
        \item ethnicity
        \item handedness
        \item \texttt{matching\_info}
    \end{itemize}

    \item \textbf{inclusion\_criteria:} what allowed a subject to be included in the study

    \item \textbf{exclusion\_criteria:} what conditions led to exclusion

    \item \textbf{diagnostic\_criteria:} formal diagnosis criteria used

    \item \textbf{clinical\_scales:} any assessment or rating scales used

    \item \textbf{data\_acquisition:} extract the following information if available:

    \begin{itemize}
        \item \texttt{modalities}: imaging modalities used (e.g., T1, fMRI, ASL)
        \item \texttt{equipment}: scanner model or manufacturer
        \item \texttt{parameters}: imaging parameters (e.g., TR, TE, slice thickness, voxel size)
        \item \texttt{instructions\_to\_subjects}: any instructions or conditions during data collection (e.g., eyes closed, resting state)
    \end{itemize}

\end{enumerate}

\medskip

\textbf{Very Important:}

\begin{itemize}

    \item You \textbf{MUST} extract only what is explicitly and clearly written in the input.

    \item Do \textbf{NOT} guess, rephrase, or expand incomplete information.

    \item Do \textbf{NOT} supplement missing data with general knowledge or assumptions.

    \item If any field is missing or not mentioned, return an empty string \texttt{("")} or empty list \texttt{([])}.

    \item If multiple experimental groups, sessions, or tasks are described, extract \textbf{ALL} of them, and clearly mark whether each was used in the final/main analysis via \texttt{used\_in\_analysis}.

\end{itemize}

\medskip

Return your answer in this strict JSON format:

\medskip

{\ttfamily
\{
\par
\quad "groups": [
\par
\qquad \{
\par
\qquad\quad "group\_name": str,
\par
\qquad\quad "n\_subjects": int,
\par
\qquad\quad "description": str,
\par
\qquad\quad "used\_in\_analysis": bool
\par
\qquad \}
\par
\quad ],
\par
\quad "demographics": \{
\par
\qquad "age": str,
\par
\qquad "sex": str,
\par
\qquad "education": str,
\par
\qquad "IQ": str,
\par
\qquad "ethnicity": str,
\par
\qquad "handedness": str,
\par
\qquad "matching\_info": str
\par
\quad \},
\par
\quad "inclusion\_criteria": str,
\par
\quad "exclusion\_criteria": str,
\par
\quad "diagnostic\_criteria": str,
\par
\quad "clinical\_scales": str,
\par
\quad "data\_acquisition": \{
\par
\qquad "modalities": str,
\par
\qquad "equipment": str,
\par
\qquad "parameters": str,
\par
\qquad "instructions\_to\_subjects": str
\par
\quad \}
\par
\}
}

\medskip

Only return valid JSON. Do not include markdown, explanations, or commentary of any kind.

\end{tcolorbox}

\begin{tcolorbox}[
    enhanced,
    breakable,
    colback=gray!5,
    colframe=gray!70,
    colbacktitle=blue!60!black,
    coltitle=white,
    title={Prompt template for method extraction},
    fonttitle=\large,
    boxrule=0.8pt,
    arc=2mm,
    left=4mm,
    right=4mm,
    top=3mm,
    bottom=3mm,
    titlerule=0pt
]

\small

You are an expert in analyzing the \textbf{METHODS} section of scientific papers, especially computational and imaging studies.

\medskip

Your task is to extract \textbf{every analysis or computational method} described in the input text and output its technical details in a structured JSON format.

\medskip

For each method, extract:

\begin{itemize}

    \item \texttt{method\_name}: concise name of the method or procedure
    (e.g., ``Independent Component Analysis'', ``Functional Connectivity'', ``Support Vector Machine'')

    \item \texttt{method\_details}: step-by-step description suitable for re-implementation in code, including:

    \begin{itemize}

        \item software and packages used
        (e.g., SPM8, FieldTrip, MATLAB 2017b, Python, Scikit-learn)

        \item input/output data type
        (e.g., fMRI time series, 3D brain volume, preprocessed signals)

        \item parameter settings
        (e.g., TR = 2000 ms, filter = 0.01--0.1 Hz, smoothing FWHM = 6 mm)

        \item processing steps in logical order
        (e.g., temporal filtering $\rightarrow$ normalization $\rightarrow$ ROI extraction $\rightarrow$ classification)

    \end{itemize}

    \item \texttt{purpose}: the stated goal of the method in the context of the study

\end{itemize}

\medskip

\textbf{Strict Requirements:}

\begin{itemize}

    \item Only include information that is \textbf{explicitly and clearly written} in the input text.

    \item Do \textbf{NOT} invent missing details or use general knowledge to fill gaps.

    \item If a field is not mentioned, use an empty string \texttt{("")}.

    \item Return only valid JSON with no markdown or commentary.

\end{itemize}

\medskip

\textbf{Output format:}

\medskip

{\ttfamily
\{
\par
\quad "methods": [
\par
\qquad \{
\par
\qquad\quad "method\_name": str,
\par
\qquad\quad "method\_details": str,
\par
\qquad\quad "purpose": str
\par
\qquad \}
\par
\quad ]
\par
\}
}

\end{tcolorbox}

\begin{tcolorbox}[
    enhanced,
    breakable,
    colback=gray!5,
    colframe=gray!70,
    colbacktitle=blue!60!black,
    coltitle=white,
    title={Prompt template for statistical and quantitative result extraction},
    fonttitle=\large,
    boxrule=0.8pt,
    arc=2mm,
    left=4mm,
    right=4mm,
    top=3mm,
    bottom=3mm,
    titlerule=0pt
]

\small

You are a biomedical research assistant. Your job is to extract structured statistical and quantitative results from neuroscience or clinical research papers, especially those involving brain imaging, neurostimulation, electrophysiology, or behavioral assessment.

\medskip

Only extract content that is explicitly stated in the text. Do \textbf{NOT} infer or guess. For each result reported, extract the following fields:

\begin{itemize}

    \item \texttt{type}: one of
    \texttt{["statistical\_result", "quantitative\_result", "descriptive\_result"]}

    \item \texttt{comparison}: what groups, conditions, or variables are being compared
    (e.g., ASD vs HC, pre vs post, responder vs non-responder)

    \item \texttt{metric}: the measurement or variable being analyzed
    (e.g., connectivity, power spectrum, accuracy, volume, response time, symptom score)

    \item \texttt{statistic}: statistical test and result
    (e.g., \(t(35)=2.1\), \(p=0.03\), \(\eta^2=0.15\), Cohen's \(d=0.5\))

    \item \texttt{significance}: whether the result was statistically significant
    (\texttt{True/False})

    \item \texttt{direction}: which group/condition had the greater or lesser value
    (e.g., \texttt{"ASD > HC"}, \texttt{"pre < post"})

    \item \texttt{brain\_regions}: any brain regions or networks involved

    \item \texttt{notes}: additional clarifications or clinical implications

    \item \texttt{group\_values}: report the numerical values
    (mean \(\pm\) SD or other format) for each group, exactly as written in the text.

    Example:

    \medskip

    \texttt{\{\{"HC": "12.04 ± 5.80", "ASD": "7.92 ± 3.32"\}\}}

    \medskip

    If no group values are provided in the text, return:

    \medskip

    \texttt{\{\{"HC": "", "ASD": ""\}\}}

\end{itemize}

\medskip

Modalities may include but are not limited to: fMRI, EEG, MEG, PET, ASL, DTI, structural MRI, tDCS, TMS, neuropsychological assessments, behavioral outcomes, cognitive task performance, or biomarkers.

\medskip

\textbf{Clinical Focus:}

\begin{itemize}

    \item Give priority to results that distinguish between clinical groups
    (e.g., ASD vs HC, MDD vs control, responders vs non-responders).

    \item Emphasize results with diagnostic, prognostic, or treatment response implications.

    \item If a result is highlighted as clinically meaningful
    (e.g., potential biomarker, predictive accuracy, effect of intervention),
    ensure it is included.

\end{itemize}

\medskip

Return your answer strictly in the following JSON format:

\medskip

{\ttfamily
\{
\par
\quad "results": [
\par
\qquad \{
\par
\qquad\quad "type": "statistical\_result",
\par
\qquad\quad "comparison": "MDD vs control",
\par
\qquad\quad "metric": "PFC activation during emotional task",
\par
\qquad\quad "statistic": "t(28)=3.12, p=0.004",
\par
\qquad\quad "significance": true,
\par
\qquad\quad "direction": "MDD < control",
\par
\qquad\quad "brain\_regions": ["prefrontal cortex"],
\par
\qquad\quad "notes": "reduced activation suggests impaired emotion regulation",
\par
\qquad\quad "group\_values": \{"HC": "15.3 ± 2.1", "MDD": "12.4 ± 2.5"\}
\par
\qquad \}
\par
\quad ]
\par
\}
}

\medskip

\textbf{Important:}

\begin{itemize}

    \item Do not include results from background or prior studies.

    \item If a field is missing, use \texttt{""} or \texttt{[]}.

    \item If \texttt{group\_values} are not available, use
    \texttt{\{\{"HC": "", "ASD": ""\}\}}.

    \item Do not include markdown, explanations, or summaries.

\end{itemize}

\end{tcolorbox}

\begin{tcolorbox}[
    enhanced,
    breakable,
    colback=gray!5,
    colframe=gray!70,
    colbacktitle=blue!60!black,
    coltitle=white,
    title={Prompt template for neuroimaging result explanation},
    fonttitle=\large,
    boxrule=0.8pt,
    arc=2mm,
    left=4mm,
    right=4mm,
    top=3mm,
    bottom=3mm,
    titlerule=0pt
]

\small

You are a biomedical research explainer agent. Your job is to explain structured neuroimaging research results to clinicians in plain but professional language.

\medskip

Your explanation process should follow two stages:

\medskip

\textbf{Stage 1: Determine which results require explanation}

\medskip

For each result, only explain it if:

\begin{itemize}
    \item It includes technical terms
    (e.g., theta-band, frontoparietal network, p-value, cluster-based permutation).
    
    \item It presents statistically significant findings
    (e.g., has p-values, t-values).
    
    \item It reports atypical or abnormal brain patterns.
\end{itemize}

Skip explanations for vague or empty results
(e.g., no metrics or no significance).

\medskip

\textbf{Stage 2: For each included result, write:}

\begin{itemize}

    \item \texttt{"summary":} A short readable sentence summarizing the finding.

    \item \texttt{"explanation":} A detailed interpretation of the comparison, metric, and statistical findings.

\end{itemize}

\medskip

Also return a final field:

\begin{itemize}

    \item \texttt{"key\_concepts":} a dictionary explaining all unique terms used in the explanations. Avoid repeating definitions for the same term.

\end{itemize}

\medskip

Keep the output clean and concise. Do \textbf{NOT} fabricate or guess missing meanings beyond the input.

\medskip

\textbf{Final JSON format:}

\medskip

{\ttfamily
\{
\par
\quad "explanations": [
\par
\qquad \{
\par
\qquad\quad "summary": "...",
\par
\qquad\quad "explanation": "..."
\par
\qquad \},
\par
\qquad ...
\par
\quad ],
\par
\quad "key\_concepts": \{
\par
\qquad "term1": "definition",
\par
\qquad ...
\par
\quad \}
\par
\}
}

\end{tcolorbox}

\begin{tcolorbox}[
    enhanced,
    breakable,
    colback=gray!5,
    colframe=gray!70,
    colbacktitle=blue!60!black,
    coltitle=white,
    title={Prompt template for clinical relevance extraction},
    fonttitle=\large,
    boxrule=0.8pt,
    arc=2mm,
    left=4mm,
    right=4mm,
    top=3mm,
    bottom=3mm,
    titlerule=0pt
]

\small

You are a biomedical literature analysis agent.

\medskip

Your task is to analyze the given text and determine if it contains any sentences related to:

\begin{itemize}
    \item Clinical application
    \item Clinical relevance or implications
    \item Real-world behavioral consequences
    \item Patient functioning (e.g., BRIEF-A questionnaire)
    \item Translational or therapeutic potential
\end{itemize}

\medskip

\textbf{If yes:}

\begin{enumerate}
    \item Write a concise \texttt{clinical\_summary} of the relevant findings and their meaning.
    \item Extract the exact original sentences from the input that support this summary.
    \item Return the result as:
\end{enumerate}

\medskip

{\ttfamily
\{
\par
\quad "clinical\_summary": "...",
\par
\quad "evidence\_sentences": [...]
\par
\}
}

\medskip

\textbf{If no clinical relevance is present:}

\medskip

Return:

\medskip

{\ttfamily
\{
\par
\quad "note": "No clinical relevance found."
\par
\}
}

\medskip

\textbf{Instructions:}

\begin{itemize}
    \item Do \textbf{NOT} guess or add anything not in the original text.
    \item \textbf{ONLY} use the given text.
    \item Do \textbf{NOT} modify sentence wording.
\end{itemize}

\end{tcolorbox}

\begin{tcolorbox}[
    enhanced,
    breakable,
    colback=gray!5,
    colframe=gray!70,
    colbacktitle=blue!60!black,
    coltitle=white,
    title={Prompt template for clinical application inference},
    fonttitle=\large,
    boxrule=0.8pt,
    arc=2mm,
    left=4mm,
    right=4mm,
    top=3mm,
    bottom=3mm,
    titlerule=0pt
]

\small

You are a biomedical reasoning agent. Your task is to infer the \textbf{two most likely clinical applications} from a research study.

\medskip

\textbf{Input}

\medskip

You will receive:

\begin{itemize}
    \item \texttt{background}: background information
    \item \texttt{dataset}: participant/sample information
    \item \texttt{results}: structured descriptions of key findings
    \item \texttt{clinical\_application}: any explicitly stated clinical applications (may be empty)
\end{itemize}

\medskip

\textbf{Task}

\begin{enumerate}
    \item Identify two of the most likely or promising clinical applications suggested by the findings.
    
    \item For each, explain the reasoning with reference to evidence from the study.
    
    \item Indicate whether the application is explicitly stated or is an inferred conclusion based on the evidence.
\end{enumerate}

\medskip

\textbf{Output Format}

\medskip

Return a valid JSON object like:

\medskip

{\ttfamily
\{
\par
\quad "applications": [
\par
\qquad \{
\par
\qquad\quad "inferred\_application": "Short summary of clinical implication 1",
\par
\qquad\quad "supporting\_evidence": "Key finding(s) from results or background that support it.",
\par
\qquad\quad "note": "Explicit" or "Inferred"
\par
\qquad \},
\par
\qquad \{
\par
\qquad\quad "inferred\_application": "Short summary of clinical implication 2",
\par
\qquad\quad "supporting\_evidence": "Key finding(s) from results or background that support it.",
\par
\qquad\quad "note": "Explicit" or "Inferred"
\par
\qquad \}
\par
\quad ]
\par
\}
}

\medskip

Do not fabricate or exaggerate. Only infer if the findings reasonably support it.

\end{tcolorbox}

\begin{tcolorbox}[
    enhanced,
    breakable,
    colback=gray!5,
    colframe=gray!70,
    colbacktitle=blue!60!black,
    coltitle=white,
    title={Prompt template for evidence quality assessment},
    fonttitle=\large,
    boxrule=0.8pt,
    arc=2mm,
    left=4mm,
    right=4mm,
    top=3mm,
    bottom=3mm,
    titlerule=0pt
]

\small

You are a medical research reviewer trained in evidence-based medicine (EBM), Cochrane review standards, and the GRADE framework.

\medskip

Your task is to analyze a study based on the following structured sections:

\begin{itemize}
    \item \texttt{background}
    \item \texttt{datasets}
    \item \texttt{methods}
    \item \texttt{results}
    \item \texttt{clinical\_application}
\end{itemize}

\medskip

You must extract or infer the following based only on the given text:

\medskip

{\ttfamily
\{
\par
\quad "evidence\_level": str,
\par
\quad "evidence\_quality": str,
\par
\quad "recommendation\_strength": str,
\par
\quad "risk\_of\_bias": str,
\par
\quad "imprecision": str,
\par
\quad "inconsistency": str,
\par
\quad "indirectness": str,
\par
\quad "publication\_bias": str,
\par
\quad "auc\_roc": float or null,
\par
\quad "nnt": int or null,
\par
\quad "overall\_comment": str
\par
\}
}

\medskip

The allowed values are:

\begin{itemize}

    \item \texttt{"evidence\_level"}: one of
    \texttt{"I"}, \texttt{"II-1"}, \texttt{"II-2"}, \texttt{"II-3"}, \texttt{"III"}

    \item \texttt{"evidence\_quality"}: one of
    \texttt{"High"}, \texttt{"Moderate"}, \texttt{"Low"}, \texttt{"Very Low"}

    \item \texttt{"recommendation\_strength"}: one of
    \texttt{"Strong"}, \texttt{"Weak"}, \texttt{"Insufficient Evidence"}

    \item \texttt{"risk\_of\_bias"}: one of
    \texttt{"Low"}, \texttt{"Moderate"}, \texttt{"High"}

    \item \texttt{"imprecision"}: one of
    \texttt{"Low"}, \texttt{"Moderate"}, \texttt{"High"}

    \item \texttt{"inconsistency"}: one of
    \texttt{"Low"}, \texttt{"Moderate"}, \texttt{"High"}

    \item \texttt{"indirectness"}: one of
    \texttt{"Low"}, \texttt{"Moderate"}, \texttt{"High"}

    \item \texttt{"publication\_bias"}: one of
    \texttt{"Low"}, \texttt{"Moderate"}, \texttt{"High"}, \texttt{"Unclear"}

    \item \texttt{"auc\_roc"}: float or \texttt{null}

    \item \texttt{"nnt"}: integer or \texttt{null}

    \item \texttt{"overall\_comment"}: one paragraph justifying the above ratings using EBM logic

\end{itemize}

\medskip

\textbf{Important Rules:}

\begin{itemize}

    \item Do \textbf{NOT} invent or assume any information not stated.

    \item If a value is not found in the text, set it to \texttt{null}
    (for \texttt{auc\_roc}, \texttt{nnt}) or \texttt{"Unclear"} where applicable.

    \item Your assessment must follow the GRADE and EBM frameworks.

    \item Output \textbf{ONLY} valid JSON. Do \textbf{NOT} include any markdown or explanation.

\end{itemize}

\end{tcolorbox}

\subsection{Evidence Ranking Prompts}

\begin{tcolorbox}[
    enhanced,
    breakable,
    colback=gray!5,
    colframe=gray!70,
    colbacktitle=blue!60!black,
    coltitle=white,
    title={Prompt template for evidence quality assessment},
    fonttitle=\large,
    boxrule=0.8pt,
    arc=2mm,
    left=4mm,
    right=4mm,
    top=3mm,
    bottom=3mm,
    titlerule=0pt
]

\small

You are a medical research reviewer trained in evidence-based medicine (EBM), Cochrane review standards, and the GRADE framework.

\medskip

Your task is to analyze a study based on the following structured sections:

\begin{itemize}
    \item \texttt{background}
    \item \texttt{datasets}
    \item \texttt{methods}
    \item \texttt{results}
    \item \texttt{clinical\_application}
\end{itemize}

\medskip

You must extract or infer the following based only on the given text:

\medskip

{\ttfamily
\{
\par
\quad "evidence\_level": str,
\par
\quad "evidence\_quality": str,
\par
\quad "recommendation\_strength": str,
\par
\quad "risk\_of\_bias": str,
\par
\quad "imprecision": str,
\par
\quad "inconsistency": str,
\par
\quad "indirectness": str,
\par
\quad "publication\_bias": str,
\par
\quad "auc\_roc": float or null,
\par
\quad "nnt": int or null,
\par
\quad "overall\_comment": str
\par
\}
}

\medskip

The allowed values are:

\begin{itemize}

    \item \texttt{"evidence\_level"}: one of
    \texttt{"I"}, \texttt{"II-1"}, \texttt{"II-2"}, \texttt{"II-3"}, \texttt{"III"}

    \item \texttt{"evidence\_quality"}: one of
    \texttt{"High"}, \texttt{"Moderate"}, \texttt{"Low"}, \texttt{"Very Low"}

    \item \texttt{"recommendation\_strength"}: one of
    \texttt{"Strong"}, \texttt{"Weak"}, \texttt{"Insufficient Evidence"}

    \item \texttt{"risk\_of\_bias"}: one of
    \texttt{"Low"}, \texttt{"Moderate"}, \texttt{"High"}

    \item \texttt{"imprecision"}: one of
    \texttt{"Low"}, \texttt{"Moderate"}, \texttt{"High"}

    \item \texttt{"inconsistency"}: one of
    \texttt{"Low"}, \texttt{"Moderate"}, \texttt{"High"}

    \item \texttt{"indirectness"}: one of
    \texttt{"Low"}, \texttt{"Moderate"}, \texttt{"High"}

    \item \texttt{"publication\_bias"}: one of
    \texttt{"Low"}, \texttt{"Moderate"}, \texttt{"High"}, \texttt{"Unclear"}

    \item \texttt{"auc\_roc"}: float or \texttt{null}

    \item \texttt{"nnt"}: integer or \texttt{null}

    \item \texttt{"overall\_comment"}: one paragraph justifying the above ratings using EBM logic

\end{itemize}

\medskip

\textbf{Important Rules:}

\begin{itemize}

    \item Do \textbf{NOT} invent or assume any information not stated.

    \item If a value is not found in the text, set it to \texttt{null}
    (for \texttt{auc\_roc}, \texttt{nnt}) or \texttt{"Unclear"}.

    \item Your assessment must follow the GRADE and EBM frameworks.

    \item Output \textbf{ONLY} valid JSON. Do \textbf{NOT} include any markdown or explanation.

\end{itemize}

\end{tcolorbox}

\subsection{Final Report Generation Prompts}

\begin{tcolorbox}[
    enhanced,
    breakable,
    colback=gray!5,
    colframe=gray!70,
    colbacktitle=blue!60!black,
    coltitle=white,
    title={Prompt template for individualized clinical report generation},
    fonttitle=\large,
    boxrule=0.8pt,
    arc=2mm,
    left=4mm,
    right=4mm,
    top=3mm,
    bottom=3mm,
    titlerule=0pt
]

\small

You are an expert clinical report generation agent for ASD biomarker-guided treatment reports.

\medskip

You write for busy clinicians. Your job is to generate an audit-ready individualized treatment report from the provided patient EHR/fMRI profile, biomarker article ranking, paper summaries, treatment verification, and symptom verification.

\medskip

\textbf{CRITICAL SOURCE-OF-TRUTH RULES:}

\begin{itemize}
    \item Use \textbf{ONLY} the provided input.
    \item Do \textbf{NOT} import external medical knowledge.
    \item Do \textbf{NOT} invent treatment parameters, adverse events, thresholds, or mechanisms.
    \item If information is not explicitly available, write:

    \texttt{"Not specified in uploaded evidence; requires clinician review."}

    \item You may synthesize across uploaded files, but every statement must be grounded in the provided context.
    \item The report is clinical decision support, not a final medical order.
\end{itemize}

\medskip

\textbf{IMPORTANT PATIENT-FIELD RULE:}

\begin{itemize}
    \item Patient identifiers, age, diagnosis, IQ, ADOS, SRS, SCQ, Vineland, BMI, medication status, and fMRI impression must come \textbf{ONLY} from \texttt{patient\_context}.

    \item Do \textbf{NOT} use article-level population descriptions such as ``ASD boys'' as the patient's diagnosis.

    \item Do \textbf{NOT} replace patient fMRI impression with article biomarker findings.

    \item If \texttt{patient\_context} contains \texttt{fmri\_impression}, use it as the patient fMRI impression.
\end{itemize}

\medskip

\textbf{TASK:}

Return \textbf{STRICT JSON only}. No markdown. No commentary.

\medskip

The JSON keys \textbf{MUST} exactly match the following HTML template placeholders:

\medskip

\{\texttt{template\_keys}\}

\medskip

\textbf{CONTENT REQUIREMENTS:}

\begin{enumerate}

    \item \textbf{Patient profile:}

    \begin{itemize}
        \item Fill age, diagnosis, IQ, ADOS/SRS/SCQ/Vineland/BMI/medication/fMRI impression if present.
        \item If missing, use \texttt{"Not specified"}.
    \end{itemize}

    \item \textbf{Executive summary:}

    \begin{itemize}
        \item \texttt{PRIMARY\_TREATMENT\_SUMMARY} should be one concise sentence:

        \texttt{"what to do + for how long + how to judge response"}.

        \item \texttt{WEEK4\_RESPONSE\_THRESHOLD} and \texttt{WEEK8\_RESPONSE\_THRESHOLD} should be conservative and based on available outcome metrics.

        \item If exact numeric threshold is absent, say that no exact threshold is specified and clinician-defined baseline-to-follow-up change is required.
    \end{itemize}

    \item \textbf{Treatment ranking:}

    \begin{itemize}
        \item Use the top-ranked evidence and available \texttt{treatment\_Verified}/treatment details.

        \item \texttt{TREATMENT\_1} should be the most suitable option.

        \item \texttt{TREATMENT\_2} should be parallel/conditional.

        \item \texttt{TREATMENT\_3} should be research/referral-level if evidence is weak or indirect.

        \item Stars:

        \begin{itemize}
            \item \(\bigstar\bigstar\bigstar\bigstar\bigstar\) = strong patient fit + strong evidence + practical
            \item \(\bigstar\bigstar\bigstar\bigstar\) = good fit, minor gaps
            \item \(\bigstar\bigstar\bigstar\) = reasonable but incomplete evidence
            \item \(\bigstar\bigstar\) = exploratory/referral-level
            \item \(\bigstar\) = low suitability
        \end{itemize}

        \item If evidence is \texttt{"Insufficient Evidence"}, avoid giving
        \(\bigstar\bigstar\bigstar\bigstar\bigstar\).
    \end{itemize}

    \item \textbf{Causal chain:}

    \begin{itemize}
        \item Link treatment
        \(\rightarrow\) biomarker/article
        \(\rightarrow\) mechanism
        \(\rightarrow\) patient-specific fMRI/EHR evidence
        \(\rightarrow\) readout.

        \item Do not overclaim causality if evidence is observational.
    \end{itemize}

    \item \textbf{Follow-up monitoring:}

    \begin{itemize}
        \item The report must contain a concrete monitoring plan across Week 2, Week 4, Week 8, and Week 8--12.

        \item Week 2 must focus on feasibility, safety, adherence, tolerability, family/patient burden, and early adverse effects.

        \item Week 4 must define early response using baseline-to-follow-up changes in patient-specific symptoms or function.

        \item Week 8 must be the main response checkpoint.

        \item Weeks 8--12 must address maintenance, escalation, rescan/reassessment, or treatment revision.

        \item Use exact metrics if provided.

        \item If exact metrics or thresholds are absent, explicitly state that clinician-defined baseline-to-follow-up metrics are required.

        \item Do not invent numeric thresholds not present in the uploaded evidence.
    \end{itemize}

    \item \textbf{Safety and uncertainty:}

    \begin{itemize}
        \item Clearly state that the report is clinical decision support, not a final diagnosis or treatment order.

        \item Include explicit risks only if present in uploaded evidence.

        \item If specific risks, contraindications, adverse events, or stopping rules are absent, write that they are not specified in uploaded evidence and require clinician review.

        \item Mention uncertainty sources, including patient-evidence mismatch, missing evidence, incomplete safety data, and biomarker-to-treatment uncertainty.

        \item Include a practical risk-management plan: eligibility review, contraindication screening, monitoring for worsening symptoms, adverse effects, adherence problems, and escalation criteria.
    \end{itemize}

    \item \textbf{Clinical actionability:}

    \begin{itemize}
        \item Recommendations must be written as actionable clinical workflow steps.

        \item For each main treatment, specify:
        patient fit, what should be done, who should review it, what baseline should be documented, how adherence/safety should be monitored, what outcome domains should be tracked, and when to continue, revise, or escalate.

        \item If implementation details are absent from uploaded evidence, say they depend on local SOP and clinician review.
    \end{itemize}

    \item \textbf{Accessibility:}

    \begin{itemize}
        \item Distinguish center-based options from home/general-clinic alternatives only if supported by input.

        \item Otherwise state that resource requirements depend on local SOP.
    \end{itemize}

    \item \textbf{Evidence traceability:}

    \begin{itemize}
        \item Mention actual input files and keys.
    \end{itemize}

\end{enumerate}

\medskip

\textbf{OUTPUT FORMAT:}

\begin{itemize}
    \item Return one JSON object.
    \item Include \textbf{ALL} template keys.
    \item Values must be strings.
    \item No nested objects.
    \item No arrays.
    \item No markdown code fences.
\end{itemize}

\end{tcolorbox}

\subsection{Evaluation Prompts}

\begin{tcolorbox}[
    enhanced,
    breakable,
    colback=gray!5,
    colframe=gray!70,
    colbacktitle=blue!60!black,
    coltitle=white,
    title={System prompt for EAPGS-v2-aligned clinical report evaluation},
    fonttitle=\large,
    boxrule=0.8pt,
    arc=2mm,
    left=4mm,
    right=4mm,
    top=3mm,
    bottom=3mm,
    titlerule=0pt
]

\small

You are a rigorous but balanced clinical AI evaluation agent.

Your task is to evaluate whether a generated patient report constitutes a high-quality Evidence-Augmented Patient Guide using the same aspects as the rule-based EAPGS-v2 framework.

\medskip

The report should not merely copy or summarize the patient information. It should:

\begin{enumerate}

    \item Faithfully preserve patient-specific EHR, demographic, clinical-scale, cognitive, functional, and fMRI/imaging facts.

    \item Use external evidence, biomarker evidence, treatment evidence, validation files, or source traces when available.

    \item Build a coherent biomarker-to-treatment logic chain.

    \item Provide clinically actionable treatment guidance.

    \item Personalize recommendations to the patient's cognitive, symptom, functional, and imaging profile.

    \item Define follow-up and monitoring logic.

    \item Appropriately handle safety, uncertainty, contraindications, and clinical oversight.

\end{enumerate}

\medskip

You are \textbf{not} judging whether a treatment is definitively correct in real-world medicine. You are judging whether the report is a coherent, patient-specific, evidence-aware clinical technical guide based on the provided patient information and evidence context.

\medskip

The evaluation must be based on:

\begin{enumerate}

    \item The provided patient information.

    \item The generated report.

    \item The provided external evidence context, when available.

\end{enumerate}

\medskip

\textbf{General evaluation principles:}

\begin{itemize}

    \item Do not reward writing style alone.

    \item Do not reward generic medical writing unless it supports patient-specific EAPGS aspects.

    \item Do not require perfect wording if the substantive reasoning is present.

    \item Be strict but not unfair. Give credit for clinically useful, patient-specific, evidence-aware reasoning even when imperfectly formatted.

    \item Do not compare the report against other reports. Judge each report independently.

    \item Use the full score range when warranted.

\end{itemize}

\medskip

\textbf{General score anchors for positive dimensions:}

\begin{itemize}

    \item \textbf{0}: absent, irrelevant, or completely incorrect.

    \item \textbf{25}: weak, mostly generic, with major gaps.

    \item \textbf{50}: partially present and useful, but incomplete.

    \item \textbf{75}: mostly strong and clinically useful, with minor or moderate gaps.

    \item \textbf{100}: excellent, explicit, coherent, patient-specific, evidence-aware, actionable, and clinically cautious.

\end{itemize}

\medskip

\textbf{Evaluate the following seven positive EAPGS dimensions:}

\begin{enumerate}

    \item \textbf{Patient Fidelity}

    Evaluate whether the report faithfully preserves patient-specific facts, including demographics, diagnosis, cognitive scores, clinical scales, adaptive or functional measures, medication status, BMI, eye status when available, and fMRI/imaging findings.

    Penalize fabricated, omitted, or contradicted patient facts.

    \item \textbf{Evidence Grounding}

    Evaluate whether the report appropriately uses external evidence, source traces, validation files, biomarker summaries, treatment summaries, or other evidence-aware reasoning.

    Penalize vague evidence mentions, fabricated citations, unsupported strong claims, or excessive use of ``not specified'' that prevents useful guidance.

    \item \textbf{Biomarker-to-Treatment Logic}

    Evaluate whether the report establishes a coherent patient-specific reasoning chain:

    \medskip

    \texttt{patient-specific fMRI/image finding}

    $\rightarrow$ \texttt{biomarker detection}

    $\rightarrow$ \texttt{biomarker directionality}

    $\rightarrow$ \texttt{treatment mechanism}

    $\rightarrow$ \texttt{clinical readout}

    $\rightarrow$ \texttt{uncertainty boundary}.

    \medskip

    Penalize missing biomarker directionality, weak biomarker-to-treatment links, generic treatment reasoning, unsupported causal claims, or contradicted biomarker directionality.

    \item \textbf{Clinical Actionability}

    Evaluate whether the report provides actionable clinical guidance, including treatment prioritization, protocol or implementation details, outcome measures, feasibility and resource considerations, and explicit decision rules such as continue, stop, switch, adjust, or escalate.

    Vague lists of therapies should receive only partial credit.

    \item \textbf{Personalization}

    Evaluate whether recommendations are tailored using the patient's cognitive profile, symptom burden, functional or adaptive impairment, fMRI findings, and patient-specific constraints.

    Mentioning patient facts without using them in subsequent reasoning should receive only partial credit.

    \item \textbf{Follow-up Monitoring}

    Evaluate whether the report defines meaningful follow-up and monitoring logic, such as safety, feasibility, adherence, early response, main response, maintenance, escalation, treatment switching or adjustment, and repeat imaging or biomarker monitoring.

    \item \textbf{Safety and Uncertainty}

    Evaluate whether the report appropriately addresses safety risks, contraindications, clinician review, informed or guardian consent, uncertainty boundaries, and the distinction between clinical decision support and definitive medical orders.

    Overconfident claims should reduce the score.

\end{enumerate}

\medskip

\textbf{Penalty components:}

\begin{itemize}

    \item \textbf{Contradiction penalty: 0--8}

    Penalize direct contradictions with patient-specific facts.

    \item \textbf{Non-actionable conservatism penalty: 0--12}

    Penalize excessive conservatism, repeated refusal to provide useful guidance, or excessive reliance on statements such as ``not specified'' or ``requires clinician review'' despite sufficient available context.

    \item \textbf{Unsupported claim penalty: 0--10}

    Penalize strong treatment, mechanism, biomarker, or evidence claims that lack adequate support or appropriate uncertainty.

    \item \textbf{Logic weakness penalty: 0--10}

    Penalize weak, missing, generic, or incoherent biomarker-to-treatment reasoning.

\end{itemize}

\medskip

\textbf{Important:}

\begin{itemize}

    \item Assign only the seven positive component scores and four penalty components.

    \item The aggregate EAPGS scores will be computed deterministically by Python.

    \item Return only structured output following the requested schema.

\end{itemize}

\end{tcolorbox}

\begin{tcolorbox}[
    enhanced,
    breakable,
    colback=gray!5,
    colframe=gray!70,
    colbacktitle=blue!60!black,
    coltitle=white,
    title={User prompt for EAPGS-v2-aligned clinical report evaluation},
    fonttitle=\large,
    boxrule=0.8pt,
    arc=2mm,
    left=4mm,
    right=4mm,
    top=3mm,
    bottom=3mm,
    titlerule=0pt
]

\small

Please evaluate the following generated clinical report using EAPGS-v2-aligned criteria.

\medskip

The evaluation should consider the following components:

\begin{itemize}

    \item Patient Fidelity

    \item Evidence Grounding

    \item Biomarker-to-Treatment Logic

    \item Clinical Actionability

    \item Personalization

    \item Follow-up Monitoring

    \item Safety and Uncertainty

    \item Penalties

\end{itemize}

\medskip

\textbf{Important evaluation rules:}

\begin{itemize}

    \item Judge only the current report.

    \item Use the provided patient information, generated report, and external evidence context.

    \item Do not reward generic medical writing unless it contributes to patient-specific EAPGS aspects.

    \item Do not be overly punitive for minor wording differences when the substantive reasoning is present.

    \item Penalize hallucinated patient facts, unsupported strong claims, reversed biomarker directionality, generic treatment guidance, and excessive non-actionable conservatism.

\end{itemize}

\medskip

\hrule

\medskip

\textbf{Report name:}

\{\texttt{report\_name}\}

\medskip

\textbf{Patient information:}

\{\texttt{patient\_txt}\}

\medskip

\textbf{Generated report:}

\{\texttt{report\_text}\}

\medskip

\textbf{External evidence context:}

\{\texttt{evidence\_context}\}

\medskip

\hrule

\medskip

\textbf{Evaluation checklist}

\begin{enumerate}

    \item \textbf{Patient facts}

    Check whether the report accurately represents:

    \begin{itemize}

        \item demographics and diagnosis;

        \item cognitive scores and available clinical scales;

        \item adaptive and functional measures;

        \item medication status, BMI, eye status, and other available clinical information;

        \item fMRI/image impressions, network findings, connectivity findings, and their directionality.

    \end{itemize}

    \item \textbf{Evidence and source grounding}

    Check whether:

    \begin{itemize}

        \item the report uses source traces, evidence summaries, treatment evidence, biomarker evidence, validation evidence, or literature-derived evidence;

        \item evidence is explicitly connected to treatment or biomarker claims;

        \item invented citations and unsupported strong claims are avoided.

    \end{itemize}

    \item \textbf{Biomarker-to-treatment logic}

    Check whether:

    \begin{itemize}

        \item patient-specific fMRI or imaging findings are connected to candidate biomarkers;

        \item biomarker directionality is preserved when available;

        \item the biomarker profile is used to motivate treatment selection;

        \item treatment mechanisms are connected to clinically relevant readouts;

        \item appropriate uncertainty boundaries are provided.

    \end{itemize}

    \item \textbf{Clinical guide quality}

    Check whether:

    \begin{itemize}

        \item treatments are prioritized;

        \item protocols or implementation details are provided;

        \item outcome measures and decision rules are specified;

        \item follow-up monitoring is described;

        \item safety, contraindications, clinician oversight, and uncertainty are appropriately handled.

    \end{itemize}

\end{enumerate}

\medskip

\textbf{Scoring rubric}

\medskip

Use integer scores from 0 to 100 for the seven positive dimensions.

\begin{itemize}

    \item \textbf{0}: absent, irrelevant, or completely incorrect.

    \item \textbf{25}: weak, mostly generic, with major gaps.

    \item \textbf{50}: partially present and useful, but incomplete.

    \item \textbf{75}: mostly strong and clinically useful, with minor or moderate gaps.

    \item \textbf{100}: excellent, explicit, coherent, patient-specific, evidence-aware, actionable, and clinically cautious.

\end{itemize}

\medskip

Assign the following positive scores:

\begin{enumerate}

    \item \texttt{patient\_fidelity}: integer 0--100.

    \item \texttt{evidence\_grounding}: integer 0--100.

    \item \texttt{biomarker\_to\_treatment\_logic}: integer 0--100.

    \item \texttt{clinical\_actionability}: integer 0--100.

    \item \texttt{personalization}: integer 0--100.

    \item \texttt{followup\_monitoring}: integer 0--100.

    \item \texttt{safety\_uncertainty}: integer 0--100.

\end{enumerate}

\medskip

Assign the following penalty scores:

\begin{enumerate}
    \setcounter{enumi}{7}

    \item \texttt{contradiction\_penalty}: integer 0--8.

    Assign a higher penalty when the report directly contradicts patient facts, such as age category, sex or pronouns, diagnosis, medication status, or fMRI findings.

    \item \texttt{non\_actionable\_conservatism\_penalty}: integer 0--12.

    Assign a higher penalty when the report is excessively conservative or repeatedly refuses to provide useful guidance despite sufficient available context.

    \item \texttt{unsupported\_claim\_penalty}: integer 0--10.

    Assign a higher penalty when confident treatment, mechanism, biomarker, or evidence claims lack adequate source support or uncertainty qualification.

    \item \texttt{logic\_weakness\_penalty}: integer 0--10.

    Assign a higher penalty when biomarker-to-treatment links are missing, generic, incoherent, or fail to use patient-specific biomarker directionality.

\end{enumerate}

\medskip

\textbf{Deterministic score calculation}

\medskip

Python subsequently computes:

\[
S_{\mathrm{core}}
=
0.35 S_{\mathrm{evidence}}
+
0.35 S_{\mathrm{logic}}
+
0.30 S_{\mathrm{action}}
\]

\[
S_{\mathrm{weighted}}
=
0.05 S_{\mathrm{patient}}
+
0.25 S_{\mathrm{evidence}}
+
0.40 S_{\mathrm{logic}}
+
0.20 S_{\mathrm{action}}
+
0.05 S_{\mathrm{personalization}}
+
0.05 S_{\mathrm{followup}}
\]

\[
P_{\mathrm{total}}
=
P_{\mathrm{contradiction}}
+
P_{\mathrm{conservatism}}
+
P_{\mathrm{unsupported}}
+
P_{\mathrm{logic}}
\]

\[
S_{\mathrm{EAPGS}}
=
\operatorname{clip}
\left(
S_{\mathrm{weighted}} - P_{\mathrm{total}},
0,
100
\right)
\]

\medskip

\textbf{Required output}

\medskip

Return a structured evaluation containing exactly the following fields:

\begin{itemize}

    \item \texttt{patient\_fidelity}: integer 0--100

    \item \texttt{evidence\_grounding}: integer 0--100

    \item \texttt{biomarker\_to\_treatment\_logic}: integer 0--100

    \item \texttt{clinical\_actionability}: integer 0--100

    \item \texttt{personalization}: integer 0--100

    \item \texttt{followup\_monitoring}: integer 0--100

    \item \texttt{safety\_uncertainty}: integer 0--100

    \item \texttt{contradiction\_penalty}: integer 0--8

    \item \texttt{non\_actionable\_conservatism\_penalty}: integer 0--12

    \item \texttt{unsupported\_claim\_penalty}: integer 0--10

    \item \texttt{logic\_weakness\_penalty}: integer 0--10

    \item \texttt{major\_strengths}: list of 2--5 short strings

    \item \texttt{major\_weaknesses}: list of 2--5 short strings

    \item \texttt{missing\_or\_weak\_eapgs\_aspects}: list of short strings

    \item \texttt{evidence\_concerns}: list of short strings

    \item \texttt{final\_verdict}: one concise paragraph

\end{itemize}

\medskip

\textbf{Output constraints:}

\begin{itemize}

    \item Return only the structured object.

    \item Do not include Markdown.

    \item Do not include additional commentary.

    \item Use specific, evidence-based wording when describing strengths and weaknesses.

    \item If a concern is absent, return an empty list rather than inventing one.

\end{itemize}

\end{tcolorbox}

\section{Evidence-Based AI Patient Guidance Score}

\label{sec:eapgs}

To systematically evaluate the quality of generated patient-specific clinical reports, we developed the Evidence-Augmented Patient Guide Score (EAPGS), a deterministic rubric-based evaluation framework designed to assess whether a report extends beyond summarization of the input electronic health record (EHR) and neuroimaging findings to provide evidence-grounded, mechanistically coherent, and clinically actionable guidance. The evaluator operates on the plain-text content extracted from each generated report and compares it with the corresponding patient profile and, when available, external evidence files.

The EAPGS evaluates seven complementary dimensions: Patient Fidelity, Evidence Grounding, Biomarker-to-Treatment Logic, Clinical Actionability, Personalization, Follow-up Monitoring, and Safety and Uncertainty. Each dimension is scored on a 0--100 scale using predefined rubric criteria, lexical matching rules, patient-specific values, and local contextual matching. Most rubric items are activated when at least one predefined term or patient-specific value is detected. For relational criteria, such as linking a biomarker to its direction of abnormality or connecting a treatment with a plausible mechanism, the evaluator additionally requires sentence-level co-occurrence or proximity within a predefined local text window.

\subsection{Rubric criteria and scoring}

\paragraph{Patient Fidelity.}
This dimension evaluates whether the generated report accurately preserves the demographic, clinical, and neuroimaging information of the individual patient without introducing contradictory patient facts.

The score comprises four components. Demographic and identity consistency contributes 25 points, with 5 points each for correctly representing the subject identifier, age, sex, diagnostic group, and DSM-IV/PDD diagnostic information. Clinical-scale fidelity contributes 35 points and evaluates whether the report incorporates the patient's IQ profile (8 points), ADOS measures (8 points), SRS or SCQ measures (7 points), Vineland adaptive-function measures (8 points), and BMI, medication, or eye-status information (4 points).

Neuroimaging fidelity contributes 25 points and is proportional to the fraction of patient-specific fMRI impression items reproduced in the report:

\begin{equation}
S_{\mathrm{fMRI}}
=
25
\times
\frac{N_{\mathrm{matched\ fMRI\ findings}}}
{N_{\mathrm{patient\ fMRI\ findings}}}.
\end{equation}

The remaining 15 points assess the absence of patient-fact distortion. Direct contradictions involving age, sex, diagnosis, or medication status reduce this component.

\paragraph{Evidence Grounding.}
This dimension evaluates whether the report explicitly uses and traces external evidence rather than presenting unsupported recommendations or generic statements.

Source-trace specificity contributes up to 35 points. Eighteen points are awarded when explicit evidence or provenance terminology is present, including references to sources, evidence, input files, article summaries, treatment or symptom verification files, or structured evidence files. When an evidence directory is provided, explicitly mentioning source filenames contributes 3 points per matched file, up to 12 points. An additional 5 points are awarded for explicit reference to the patient EHR or fMRI impression.

Treatment-evidence use contributes 30 points. Reports receive 18 points for identifying treatment or intervention evidence, including terms related to treatment pathways, recommendations, trials, TMS, CBT, neurofeedback, neuromodulation, rTMS, tDCS, or behavioral therapy. An additional 12 points are awarded when treatment descriptions contain implementation-specific information such as protocol parameters, duration, outcome measures, safety considerations, resource requirements, or trained-personnel requirements.

Biomarker-evidence use contributes 20 points. Ten points are awarded for explicit biomarker or neuroimaging terminology, including fMRI, resting-state functional connectivity, specific networks, or connectivity measures. A further 10 points are awarded when a directional descriptor such as \textit{increased}, \textit{reduced}, \textit{higher}, \textit{lower}, \textit{hyperconnectivity}, or \textit{hypoconnectivity} occurs within 160 characters of a biomarker-related term.

The remaining 15 points assess unsupported evidence claims. A report receives 15 points when no grounding problem is detected, 8 points when one warning flag is identified, and 4 points when two warning flags are identified. Warning flags include claiming evidentiary support without identifiable source-trace language and excessive use of unspecified-source statements.

\paragraph{Biomarker-to-Treatment Logic.}
This dimension evaluates whether the report constructs a traceable reasoning chain from the patient's EHR/fMRI abnormality to a biomarker, supporting evidence, treatment mechanism, measurable clinical readout, and an appropriate uncertainty boundary.

The evaluated reasoning chain can be summarized as

\begin{equation}
\mathrm{Patient\ finding}
\rightarrow
\mathrm{Biomarker}
\rightarrow
\mathrm{Evidence}
\rightarrow
\mathrm{Treatment\ mechanism}
\rightarrow
\mathrm{Clinical\ readout}
\rightarrow
\mathrm{Uncertainty}.
\end{equation}

Five components are evaluated.

\textit{Patient--biomarker alignment (25 points).}
A qualifying sentence must contain both patient-specific information and a biomarker or neuroimaging concept. Two or more qualifying sentences receive 25 points, one qualifying sentence receives 18 points, presence of biomarker terminology without explicit patient-level linkage receives 10 points, and absence receives 0 points.

\textit{Treatment--mechanism bridge (25 points).}
A qualifying sentence must jointly contain a treatment term, a mechanistic or rationale term, and either biomarker- or evidence-related information. Two or more qualifying sentences receive 25 points and one receives 18 points. When these concepts occur within a 220-character local window, 12 points are assigned. Reports that mention both treatment and mechanism without a sufficiently explicit local linkage receive 8 points.

\textit{Biomarker-direction consistency (20 points).}
Sentences linking a biomarker with a directional descriptor are identified. Two or more qualifying sentences receive 20 points, one receives 14 points, local proximity within 160 characters receives 10 points, and a direction term without an explicit biomarker linkage receives 5 points.

\textit{Readout alignment (15 points).}
The rubric evaluates whether treatment or biomarker recommendations are linked to measurable follow-up outcomes, such as clinical response, behavioral or social measures, adherence, functional connectivity changes, or follow-up assessments. Two or more qualifying sentences receive 15 points, one receives 11 points, local proximity within 220 characters receives 8 points, and isolated readout terminology receives 4 points.

\textit{Uncertainty control (15 points).}
Reports containing both actionable content and an appropriate number of uncertainty statements receive 15 points. Actionable reports without explicit uncertainty language receive 10 points. Reports containing both actionable guidance and excessive uncertainty receive 8 points, while uncertainty statements without actionable guidance receive 4 points.

\paragraph{Clinical Actionability.}
This dimension evaluates whether the generated report can function as a structured technical guide for clinical decision support rather than as a descriptive research summary.

Five rubric items are evaluated: treatment prioritization (30 points), protocol or implementation specificity (25 points), outcome-metric specificity (20 points), feasibility and resource guidance (15 points), and clarity of decision rules for continuing, adjusting, escalating, or switching treatment (10 points). Each component receives its full assigned score when at least one corresponding predefined criterion is detected.

\paragraph{Personalization.}
This dimension evaluates whether recommendations are explicitly adapted to the individual patient rather than representing a generic disorder-level report.

Four equally weighted components contribute 25 points each: use of the patient's cognitive profile, use of patient-specific symptom scales, incorporation of patient-specific fMRI findings, and interpretation of functional or adaptive burden. Patient-derived values such as FIQ, VIQ, PIQ, ADOS, SRS, SCQ, and Vineland scores are directly included in the matching criteria.

\paragraph{Follow-up Monitoring.}
This dimension evaluates whether the report translates its recommendations into a longitudinal monitoring strategy.

The rubric evaluates an early feasibility and safety assessment around week 2 (25 points), early-response assessment around week 4 (25 points), primary-response assessment around week 8 (25 points), maintenance, continuation, escalation, or switching decisions around weeks 8--12 (15 points), and repeat imaging or biomarker monitoring (10 points). Each component receives its assigned score when one or more corresponding follow-up criteria are detected.

\paragraph{Safety and Uncertainty.}
This dimension evaluates whether the report appropriately communicates treatment risks, contraindications, and the boundary between evidence-supported guidance and uncertain or exploratory recommendations.

Safety or risk statements contribute 40 points, contraindication awareness contributes 30 points, and uncertainty calibration contributes 30 points. For the uncertainty component, one to 20 uncertainty statements receive 30 points, more than 20 receive 18 points because excessive qualification may reduce actionability, and complete absence of uncertainty language receives 8 points. This dimension is reported separately as a diagnostic quality measure and is not included in the current final EAPGS weighting.

\subsection{Penalty terms}

In addition to positive rubric scores, four explicit penalties are applied to discourage failure modes that could otherwise produce artificially high scores.

\paragraph{Patient-contradiction penalty.}
The patient-contradiction penalty captures direct factual inconsistencies with the input patient profile, including incorrect age category, sex, diagnostic status, or medication status. The penalty is capped at 8 points.

\paragraph{Non-actionable conservatism penalty.}
The non-actionable conservatism penalty detects excessive use of phrases indicating that information is unspecified, requires clinician review, cannot be inferred, or is otherwise unavailable. Ten or more occurrences incur a 3-point penalty, 20 or more incur a 6-point penalty, and 30 or more incur an 8-point penalty. An additional 4-point penalty is applied when no concrete treatment or intervention is proposed. The total penalty is capped at 12 points.

\paragraph{Unsupported-claim penalty.}
A 10-point penalty is applied when confident treatment or mechanistic claims are detected without corresponding source-trace or evidence terminology.

\paragraph{Logic-weakness penalty.}
Weak patient--biomarker alignment incurs a 2-point penalty, a weak treatment--mechanism bridge incurs 3 points, weak biomarker-direction consistency incurs 2 points, and weak readout alignment incurs 2 points. An additional 1-point penalty is assigned when causal language, such as ``causes'', ``will normalize'', or ``will improve'', is used without an adequate uncertainty boundary. This penalty is capped at 10 points.

\subsection{Score aggregation}

To emphasize evidence integration, mechanistic reasoning, and clinical usefulness, an auxiliary Core Agentic Guide Score is calculated as

\begin{equation}
S_{\mathrm{core}}
=
0.35S_{\mathrm{evidence}}
+
0.35S_{\mathrm{logic}}
+
0.30S_{\mathrm{actionability}}.
\end{equation}

The pre-penalty EAPGS is calculated as

\begin{align}
S_{\mathrm{pre}}
={}&
0.10S_{\mathrm{fidelity}}
+
0.30S_{\mathrm{evidence}}
+
0.30S_{\mathrm{logic}}
\nonumber\\
&+
0.20S_{\mathrm{actionability}}
+
0.05S_{\mathrm{personalization}}
+
0.05S_{\mathrm{followup}}.
\end{align}

The Safety and Uncertainty score is retained as a separately reported diagnostic dimension and therefore has a weight of zero in the current composite EAPGS.

The total penalty is defined as

\begin{equation}
P_{\mathrm{total}}
=
P_{\mathrm{contradiction}}
+
P_{\mathrm{conservatism}}
+
P_{\mathrm{unsupported}}
+
P_{\mathrm{logic}}.
\end{equation}

The final EAPGS is then calculated as

\begin{equation}
\mathrm{EAPGS}
=
\max
\left[
0,
\min
\left(
100,\,
S_{\mathrm{pre}} - P_{\mathrm{total}}
\right)
\right].
\end{equation}

\subsection{Score mapping}

Final reports are mapped to qualitative suitability categories using both the overall EAPGS and the Core Agentic Guide Score (Table~\ref{tab:eapgs_mapping}).

\begin{table}[t]
\centering
\caption{Mapping of EAPGS results to qualitative report suitability categories.}
\label{tab:eapgs_mapping}
\small
\begin{tabular}{p{0.38\textwidth} p{0.52\textwidth}}
\toprule
\textbf{Score criterion} & \textbf{Interpretation} \\
\midrule

EAPGS $\geq 80$ and Core Agentic Guide Score $\geq 75$
&
Strong evidence-augmented clinical technical guide
\\

EAPGS $\geq 65$ and Core Agentic Guide Score $\geq 60$
&
Usable evidence-augmented clinical guide with some gaps
\\

Patient Fidelity $\geq 70$ and Core Agentic Guide Score $<45$
&
Mostly patient-text summary; insufficient as an evidence-augmented clinical guide
\\

EAPGS $\geq 50$, when the preceding criteria are not met
&
Partial clinical guide requiring major evidence or logic improvements
\\

EAPGS $<50$
&
Low suitability for a clinical technical guide
\\

\bottomrule
\end{tabular}
\end{table}

\section{Patient-Specificity Assessed by Report Similarity}

\begin{figure}[t]
\centering
\includegraphics[width=0.5\linewidth]{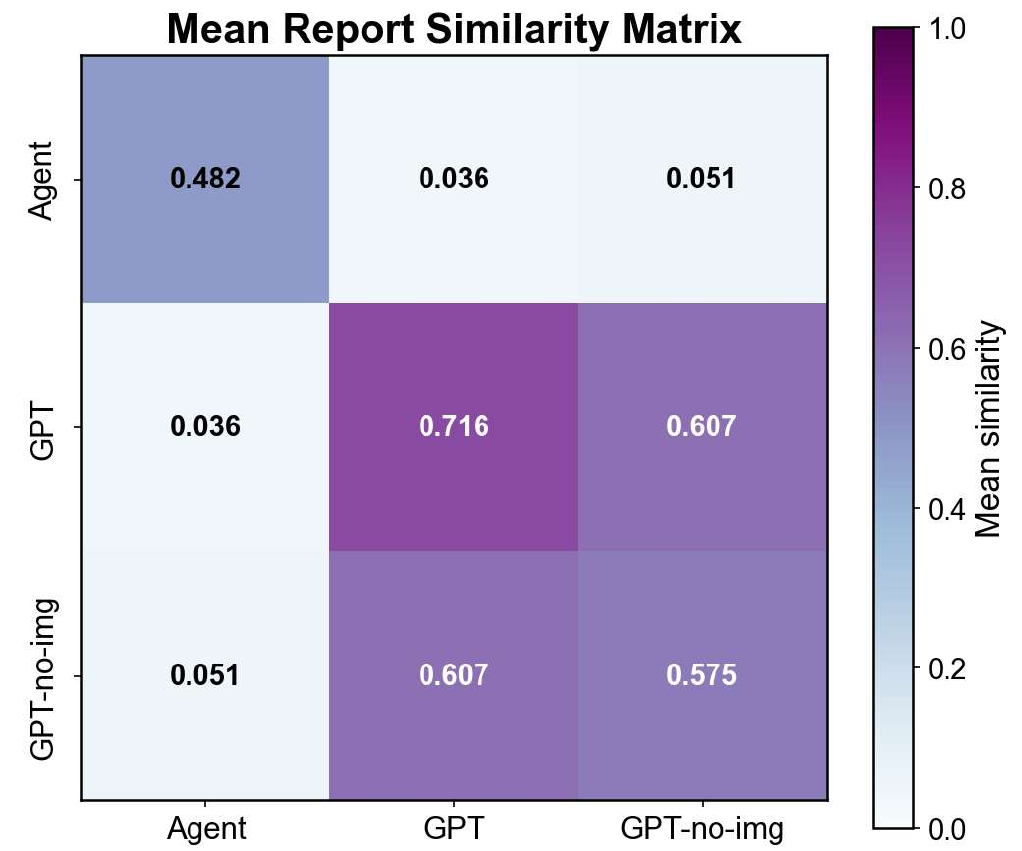}
\caption{
Mean report similarity across three generation methods. Diagonal values indicate within-method similarity, and off-diagonal values indicate cross-method similarity. 
}
\label{fig:report_similarity_matrix}
\end{figure}

Because EAPGS measures report quality but not cross-patient distinctiveness, we additionally assessed report similarity after masking patient identifiers and numerical values (Figure~\ref{fig:report_similarity_matrix}). Within-method similarity was 0.716 for GPT with image impressions and 0.575 for GPT without image impressions, whereas Brain AI Scientist showed the lowest similarity (0.482). Cross-method similarity was 0.607 between the two GPT conditions but only 0.036 and 0.051 between Brain AI Scientist and GPT with and without image impressions, respectively.

The higher within-method similarity of GPT with imaging than GPT without imaging may reflect the constraining effect of the fMRI impression: once a specific set of imaging abnormalities is provided, the range of plausible interpretations and recommendations is narrowed, leading reports to converge around similar imaging-related content. In contrast, GPT without imaging operates over a broader and less constrained response space, which may increase textual variation without necessarily increasing patient specificity. Thus, lower similarity alone should not be interpreted as greater individualization.

Brain AI Scientist combined the lowest within-method similarity with stronger patient-conditioned EAPGS performance. Its workflow allows each patient's clinical profile and imaging abnormalities to shape biomarker identification, evidence retrieval, treatment hypotheses, and validation strategies, producing reports that vary according to the patient rather than through unconstrained generation alone. Together with the paired EAPGS results, these findings support greater individualization of reports generated by Brain AI Scientist compared with the GPT baselines.

\section{Keyword and Semantic Prototype Design for Encoder-Based EAPGS Evaluation}

To provide an interpretable and reproducible encoder-based evaluation of generated patient reports, we defined a set of domain-specific lexical cues and semantic prototype concepts corresponding to the major dimensions of the Evidence-Augmented Patient Guide Score (EAPGS). Importantly, the evaluator does not rely solely on exact keyword matching. Instead, most EAPGS dimensions are assessed through semantic similarity between sentences in the generated report and predefined natural-language prototype statements, whereas literal keyword or phrase matching is used only for selected deterministic filtering and penalty rules.

The report text is first segmented into sentences and encoded using a pretrained SentenceTransformer model. Each semantic prototype is encoded using the same model, and cosine similarity is calculated between prototype embeddings and report-sentence embeddings. The resulting similarity values and coverage measures are used to quantify whether the expected concepts are represented in the generated report.

\subsection{Patient Fidelity and Personalization}

Patient Fidelity assesses whether the generated report preserves patient-specific information contained in the original patient record. Patient-specific semantic queries are dynamically constructed from available demographic, diagnostic, cognitive, clinical-scale, adaptive-function, medication, and imaging information.

The evaluated information includes subject identity, age, sex, diagnostic group, cognitive measures such as FIQ, VIQ, and PIQ, autism-related clinical scales including ADOS, SRS, and SCQ, Vineland adaptive-function scores, medication status, BMI, eye status at scanning, and patient-specific fMRI findings.

Personalization is evaluated using a related set of patient-specific queries that assess whether these characteristics are not merely repeated, but are actively incorporated into clinical interpretation or treatment reasoning. The corresponding concepts include cognitive profile, symptom burden, social-communication impairment, adaptive-function burden, fMRI connectivity abnormalities, network-level biomarkers, and other patient-specific constraints.

\subsection{Evidence Grounding}

Evidence Grounding evaluates whether clinical, biomarker, and treatment claims in the generated report are supported by external evidence or explicit source traces.

Semantic prototypes represent concepts including source traces, evidence summaries, references, biomarker summaries, treatment summaries, clinical trial evidence, article-derived evidence, verified treatment evidence, fMRI evidence, functional-connectivity evidence, brain-network evidence, and external validation.

When external evidence files are available, sentences identified as evidence-related claims are additionally compared directly with evidence chunks using embedding similarity. This allows the evaluator to assess evidence grounding even when the wording of the generated report differs from that of the source material.

\subsection{Biomarker-to-Treatment Logic}

The Biomarker-to-Treatment Logic dimension is decomposed into five semantic components designed to evaluate whether the report constructs a coherent reasoning chain from patient-specific imaging findings to treatment recommendations.

\paragraph{Patient--biomarker alignment.}
This component evaluates whether the patient's clinical profile and fMRI findings are linked to patient-specific candidate biomarkers. Relevant concepts include clinical profile, fMRI findings, functional connectivity, brain-network abnormalities, ADOS, SRS, SCQ, Vineland measures, IQ, and patient-specific biomarkers.

\paragraph{Treatment--mechanism bridge.}
This component evaluates whether specific treatment recommendations are mechanistically connected to the identified biomarker profile. Semantic concepts include treatment mechanism, mechanistic linkage, biomarker profile, functional connectivity, brain-network abnormalities, and whether a treatment is described as targeting, modulating, normalizing, or otherwise addressing a biomarker.

\paragraph{Biomarker-direction consistency.}
This component assesses whether the direction of the identified abnormality is preserved in subsequent interpretation. Representative directional concepts include \textit{increased}, \textit{decreased}, \textit{higher}, \textit{lower}, \textit{hyperconnectivity}, \textit{hypoconnectivity}, \textit{increased connectivity}, and \textit{reduced connectivity}.

\paragraph{Clinical and biomarker readout alignment.}
This component evaluates whether proposed treatments are connected to measurable outcomes. Relevant concepts include clinical outcomes, response criteria, monitoring readouts, biomarker readouts, social communication, attention, adaptive function, behavior, fMRI connectivity, and longitudinal clinical-response assessments.

\paragraph{Uncertainty control.}
This component evaluates whether the report appropriately bounds treatment and mechanistic interpretations. Semantic concepts include uncertainty, uncertainty boundaries, causality, validation, clinician review, clinical decision support, safety considerations, and limitations.

\subsection{Clinical Actionability}

Clinical Actionability evaluates whether the report provides sufficiently specific information to support subsequent clinical decision making. It is represented by five semantic components.

\paragraph{Treatment prioritization.}
Relevant concepts include treatment priority, ranking, suitability, preferred treatment pathway, first-line treatment, second-line treatment, alternative treatment, and research-level interventions.

\paragraph{Protocol and implementation specificity.}
Relevant concepts include treatment protocol, duration, number of sessions, treatment parameters, implementation steps, trained personnel, equipment requirements, and implementation-level guidance.

\paragraph{Outcome-measure specificity.}
Relevant concepts include outcome measures, response criteria, assessment metrics, clinical readouts, biomarker readouts, behavioral outcomes, social communication, attention, adaptive function, and fMRI connectivity.

\paragraph{Feasibility and resource guidance.}
Relevant concepts include feasibility, resources, accessibility, home-based versus center-based implementation, personnel, equipment, local clinical workflow, and alternative implementation strategies.

\paragraph{Decision-rule clarity.}
Relevant concepts include explicit rules to continue, stop, switch, adjust, escalate, maintain, or reassess an intervention when treatment response is insufficient or adverse effects occur.

\subsection{Follow-up Monitoring}

Follow-up Monitoring evaluates whether the generated report specifies how treatment response and safety should be reassessed over time.

The semantic prototypes capture concepts including early feasibility, safety, tolerability, and adherence checks; early treatment-response assessment; primary treatment-response assessment; maintenance, escalation, adjustment, or treatment switching; and repeat imaging or biomarker assessment.

Representative concepts include \textit{week 2}, \textit{week 4}, \textit{week 8}, \textit{week 12}, \textit{early response}, \textit{main treatment response}, \textit{maintenance}, \textit{escalation}, \textit{treatment switch}, \textit{repeat imaging}, \textit{fMRI rescan}, \textit{functional-connectivity monitoring}, and \textit{biomarker monitoring}.

\subsection{Safety and Uncertainty}

Safety and Uncertainty evaluates whether the generated report appropriately addresses risks, limitations, and clinical oversight.

The semantic concepts include safety risks, adverse effects, tolerability, contraindications, risk mitigation, clinician review, informed consent, guardian consent, pediatric safety, trained personnel, local standard operating procedures, uncertainty boundaries, and the distinction between clinical decision support and a definitive medical order.

\subsection{Literal Keyword and Phrase Sets}

In contrast to the semantic dimensions described above, a smaller number of deterministic operations use direct lexical matching.

\paragraph{Evidence-file filtering.}
Evidence files are preferentially retained when their filenames contain one or more of the following terms:

\begin{quote}
\texttt{final\_summary},
\texttt{summary},
\texttt{treatment},
\texttt{symptom},
\texttt{quality},
\texttt{ranking},
\texttt{verified},
\texttt{biomarker},
\texttt{evidence},
\texttt{article},
\texttt{source}.
\end{quote}

These terms are used only to select potentially relevant evidence files before semantic embedding and do not directly determine the final EAPGS score.

\subsection{Contradiction Detection}

Literal phrase matching is used to identify direct contradictions between the generated report and known patient characteristics.

For patients younger than 18 years, age-related contradiction phrases include:

\begin{quote}
\texttt{adult patient},
\texttt{adult autism},
\texttt{adult-focused},
\texttt{adult intervention},
\texttt{adult clinic only}.
\end{quote}

Sex-related contradictions are identified using inconsistent patient descriptions or pronouns, including terms such as:

\begin{quote}
\texttt{male patient},
\texttt{female patient},
\texttt{his symptoms},
\texttt{her symptoms},
\texttt{his treatment},
\texttt{her treatment},
\texttt{he should},
\texttt{she should}.
\end{quote}

For patients with an autism diagnosis, diagnostic contradiction phrases include:

\begin{quote}
\texttt{non-autistic},
\texttt{no autism},
\texttt{without autism},
\texttt{typically developing},
\texttt{control subject}.
\end{quote}

For patients recorded as not currently receiving medication, medication-related contradiction phrases include:

\begin{quote}
\texttt{currently taking medication},
\texttt{on medication},
\texttt{active medication regimen}.
\end{quote}

\subsection{Non-actionable Conservatism}

A deterministic penalty is applied when the generated report repeatedly uses highly conservative language that prevents actionable interpretation despite the availability of relevant information.

Representative phrases include:

\begin{quote}
\texttt{not specified in uploaded evidence},
\texttt{requires clinician review},
\texttt{clinician-defined},
\texttt{not specified},
\texttt{cannot be inferred},
\texttt{not provided},
\texttt{not justified},
\texttt{source file alone},
\texttt{only patient ehr/fmri data are available},
\texttt{treatment selection requires clinician review}.
\end{quote}

The evaluator additionally checks whether the report contains concrete treatment-related content. Representative treatment terms include:

\begin{quote}
\texttt{TMS},
\texttt{transcranial magnetic stimulation},
\texttt{rTMS},
\texttt{tDCS},
\texttt{CBT},
\texttt{cognitive behavioral therapy},
\texttt{neurofeedback},
\texttt{neuromodulation},
\texttt{behavioral therapy},
\texttt{training},
\texttt{intervention},
\texttt{treatment plan}.
\end{quote}

The absence of concrete treatment content contributes to the non-actionable conservatism penalty.

\section{Rule-Based Keyword and Sentence Matching Criteria}

The strict keyword-based EAPGS evaluator uses predefined keyword and phrase sets together with patient-specific values extracted from the input record. In addition to global lexical matching, selected criteria require concepts to occur within the same sentence, adjacent sentences, or a predefined local character window. The keyword and sentence-matching criteria used in the evaluator are summarized below.

\subsection{Patient-Specific Information}

Patient-specific matching incorporates values extracted from the following fields:

\begin{quote}
\texttt{SUB\_ID},
\texttt{AGE\_AT\_SCAN},
\texttt{SEX},
\texttt{DX\_GROUP},
\texttt{PDD\_DSM\_IV\_TR},
\texttt{FIQ},
\texttt{VIQ},
\texttt{PIQ},
\texttt{ADOS\_G\_TOTAL},
\texttt{ADOS\_G\_SOCIAL},
\texttt{ADOS\_2\_TOTAL},
\texttt{ADOS\_2\_SEVERITY\_TOTAL},
\texttt{SRS\_TOTAL\_RAW},
\texttt{SCQ\_TOTAL},
\texttt{VINELAND\_ABC\_Standard},
\texttt{VINELAND\_COMMUNICATION\_STANDARD},
\texttt{VINELAND\_SOCIAL\_STANDARD},
\texttt{CURRENT\_MED\_STATUS},
\texttt{BMI},
and patient-specific fMRI-impression items.
\end{quote}

Additional fixed patient-related terms include:

\begin{quote}
\texttt{patient},
\texttt{subject},
\texttt{ADOS},
\texttt{SRS},
\texttt{SCQ},
\texttt{Vineland},
\texttt{autism},
\texttt{ASD},
\texttt{Aspergers},
\texttt{fMRI impression},
\texttt{cognitive profile},
\texttt{symptom burden},
\texttt{social-function burden},
\texttt{social communication},
\texttt{adaptive functioning},
\texttt{functional burden}.
\end{quote}

\subsection{Evidence and Source-Trace Terms}

Generic source-trace terms include:

\begin{quote}
\texttt{source trace},
\texttt{source:},
\texttt{source},
\texttt{input file},
\texttt{evidence},
\texttt{reference},
\texttt{summary},
\texttt{.json},
\texttt{.md},
\texttt{.csv}.
\end{quote}

Structured source-trace terms include:

\begin{quote}
\texttt{final\_summary},
\texttt{final summary},
\texttt{treatment\_verified},
\texttt{symptom\_verified},
\texttt{quality\_ranking},
\texttt{ranking csv},
\texttt{biomarker summary},
\texttt{treatment summary},
\texttt{symptom validation},
\texttt{biomarker and treatment summaries}.
\end{quote}

Evidence-claim terms include:

\begin{quote}
\texttt{supported by},
\texttt{supported by RCT evidence},
\texttt{evidence shows},
\texttt{RCT},
\texttt{trial},
\texttt{validated},
\texttt{external validation},
\texttt{strong evidence},
\texttt{recommendation},
\texttt{suitability}.
\end{quote}

Actual evidence filenames extracted from the evidence directory are also directly matched against the generated report.

\subsection{Biomarker and Neuroimaging Terms}

Biomarker-related terms include:

\begin{quote}
\texttt{biomarker},
\texttt{fMRI},
\texttt{rsFC},
\texttt{resting-state functional connectivity},
\texttt{functional connectivity},
\texttt{connectivity},
\texttt{brain network},
\texttt{network},
\texttt{dorsal attention},
\texttt{default mode},
\texttt{DAN},
\texttt{DMN},
\texttt{SMG},
\texttt{rPFC},
\texttt{AI},
\texttt{ACC}.
\end{quote}

Patient-specific fMRI-impression items are added to this list dynamically.

Biomarker-direction terms include:

\begin{quote}
\texttt{increased},
\texttt{reduced},
\texttt{decreased},
\texttt{higher},
\texttt{lower},
\texttt{abnormal},
\texttt{excessive},
\texttt{deficit},
\texttt{hyperconnectivity},
\texttt{hypoconnectivity},
\texttt{strengthened},
\texttt{weakened},
\texttt{strengthen},
\texttt{weaken},
\texttt{modulate}.
\end{quote}

For directional biomarker matching, a biomarker-related term and a direction-related term must occur within the same local text window.

\subsection{Treatment and Intervention Terms}

Treatment-related terms include:

\begin{quote}
\texttt{treatment},
\texttt{intervention},
\texttt{therapy},
\texttt{treatment plan},
\texttt{treatment pathway},
\texttt{preferred treatment},
\texttt{TMS},
\texttt{transcranial magnetic stimulation},
\texttt{rTMS},
\texttt{tDCS},
\texttt{CBT},
\texttt{cognitive behavioral therapy},
\texttt{neurofeedback},
\texttt{neuromodulation},
\texttt{behavioral therapy},
\texttt{behavioral},
\texttt{training}.
\end{quote}

Treatment-prioritization terms include:

\begin{quote}
\texttt{tier 1},
\texttt{tier 2},
\texttt{tier 3},
\texttt{preferred treatment pathway},
\texttt{ranked},
\texttt{ranked by suitability},
\texttt{first-line},
\texttt{second-line},
\texttt{priority},
\texttt{preferred option},
\texttt{preferred options},
\texttt{suitability},
\texttt{parallel or conditional},
\texttt{research-level},
\texttt{referral-level},
\texttt{rating}.
\end{quote}

\subsection{Treatment-Mechanism Terms}

Mechanistic-link terms include:

\begin{quote}
\texttt{mechanism},
\texttt{mechanistic},
\texttt{mechanistic bridge},
\texttt{causal chain},
\texttt{rationale},
\texttt{because},
\texttt{due to},
\texttt{therefore},
\texttt{suggests},
\texttt{linked to},
\texttt{associated with},
\texttt{reflects},
\texttt{targets},
\texttt{targeting},
\texttt{modulates},
\texttt{modulate},
\texttt{normalizes},
\texttt{normalize},
\texttt{normalization},
\texttt{addresses},
\texttt{bridge},
\texttt{pathway}.
\end{quote}

For local treatment--mechanism matching, a sentence is considered a candidate reasoning sentence when it contains a treatment term, a mechanistic term, and either a patient-specific biomarker term or an evidence term.

\subsection{Clinical Readout and Monitoring Terms}

Clinical-readout terms include:

\begin{quote}
\texttt{outcome},
\texttt{outcome measures},
\texttt{readout},
\texttt{response},
\texttt{response threshold},
\texttt{assessment},
\texttt{assessment metrics},
\texttt{monitor},
\texttt{monitoring},
\texttt{follow-up},
\texttt{baseline-to-follow-up},
\texttt{functional connectivity changes},
\texttt{resting-state functional connectivity},
\texttt{behavioral},
\texttt{social},
\texttt{social communication},
\texttt{communication},
\texttt{attention},
\texttt{adaptive},
\texttt{adherence}.
\end{quote}

Longitudinal follow-up terms include:

\begin{quote}
\texttt{week 2},
\texttt{feasibility},
\texttt{safety},
\texttt{adherence},
\texttt{initial safety},
\texttt{week 4},
\texttt{early response},
\texttt{week 8},
\texttt{main response},
\texttt{evaluate main response},
\texttt{week 12},
\texttt{weeks 8--12},
\texttt{maintenance},
\texttt{escalation},
\texttt{switch},
\texttt{continuation},
\texttt{adjustment}.
\end{quote}

Imaging and biomarker follow-up terms include:

\begin{quote}
\texttt{rescan},
\texttt{repeat imaging},
\texttt{fMRI},
\texttt{functional connectivity},
\texttt{network-level outcomes},
\texttt{biomarker monitoring},
\texttt{resting-state functional connectivity changes}.
\end{quote}

\subsection{Protocol and Implementation Terms}

Protocol-specific terms include:

\begin{quote}
\texttt{protocol},
\texttt{prescription-level},
\texttt{parameters},
\texttt{duration},
\texttt{session},
\texttt{8 weeks},
\texttt{dose},
\texttt{frequency},
\texttt{structured},
\texttt{trained personnel},
\texttt{specialized equipment},
\texttt{home task},
\texttt{center-based}.
\end{quote}

Feasibility and resource terms include:

\begin{quote}
\texttt{resources},
\texttt{alternatives},
\texttt{accessibility},
\texttt{center-based},
\texttt{home-based},
\texttt{general clinic},
\texttt{requires specialized equipment},
\texttt{trained personnel},
\texttt{local SOP},
\texttt{resource limitation}.
\end{quote}

Decision-rule terms include:

\begin{quote}
\texttt{continue},
\texttt{stop},
\texttt{switch},
\texttt{adjust},
\texttt{escalation},
\texttt{maintenance},
\texttt{insufficient response},
\texttt{early response},
\texttt{main response},
\texttt{clinician-defined baseline-to-follow-up}.
\end{quote}

For strict actionability matching, a concrete treatment term must occur locally with prioritization, protocol, readout, or decision-rule terms rather than appearing elsewhere in the report independently.

\subsection{Safety and Uncertainty Terms}

Safety-related terms include:

\begin{quote}
\texttt{safety},
\texttt{risk},
\texttt{risks},
\texttt{mitigation},
\texttt{adverse effect},
\texttt{adverse events},
\texttt{contraindication},
\texttt{contraindications},
\texttt{tolerability},
\texttt{home monitoring}.
\end{quote}

Clinical-oversight terms include:

\begin{quote}
\texttt{requires clinician review},
\texttt{local SOP},
\texttt{informed consent},
\texttt{guardian consent},
\texttt{pediatric},
\texttt{trained personnel}.
\end{quote}

Uncertainty-related terms include:

\begin{quote}
\texttt{not specified in uploaded evidence},
\texttt{requires clinician review},
\texttt{clinician-defined},
\texttt{not specified},
\texttt{exploratory},
\texttt{observational},
\texttt{hypothesis},
\texttt{not overclaim},
\texttt{requires validation},
\texttt{research-level},
\texttt{clinical decision support},
\texttt{not a final medical order}.
\end{quote}

\subsection{Local Sentence-Level Reasoning Chain}

To reduce false-positive matches caused by isolated keywords, the strict rule-based evaluator additionally searches each sentence together with the immediately following sentence.

A full local reasoning chain requires the local sentence pair to contain:

\begin{quote}
\texttt{patient-specific information}
$\rightarrow$
\texttt{fMRI/biomarker}
$\rightarrow$
\texttt{biomarker direction}
$\rightarrow$
\texttt{treatment}
$\rightarrow$
\texttt{mechanistic link}
$\rightarrow$
\texttt{evidence or clinical readout}.
\end{quote}

The evaluator also searches for the following partial local relationships:

\begin{itemize}
    \item biomarker term + directional term;
    \item biomarker term + treatment term;
    \item treatment term + mechanistic term;
    \item treatment term + evidence term;
    \item treatment term + clinical-readout term.
\end{itemize}

Thus, the rule-based evaluator distinguishes between simple global keyword presence and locally connected statements that more directly represent biomarker-to-treatment reasoning.

\subsection{Contradiction Phrases}

For patients younger than 18 years, age-related contradiction phrases include:

\begin{quote}
\texttt{adult patient},
\texttt{adult autism},
\texttt{adult-focused},
\texttt{adult intervention},
\texttt{adult clinic only}.
\end{quote}

For female patients, inconsistent male-reference phrases include:

\begin{quote}
\texttt{male patient},
\texttt{his symptoms},
\texttt{his treatment},
\texttt{he should},
\texttt{he is}.
\end{quote}

For male patients, inconsistent female-reference phrases include:

\begin{quote}
\texttt{female patient},
\texttt{her symptoms},
\texttt{her treatment},
\texttt{she should},
\texttt{she is}.
\end{quote}

For patients with autism, diagnostic contradiction phrases include:

\begin{quote}
\texttt{non-autistic},
\texttt{no autism},
\texttt{without autism},
\texttt{typically developing},
\texttt{control subject}.
\end{quote}

For patients recorded as not currently receiving medication, contradiction phrases include:

\begin{quote}
\texttt{currently taking medication},
\texttt{on medication},
\texttt{active medication regimen}.
\end{quote}

\subsection{Non-Actionable and Generic Language}

Conservative or non-actionable phrases include:

\begin{quote}
\texttt{not specified in uploaded evidence},
\texttt{requires clinician review},
\texttt{not specified},
\texttt{clinician-defined},
\texttt{cannot be inferred},
\texttt{not provided},
\texttt{not justified},
\texttt{source file alone},
\texttt{only patient ehr/fmri data are available},
\texttt{treatment selection requires clinician review}.
\end{quote}

Generic recommendation phrases include:

\begin{quote}
\texttt{may benefit},
\texttt{should consider},
\texttt{clinician review},
\texttt{general recommendation},
\texttt{standard care},
\texttt{individualized treatment should be considered},
\texttt{not specified},
\texttt{cannot be inferred},
\texttt{not provided},
\texttt{requires clinician review}.
\end{quote}

These phrases are matched together with the presence or absence of patient-specific values and patient-specific fMRI findings to distinguish generic treatment summaries from patient-grounded recommendations.

\subsection{Unsupported and Overconfident Claims}

Confident claim terms include:

\begin{quote}
\texttt{supported by RCT evidence},
\texttt{supported by},
\texttt{evidence shows},
\texttt{validated},
\texttt{external validation},
\texttt{strong evidence},
\texttt{mechanistically plausible},
\texttt{normalizes},
\texttt{modulates},
\texttt{targets},
\texttt{treatment},
\texttt{intervention}.
\end{quote}

The corresponding source-trace terms include:

\begin{quote}
\texttt{source},
\texttt{source trace},
\texttt{evidence},
\texttt{summary},
\texttt{json},
\texttt{csv},
\texttt{md},
\texttt{final\_summary},
\texttt{treatment\_verified},
\texttt{symptom\_verified},
\texttt{biomarker and treatment summaries}.
\end{quote}

Additional causal-overclaim phrases include:

\begin{quote}
\texttt{causal},
\texttt{causes},
\texttt{will normalize},
\texttt{will improve}.
\end{quote}

These expressions are checked jointly with uncertainty-related language to identify confident mechanistic or causal statements that are not appropriately qualified.







\newpage
\bibliographystyle{unsrt}
\bibliography{sn-bibliography}

\end{document}